# Atomistic investigation of deformation and fracture of individual structural components of metal matrix composites

Marcin Maździarz, Szymon Nosewicz*

*Institute of Fundamental Technological Research Polish Academy of Sciences, Pawińskiego 5B, 02-106 Warsaw, Poland*

## Abstract

This paper focuses on the development of the atomistic framework for determining the lower scale mechanical parameters of single components of a metal matrix composite for final application to a micromechanical damage model. Here, the deformation and failure behavior of NiAl–$Al_2O_3$ interfaces and their components, metal and ceramic, are analyzed in depth using molecular statics calculations. A number of atomistic simulations of strength tests, uniaxial tensile, uniaxial compressive and simple shear, have been performed in order to obtain a set of stiffness tensors and strain–stress characteristics up to failure for 30 different crystalline and amorphous systems. Characteristic points on the strain–stress curves in the vicinity of failure are further analyzed at the atomistic level, using local measures of lattice disorder. Numerical results are discussed in the context of composite damage at upper microscopic scale based on images of the fracture surface of NiAl–$Al_2O_3$ composites.

*Corresponding author

*Email address:* snosew@ippt.pan.pl (Szymon Nosewicz)

modelling, Mechanical testing, Molecular Statics

## 1. Introduction

Metal matrix composites (MMCs) are advanced materials that have extensive applications in aerospace, the automotive industry, defense, and various fields of engineering. MMCs can be customized to exhibit exceptional characteristics, including improved performance at high temperatures, high specific strength and stiffness, improved wear resistance, as well as superior thermal and mechanical fatigue properties, surpassing those of alloys that lack reinforcement [1]. Bearing in mind their possible industrial application, their deformation and fracture behavior is one of the most crucial issues in the context of the durability and long-term performance of MMCs [2].

The current state of knowledge points to the necessity of research towards a better understanding of the relations between the microstructure and the material properties of MMCs at different scales [3]. Unlike conventional methods that concentrate on a single scale, multiscale analysis concurrently encompasses models at various scales, sharing the efficiency of macroscopic scales/models with the precision of microscopic ones. Such an approach harnesses the benefits of computations at the lowest scale, exemplified by molecular simulations, which provide insights into atomic-level phenomena over short time intervals, along with macroscopic simulations, which facilitate investigations over significantly extended temporal scales. Atomistic simulations become useful to provide a better understanding of the conditions of the interface and the mechanical properties of the matrix and reinforcement mono- and polycrystals [4].

The most recent studies and scientific challenges encountered in the numerical simulation of metal/oxide interfaces at the nano/micro scale, as well as microscopic analysis related to the microstructure, are nicely reviewed in [5], where nearly 200 papers are cited. However, we will limit ourselves here to systems somewhat similar to those studied in the present paper, in which attention is focused on NiAl/$Al_2O_3$ composite, which belongs to the class of intermetallic-matrix composites reinforced with ceramics, and the mechanical properties of its individual structural components (matrix, ceramic reinforcement, (inter)metallic–ceramic interface) at an atomistic nanoscale.

The structural stability of $(5\times2)\beta-Ni_{1-x}Al_x(110)/(3\times\sqrt{3})Al_2O_3(0001)$ interface and work of separating such interface, pure and alloyed, was examined by an *ab initio* study in [6, 7]. The NiAl(110)/$Al_2O_3$(0001) interface, pure and doped, was uniaxially tensioned by an *ab initio* simulation in [8]. The stress–strain curve determined there made it possible to determine the strength and Young's modulus for this interface. The analysis of the Al(111)/$Al_2O_3$(0001) interface, pure and doped, by an *ab initio* study in [9] allowed concluding that there is no straightforward relation between the interface energy, the work of adhesion, or the tensile properties and the type of termination of corundum, Al or O. An Al-terminated Al/$\alpha$-$Al_2O_3$ interface was uniaxially tensioned using molecular dynamics (MD) simulations in [10]. Using first-principles calculations, the ideal strength and elastic behavior under the tensile and shear loadings of differently oriented ideal NiAl crystals were analyzed in [11]. Monocrystalline NiAl nanowires with different cross-sectional dimensions were uniaxially tensioned along the [100], [110], and [111] orientations by the use of the molecular dynamics simulations. Similar simulations were used to examine the deformation of NiAl nanowires subjected to uniaxial tensile strain at different strain rates and temperatures in [12]. In [13], the Ni/NiAl interfaces were subjected to tensile tests along the [100], [110], and [111] orientations under uniaxial stress conditions using MD simulations. The Hall–Petch relation for nanocrystalline aluminum with different grain sizes was examined by molecular simulations in [14].

Unlike the mentioned papers concerning the modeling of the individual components separately, the present paper focuses on a comprehensive investigation of the deformation and damage behavior of all single structural components of a NiAl–$Al_2O_3$ composite (Fig. 1) within one molecular dynamics framework. This kind of atomistic study allows comparing the numerical results from strength test simulations of the (inter)metallic matrix, the ceramic reinforcement, and, finally, the NiAl–$Al_2O_3$ interface, in the context of a microscopic composite fracture.

Last but not least, this paper aims at determining the parameters for the upper scale modeling atomistically. The elastic properties and strength necessary for the performance of micromechanical models have been evaluated for single structural components in the form of monocrystals and amorphous ones, representing grain boundaries. Simulations of uniaxial tensile, compressive and shear tests have been performed in order to evaluate the corresponding properties. The effect of the crystallographic orientations and their impact on elastic/damage properties have also been studied.

A detailed formulation of the problem, the motivation for this work, and the method of modeling are presented in Section 2.1.

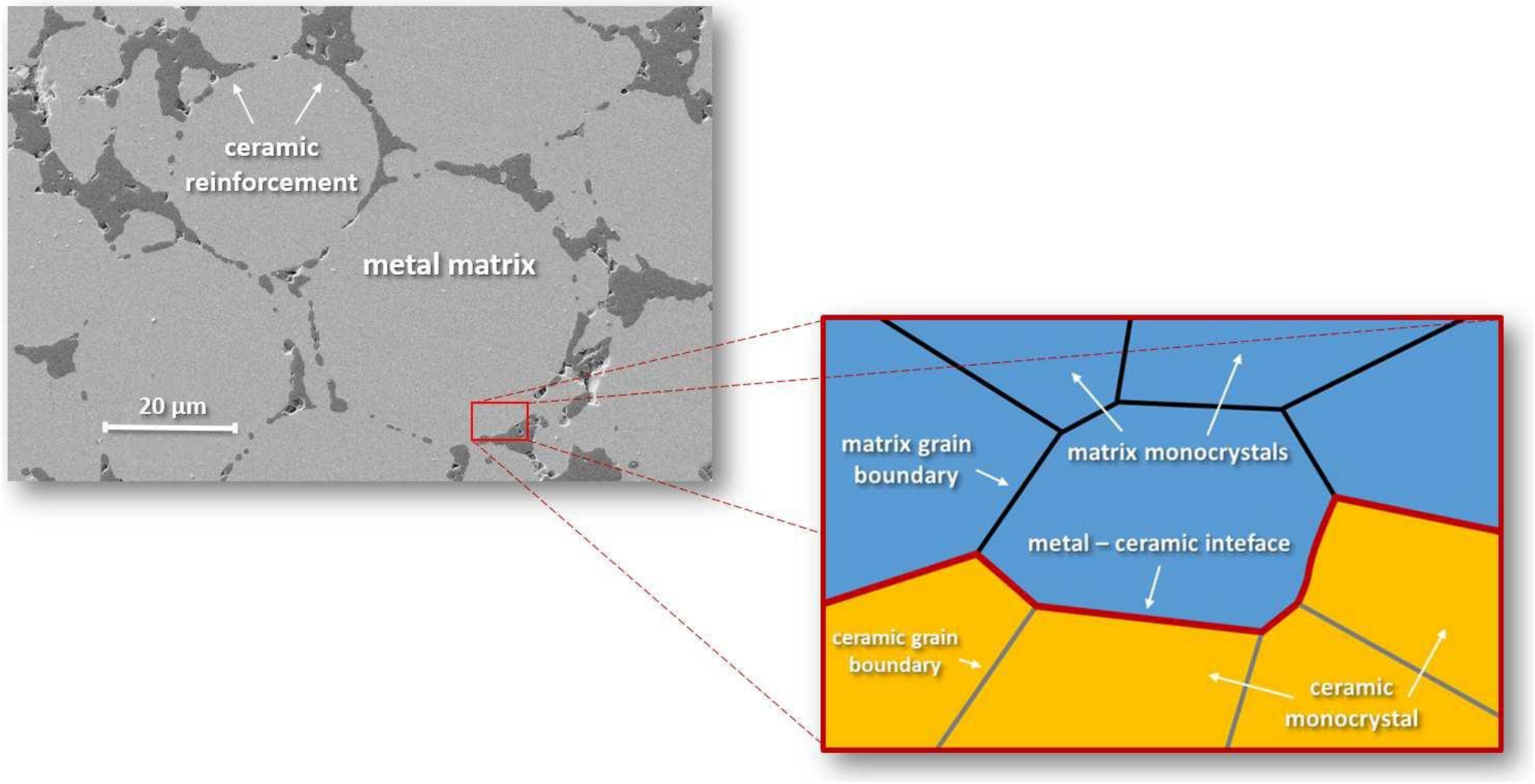

Figure 1: Scanning electron microscopy image of metal matrix composite reinforced by ceramic particles with selection of the main microstructural components.

## 2. Fundamentals of the method

### *2.1. Formulation and motivation of the problem*

#### *2.1.1. Structural components and damage/fracture behavior of metal matrix composites across various length scales*

Following the multiscale description of materials, the macroscopic mechanical properties of MMCs arise from material effects occurring at the microscopic and atomistic scales. The deformation and damage behaviour of particle-reinforced composites depends on the sort of matrix material, as well as on the type, morphology, dimensions, volume fraction, orientation, and spatial distribution of the reinforcing ceramic phase within the composite. In addition to these microscopic attributes, the quality of the interfacial bonding between the metal and ceramic components has significant importance, as highlighted in [15]. In the case of MMCs, various bonding mechanisms come into play, namely: (i) mechanical bonding, (ii) chemically reactive bonding occurring at the interface between the components, (iii) diffusion bonding, and (iv) adhesive bonding. The specific type of bonding exhibited imparts distinctive properties to the interface, consequently influencing the overall characteristics of the composite material [16]. An example of the different types of interface investigated by transmission electron microscopy has been presented in Fig. 2 as representatives of the adhesive and diffusive types of metal–ceramic bonding.

a) b)

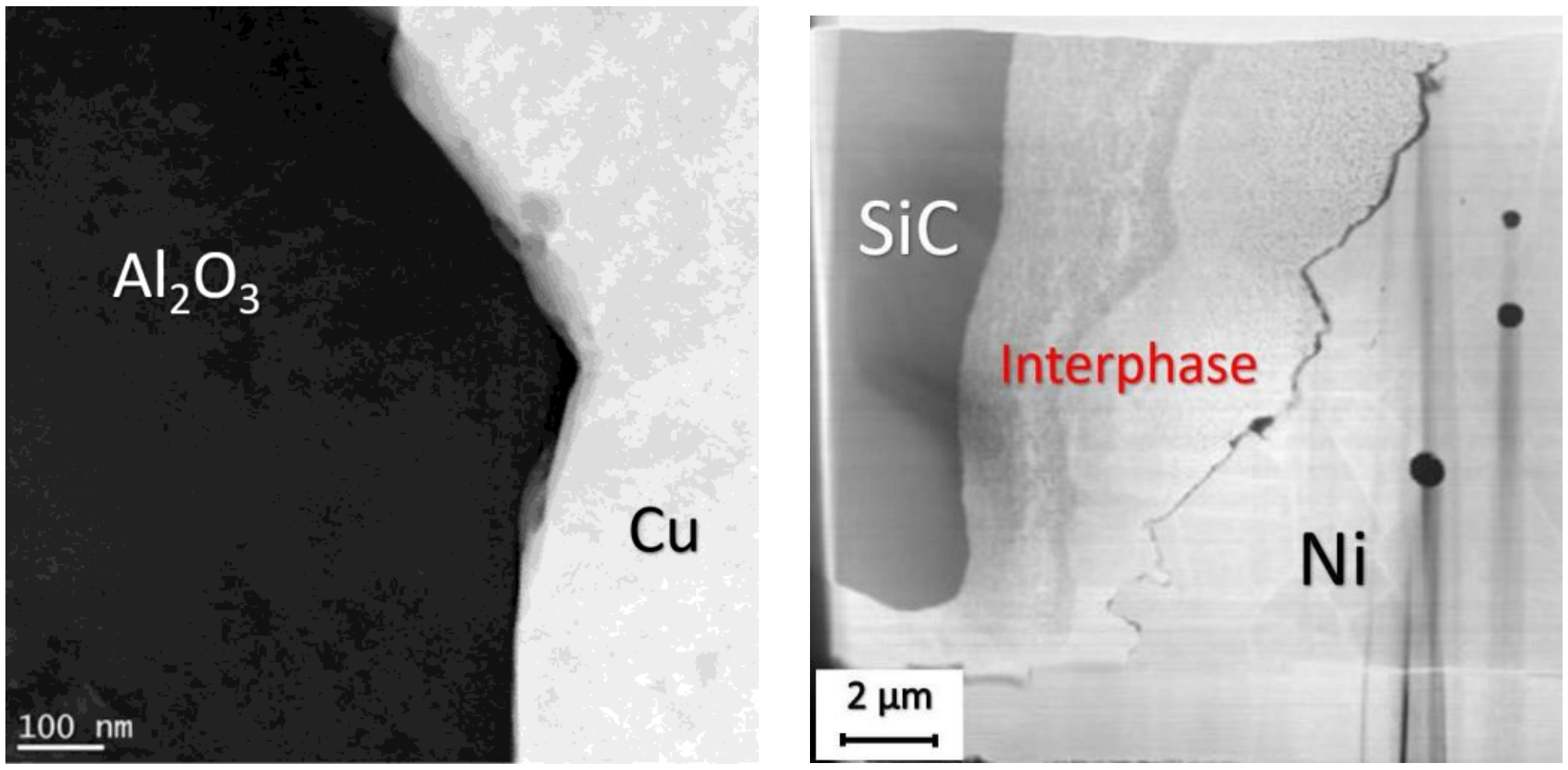

Figure 2: TEM images of various type of metal–ceramic interfaces: a) Cu-$Al_2O_3$ (adhesive) [17], b) Ni-SiC (diffusive) [18].

Depending on the microstructural properties of the composite (the quality of the interface), a variety of damage modes at the microscopic level have been identified for such materials: reinforcement fracture, matrix/reinforcement interface debonding, and matrix cracking [19]. Fracture can occur either in the presence or absence of a substantial amount of plastic strain, depending on the properties of the phases (both the reinforcement and the matrix) and the cohesive strength of the interface between them, indicating the ductile or brittle type of deformation, respectively. The brittle type occurs in composites with a weak cohesion force at the particle/matrix interface [20], a large amount of ceramic reinforcement [21] and/or a brittle matrix [22]. Ductile deformation is characteristic of composites with a plastic matrix with a relatively small amount of reinforcement and relatively good bonding of metal/ceramic particles [21].

Going deeper into the damage characteristics of MMCs, a fracture can occur intergranularly or/and transgranularly, through the main components of each phase—the grain boundaries and/or the grains themselves (Fig. 1). The initial category, which encompasses intergranular fracture, intergranular stress corrosion cracking, fatigue, and liquid metal embrittlement, among others, remains among the most critical challenges in materials engineering, as referenced in [23]. While the specific failure mechanisms may vary with the various forms of intergranular degradation, a shared characteristic among them all is the propagation of damage along the grain boundaries within the material. Intergranular fracture frequently occurs in metals harboring a substantial concentration of brittle particles situated along the grain boundaries. These particles create a pathway for the propagation of a crack, subsequently diminishing the material's fracture toughness and damage tolerance [24].

Transgranular fracture refers to the propagation of a crack through the grains following a pathway with the greatest intensity of stress. In the case of MMCs, when the ceramic reinforcement is under compression, the whole matrix is likely to be under tension, which provides a path for the propagation of a crack through the matrix grains alone (type I transgranular fracture, Fig. 3a) [25]. As the thermal expansion coefficient of the matrix is lower than that of the ceramic phase, the particle reinforcement will be under tensile stress (and the matrix under compressive stress). Thus, the path of propagation of a crack may occur through the reinforcement or along the interface between the matrix and the reinforcement (type II transgranular fracture, Fig. 3b). The fracture mode can also be referred to as *interface failure*.

a) b)

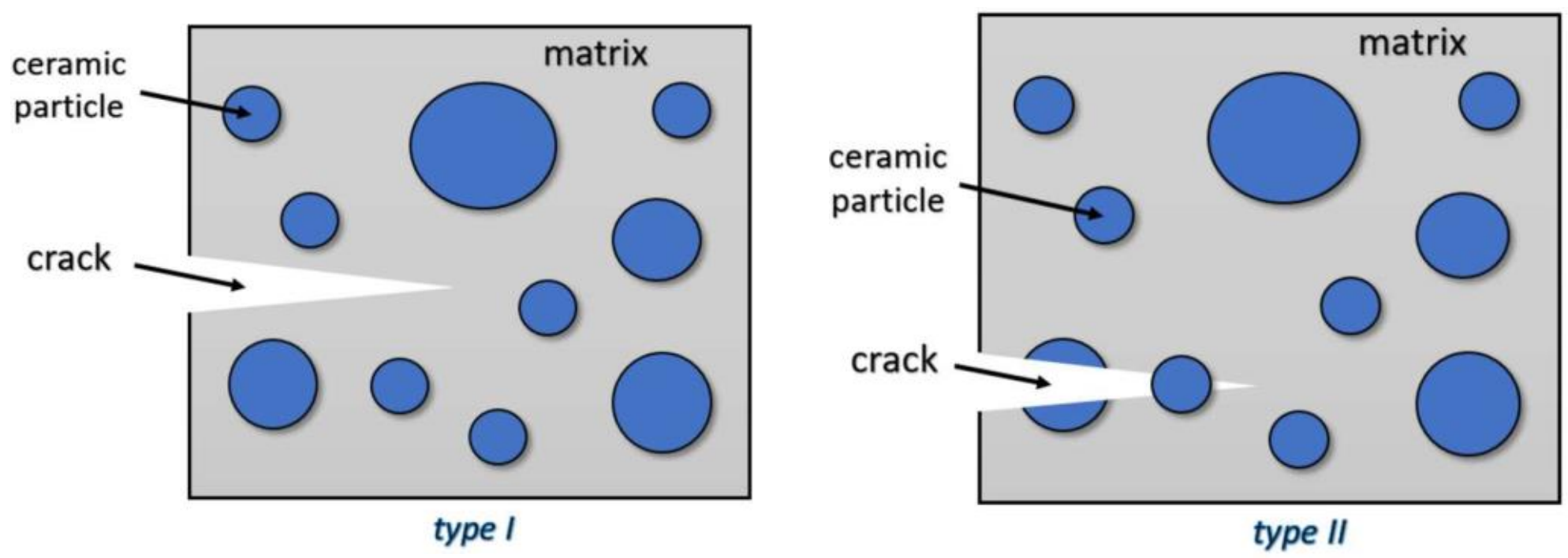

Figure 3: Two types of transgranular fracture mode: a) type I, b) type II.

Finally, the fracture behavior of MMCs has its source at the atomistic scale. The interfaces between two adjacent grains can be categorized into three types: coherent, semi-coherent and incoherent - based on the lattice structures and parameters of the materials involved [26]. The first one typically forms when two metals possess the same lattice structure, such as FCC (Face-Centered Cubic) or BCC (Body-Centered Cubic), and there is a relatively small difference between their lattice parameters. Semi-coherent interfaces are typically observed when both materials on either side have the same lattice type, but the difference in lattice parameters is considerable. Alternatively, a semi-coherent interface may exist when the lattice parameter mismatch is small, but the thickness of each layer exceeds a particular threshold value [27]. Incoherent interfaces are characteristic of two materials with different lattice types, such metal–ceramic bonding. These interfaces usually exhibit low shear strength, earning them the designation of a weak interface. Beyond the atomistic properties of the interface, the macroscopic mechanical properties of MMCs are significantly influenced by factors like the density of the material defects, the grain orientations, and the type of grain boundaries.

#### *2.1.2. Determination of upper scale parameters from atomistic modeling using a multiscale approach*

As was presented above, it is not possible to fully capture the wide range of material effects that occur at various levels of scale during composite deformation using a single-scale approach. Macroscopic models do not explicitly take into account the microstructure of the composite material and describe it by establishing complex constitutive relations, just like the Gurson–Tvergaard–Needleman (GTN) model, which is currently among the primary material damage models employed for assessing the load-bearing capacity of metal engineering structures. [21, 28, 29]. The GTN model is a complicated one, and it needs to be provided with a number of the input mechanical and damage parameters (up to 10), which are usually difficult to estimate for complex materials, such a MMCs.

Alternatively, the deformation and failure of a composite can be modeled by micromechanical discrete models, such as the discrete element method [30, 31, 32]. In the context of the discrete element method (DEM), the fundamental assumption is that a material can be effectively depicted as a collection of rigid particles that interact with each other [33] (Fig. 4). Despite their great capabilities, the primary challenge associated with employing the DEM lies in selecting an appropriate interparticle contact model and determining suitable model parameters that result in the desired macroscopic material behavior. In this context, the contact stiffness and bond strength are typically considered the most critical parameters that influence the pre-critical behavior and eventual failure of the material.

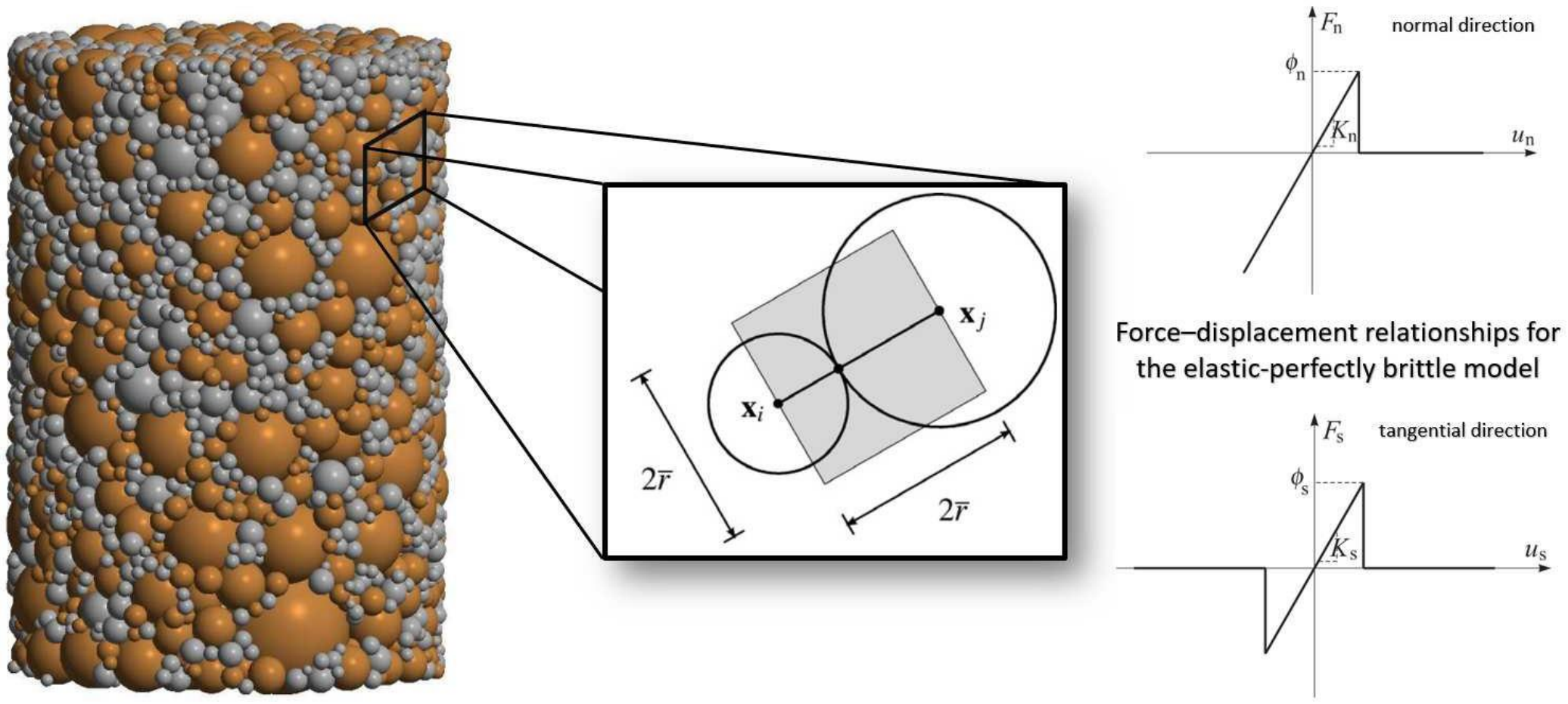

Figure 4: Discrete element framework with contact laws for the normal and tangential direction in the elastic-perfectly brittle model.

Brittle materials (such as NiAl/$Al_2O_3$) can be modeled using the elastic-perfectly brittle model of contact interaction (Fig. 4). Such a model assumes an initial bonding between neighboring particles which can be cracked under excessive load. This feature reproduces an initiation and propagation of material fracture within DEM modeling. As a pair of discrete elements are connected by a bond, the contact forces in both the normal and tangential directions are derived using linear constitutive relations:

$$F_n = K_n u_n \tag{1}$$

$$\backslash F_s \backslash = K_s \backslash u_s \backslash \tag{2}$$

where $K_n$, $K_s$ are the contact stiffnesses in the normal and tangential direction, respectively, $u_n$ is the penetration of the two particles, and $\backslash u_s \backslash$ is the relative displacement at the contact point in the tangential direction.

The contact bond between two discrete elements can be conceptualized as an elongated bar with a certain length $L = 2\bar{r}$ and uniform cross-sectional area $A = (2\bar{r})^2$. Considering the simple geomtrical relations presented, the stiffness modulus $K_n$ is given by the following expression:

$$K_n = 2E_c\bar{r} \tag{3}$$

with the contact stiffness modulus $E_c$ as a certain scaling constant correlated with the Young's modulus of the equivalent continuum material $E$ [33].

If we denote the maximum tensile and shear stresses in the bar connecting a pair of particles by $\sigma_c$ and $\tau_c$, respectively, we can represent the corresponding strengths of the bond as $\varphi_n$ and $\varphi_s$ using the following expression:

$$\varphi_n = \sigma_c(2\bar{r})^2 \tag{4}$$

$$\varphi_s = \tau_c(2\bar{r})^2 \tag{5}$$

Due to the troublesome issue of how to determine the input parameters for micro- and macroscopic models, multiscale numerical modeling has seen widespread application in various scientific and engineering disciplines. In the work presented in the present paper, the numerical analysis at a lower level will provide parametric data to the upper level in a similar way as [34]. The input parameters of the microscopic models (just like DEM) have been determined via a simulation from the lower scale. The molecular dynamics framework provides valuable insights and guidance to describe the deformation of composites at the atomistic scale accurately.

In the present paper, the theory of linear-elastic fracture mechanics (LEFM) has been employed due to the expected brittle/semi-brittle deformation and failure character of NiAl–$Al_2O_3$ composites. LEFM has been successfully applied in qualitatively as well as quantitatively determining fracture properties, such as the fracture strength or fracture toughness [35]. Since the micromechanical model of brittle polycrystal composites based on LEFM assumptions (such as NiAl–$Al_2O_3$) requires the elastic constants and fracture strength for different kinds of contact interaction models, various types of atomistic analyses have been employed.

The deformation and damage behavior of the pure NiAl matrix and the pure ceramic $Al_2O_3$ inclusion have been investigated separately at the atomistic scale using two different states: *monocrystal* and *amorphous*. As the NiAl monocrystals reveal relatively high cubic anisotropy effects [36], the tensile, compressive, and shear properties of the NiAl grains have been evaluated for three different orientations. The final result of the simulation of the NiAl monocrystals is the determination of the elastic stiffness tensor, which can then be transferred to a micromechanical model (such as DEM) as the representation of the elastic behavior of a cubic anisotropic material, similarly to [37]. The linear-elastic stress–strain relation for a material with cubic symmetry characterized by a three-fold rotational symmetry with respect to each of the vectors (1 1 1), (-1 1 1), (1 -1 1) and (1 1 -1) is given in Voigt notation for crystal orientation $X$=[100] $Y$ =[010] $Z$ =[001] by

$$\begin{bmatrix} \sigma_{11} \\ \sigma_{22} \\ \sigma_{33} \\ \sigma_{12} \\ \sigma_{23} \\ \sigma_{13} \end{bmatrix} = \begin{bmatrix} C_{11} & C_{12} & C_{12} & 0 & 0 & 0 \\ C_{12} & C_{11} & C_{12} & 0 & 0 & 0 \\ C_{12} & C_{12} & C_{12} & 0 & 0 & 0 \\ 0 & 0 & 0 & C_{44} & 0 & 0 \\ 0 & 0 & 0 & 0 & C_{44} & 0 \\ 0 & 0 & 0 & 0 & 0 & C_{44} \end{bmatrix} \cdot \begin{bmatrix} \varepsilon_{11} \\ \varepsilon_{22} \\ \varepsilon_{33} \\ 2\varepsilon_{12} \\ 2\varepsilon_{23} \\ 2\varepsilon_{13} \end{bmatrix}, \qquad (6)$$

where $C_{11}$, $C_{12}$ and $C_{44}$ are three independent elastic constants. In the case of an isotropic material, these constants become $C_{11} = 2\mu + \lambda$, $C_{12} = \lambda$ and $C_{44} = \mu$, with $\mu$ and $\lambda$ being the Lamé constants. The degree of deviation of a cubic material's behavior from an isotropic one is frequently quantified by the Zener ratio:

$$Z = \frac{2C_{44}}{C_{11} - C_{12}}, \qquad (7)$$

which is equal to 1 in the isotropic case.

Moreover, in order to evaluate the averaged representation of the fracture characteristics of the grain boundaries of the NiAl matrix and the ceramic $Al_2O_3$, an amorphous sample has been investigated [38], where the aim was to obtain the isotropic averaged response of the NiAl grain boundaries with certain values of the Young's modulus $E$ and fracture strengths $\sigma_c$, $\tau_c$ of each test (tensile, compressive, shear) in order to apply them to a micromechanical model.

Finally, the mechanical properties of the metal–ceramic interface have been determined by simulation of the atomistic strength tests of two bonded monocrystals [39]. Two generated samples representing the real structure of metal and ceramic interface with crystallographic features have been simulated. For the final stage of the atomistic simulations, the mechanical properties of two bonded amorphous crystals (metallic and ceramic) have been simulated. In this way, we can determine the elastic constants and fracture strength parameters of the whole contact interactions within metal–ceramic composites. The details of the atomistic modeling are presented below in Section 2.2.

### *2.2. Computational methods*

Nickel-aluminium (B2-NiAl) alloy crystallizes in the cubic $Pm\bar{3}m$ space group, where the crystallographic axes of the crystal lattice are oriented in the $X$=[100], $Y$=[010] and $Z$=[001] directions, see Fig. 5 a), with lattice constants $a = b = c = 2.93$ Å, when oriented in the $X$=[110], $Y$ =[-110] and $Z$ =[001] directions, see Fig. 5 b), $a = b = 4.14$ Å and $c = 2.93$ Å, when oriented in the $X$ =[111], $Y$ =[-1-12] and $Z$ =[1-10] directions, see Fig. 5 c), $a = 5.075$ Å, $b = 7.177$ Å and $c = 4.14$ Å. For these three orientations of the NiAl monocrystal, the computational region was chosen to be approximately cubic, with volume $V \approx 1500$ Å$^3$. The NiAl amorphous crystal, see Fig. 5 d), was generated using the Atomsk [40] code and the method of generation follows that of [38, 41]. The polycrystal has 128 grains and the sample size is chosen so that structurally it is amorphous, this was achieved at $V \approx 180000$ Å$^3$.

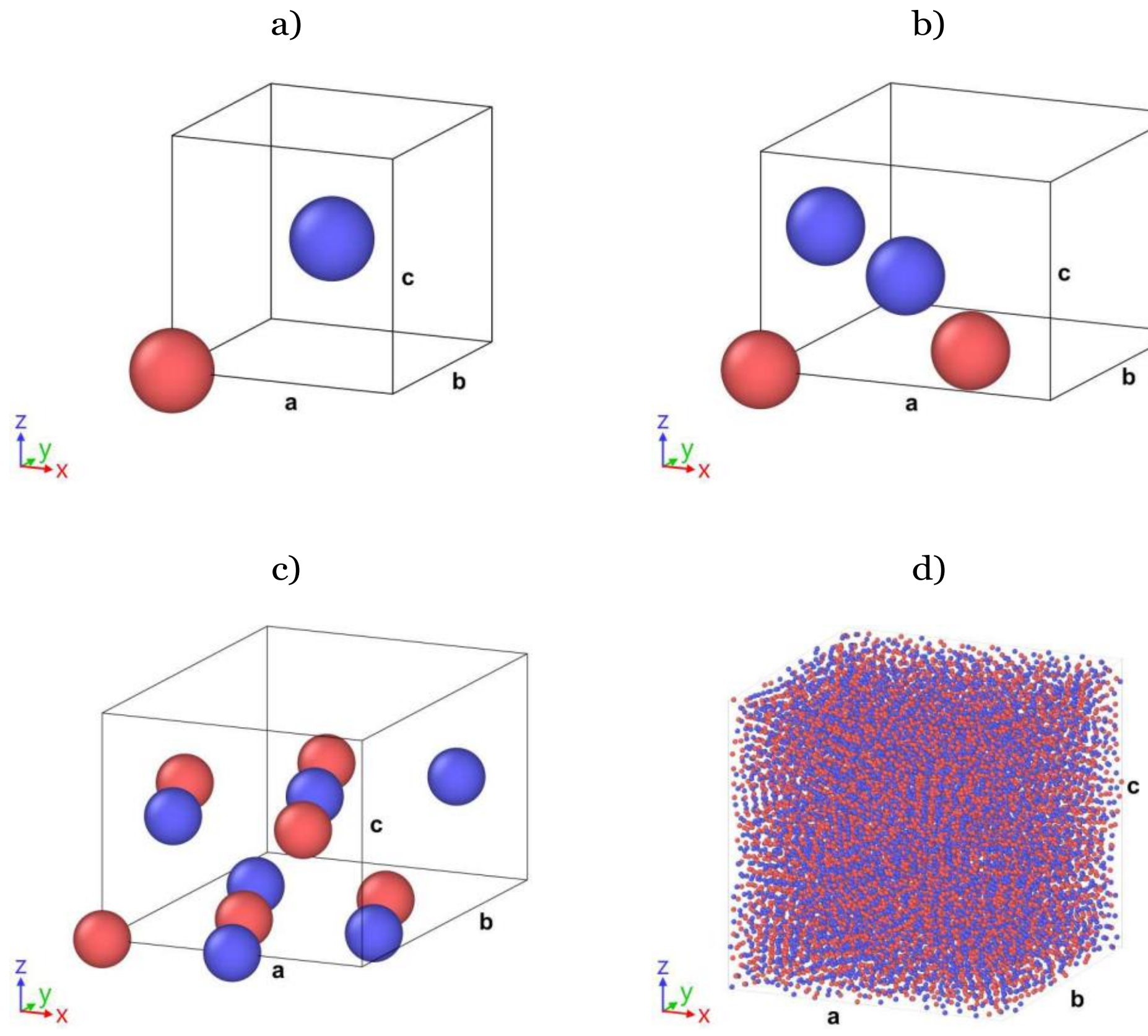

Figure 5: NiAl: a) basic cell X=[100] Y=[010] Z=[001], b) basic cell X=[110] Y=[-110] Z=[001], c) basic cell X=[111] Y=[-1-12] Z=[1-10], d) amorphous (The red and blue atoms represent Ni and Al, respectively).

Corundum is a crystalline form of aluminum oxide ($\alpha$-$Al_2O_3$) and crystallizes in the trigonal R$\bar{3}$c space group; for the convectional unit cell, see Fig. 6 a), $a = b = 4.758$ Å and $c = 12.99$ Å, for the orthorhombic basic cell, see Fig. 6 b), $a = 4.758$ Å, $b = 8.24$ Å and $c = 12.99$ Å. Only one orientation of the $Al_2O_3$ monocrystal was analyzed; the computational sample was approximately cubic with volume $V \approx 1100$ Å$^3$. The generation of the amorphous corundum was carried out similarly to that for NiAl and the approximately cubic region, see Fig. 6 c), has a volume of $V \approx 180000$ Å$^3$.

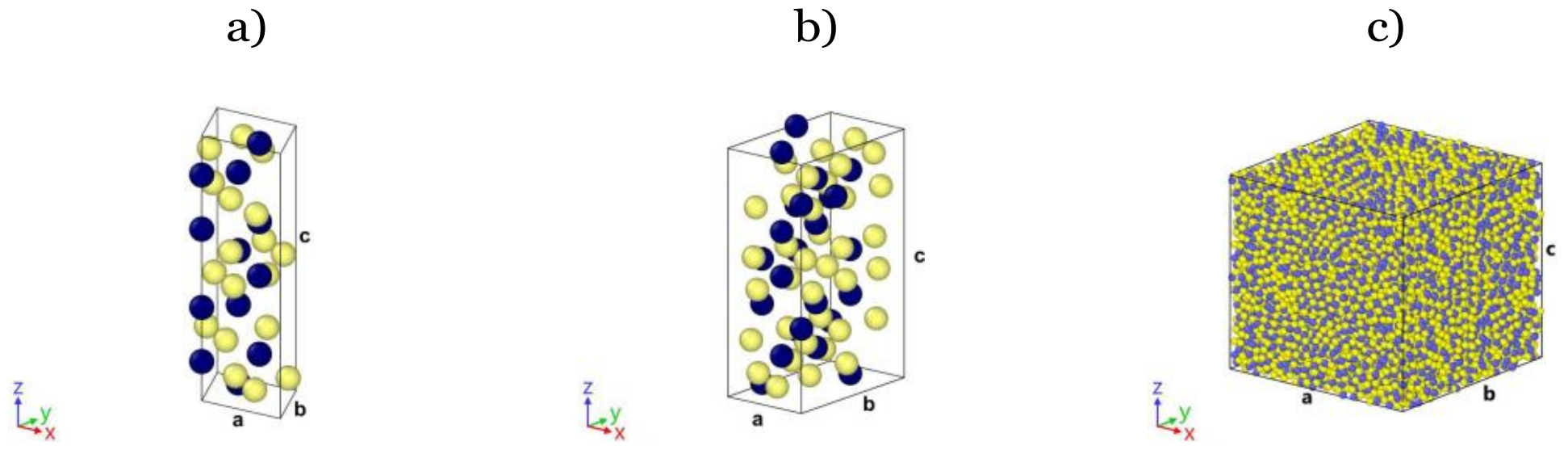

Figure 6: $Al_2O_3$: a) hexagonal, b) orthorhombic basic cell and c) amorphous (The yellow and blue atoms represent O and Al, respectively).

The lattice mismatch between NiAl and $Al_2O_3$ ranges from 15% to 74% depending on the mutual orientation, and it is unreasonable to assume that one lattice will stretch to the other and provide a coherent interface. So, incoherent interfaces were built of such sizes that $N_X \times N_Y \times N_Z$ of the $Al_2O_3$ basic cell equals approximately $M_X \times M_Y \times M_Z$ times the NiAl basic cell. A similar approach but much smaller supercells, i.e. 5×2 for NiAl and $3\times\sqrt{3}$ for $Al_2O_3$, were used for DFT calculations in [6]. This has been obtained for 12×7×4 orthorhombic $Al_2O_3$ basic cells and 20×20×18 NiAl basic cells when $X$ =[100], $Y$ =[010] and $Z$ =[001], see Fig. 7 a), 12×7×4 orthorhombic $Al_2O_3$ basic cells and 14×14×18 NiAl basic cells when $X$ =[110], $Y$ =[-110] and $Z$ =[001], see Fig. 7 b), and 12×7×4 orthorhombic $Al_2O_3$ basic cells and 11×8×13 NiAl basic cells when $X$ =[111], $Y$ =[-1-12] and $Z$ =[1-10], see Fig. 7 c). To achieve an interface between amorphous corundum and amorphous NiAl, the sample was generated similarly to the pure components, except that the height was reduced twice in the $Z$ direction, see Fig. 7 d). All samples are approximately cubic with volume $V \approx 180000\,\text{Å}^3$

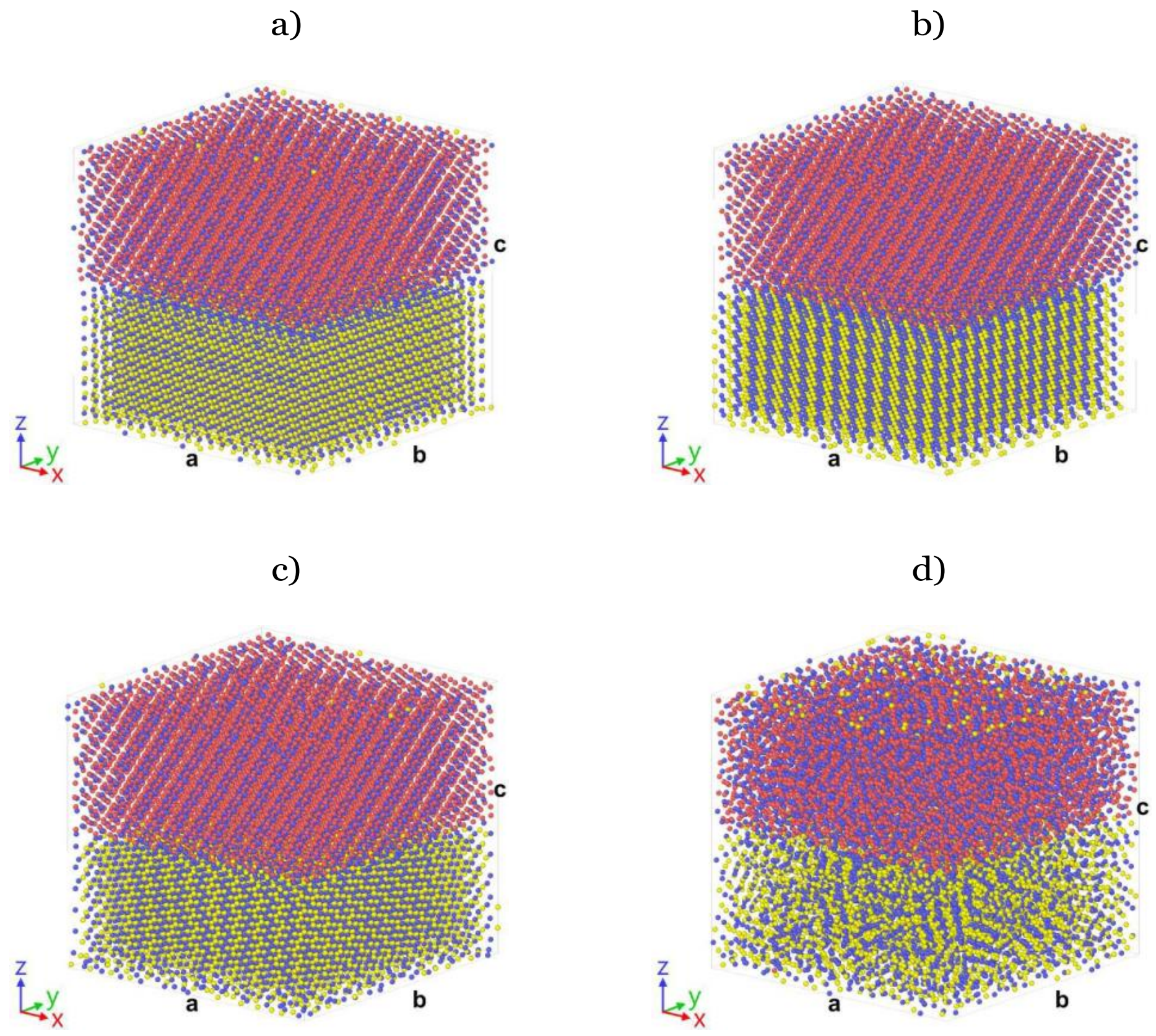

Figure 7: $Al_2O_3$-NiAl: a) 12×7×4 orthorhombic $Al_2O_3$ basic cells and 20×20×18 NiAl basic cells $X$ =[100] $Y$ =[010] $Z$ =[001], b) 12×7×4 orthorhombic $Al_2O_3$ basic cells and 14×14×18 NiAl basic cells $X$=[110] $Y$ =[-110] $Z$ =[001], c) 12×7×4 orthorhombic $Al_2O_3$ basic cells and 11×8×13 NiAl basic cells $X$ =[111] $Y$ =[-1-12] $Z$ =[1-10], d) $Al_2O_3$ amorphous and NiAl amorphous. (The red, yellow and blue atoms represent Ni, O and Al, respectively.)

All molecular statics (MS) [42] simulations were carried out using the Large-scale Atomic/Molecular Massively Parallel Simulator (LAMMPS) [43]. For the Ni–Al system, the embedded-atom method (EAM) potential [44], for $Al_2O_3$ and $Al_2O_3$-NiAl the charge optimized many-body (COMB) potentials [45, 46] were used, respectively.

To obtain the components of the elasticity tensor, $C_{IJ}$ , for all pre-relaxed structures, the stress–strain method with a maximum strain amplitude of $10^{-4}$ was employed [43, 47]. The isotropised bulk modulus $B$, the shear modulus $G$, Young's modulus $E$, and Poisson's ratio $\nu$ were determined using a Voigt–Reuss–Hill average [48], whereas the universal elastic anisotropy index $A^U$ was calculated according to [49].

To obtain stress–strain profiles, three numerical molecular homogeneous deformation tests were performed using the MS approach [50]: these selected tests are, namely, the uniaxial strain (US) in $Z$ direction, simple shear (SS) in the $XZ$ direction and in the $YZ$ direction. If we analyze the components of the composite, NiAl and $Al_2O_3$, separately, we keep their orientations as in the composite. Each test was divided into 50 steps and the results were recorded after minimizing the energy and the forces. The deformation gradient F for uniaxial strain in the $Z$ direction without perpendicular deformations is defined by

$$F_Z^{US} \rightarrow \begin{bmatrix} 1 & 0 & 0 \\ 0 & 1 & 0 \\ 0 & 0 & \lambda \end{bmatrix}, \tag{8}$$

where $\lambda = L/L_0$ is the principal stretch/compression ratio. The simulation box was stretched by 40%, returned along the same path to the initial configuration, then compressed 40%, and again returned along the same path to the initial configuration.

The deformation gradient F for simple shear in the $XZ$ direction can be written as

$$F^{SS}_{XZ} \rightarrow \begin{bmatrix} 1 & 0 & \gamma \\ 0 & 1 & 0 \\ 0 & 0 & 1 \end{bmatrix}, \tag{9}$$

whereas the deformation gradient F for simple shear in the $YZ$ direction is

$$F^{SS}_{YZ} \rightarrow \begin{bmatrix} 1 & 0 & 0 \\ 0 & 1 & \gamma \\ 0 & 0 & 1 \end{bmatrix}, \tag{10}$$

where $\gamma = \tan(\Phi)$ and $\Phi$ is the angular change. The simulation box was sheared by $\gamma$=40% and returned along the same path to the initial configuration.

Since the deformations used in the simulations are significant, the Biot strain tensor, $E_{Biot} = (F^T F)^{1/2} - I$, is used in the figures, it provides a correct description of the finite deformations and at the same time is the closest to the small strain tensor $\varepsilon$, see [51].

To visualize the studied structures on an atomistic level, the OVITO [52] program was used. To measure the local lattice disorder, the cohesive energy per atom ($E_c$/atom) and the centrosymmetry parameter (CSP) [43] were used.

## 3. Numerical results and discussion

All the results obtained for the 30 (4×NiAl×3 + 2×$Al_2O_3$×3 + 4×$Al_2O_3$-NiAl×3) simulations are available in the Appendix A. The findings include the determined initial stiffness tensors and stress (Cauchy) – strain (Biot) curves for uniaxial tensile-compressive, simple shear in the $XZ$ direction and simple shear in the $YZ$ direction tests.

Among these tests, we selected those for which pronounced damage was obtained and additionally analyzed them at the atomistic level. These selected results are presented below.

### *3.1. Metal matrix*

To assess the reliability of the results obtained, we will compare those obtained here with those available from other authors. Analyzing the stress–strain relations for NiAl in Figs. 8a)-c), it can be seen that the behavior of the material greatly depends on the orientation of the crystal and whether it is crystalline or amorphous. Similar observations have been made by other authors as well [11]. Thus, from *ab initio* calculations, it came out that during NiAl stretching for different crystal orientations, the ideal strength varies between 17.3 GPa and 24.9 GPa. The results obtained here are quite similar, with maximum axial tensile stress ranging from about 15 GPa to 30 GPa, see Fig. 8a). It is interesting to note that the crystal that is the stiffest in tension, i.e., orientation Fig. 5c), is the least stiff in shear.

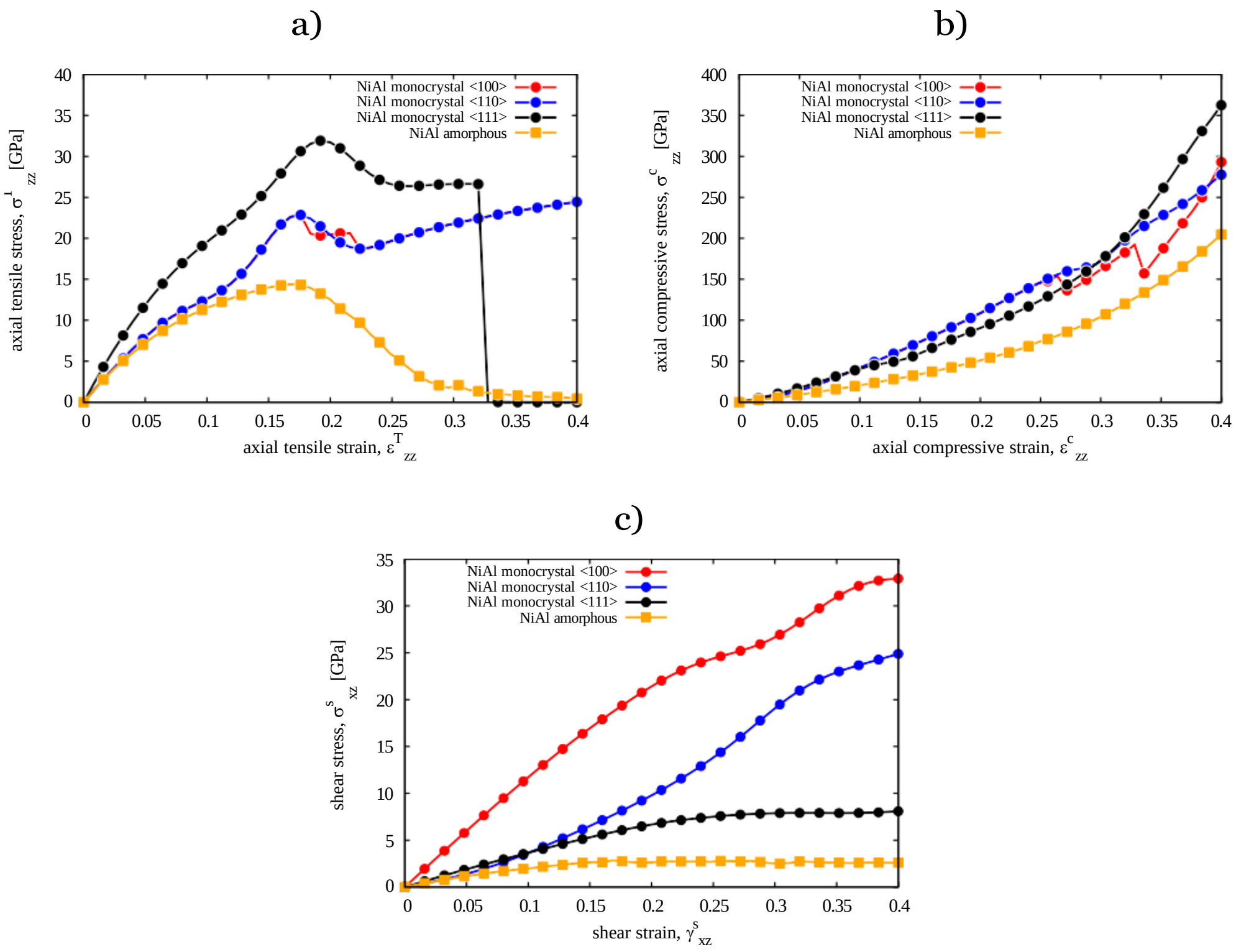

Figure 8: Stress (Cauchy) - strain (Biot) results of: a) uniaxial tensile test, b) uniaxial compressive test and c) simple shear test of NiAl monocrystal and amorphous.

Amorphous NiAl has about half the tensile strength of the strongest crystalline NiAl, while its shear strength is up to ten times lower. During tension we have not only a quantitative but also a qualitative difference between the behavior of crystalline and amorphous NiAl. We try to explain this difference at the atomistic level. An analysis of Fig. 9 for the NiAl monocrystal shows that between the two deformation steps there is a stepwise but uniform increase in the cohesive energy per atom and the centrosymmetry parameter, with bonds breaking uniformly across the section. Moreover, the sudden drop of stress has been registered in the stress-strain curve, which refers to the fracture with brittle manner. For amorphous NiAl, Fig. 10, the cohesive energy per atom and CSP also increase but not suddenly; in cross-section, the bonds break gradually. This effects has been also revealed by the stress-strain curve, which can be characterized by relatively long range of softening regime. This explains why we have brittle behavior in one case and more ductile behavior in the other.

The stiffness tensors for monocrystalline and amorphous NiAl are collected in Tables 11–15. The present calculated elastic constants of NiAl depicted in Fig. 5 are in pretty good agreement with those coming from *ab initio* calculations. For the first orientation of the monocrystal, see Fig. 5a), we obtained the following elastic constants: $C_{11}$ = 190.87 GPa, $C_{12}$ = 142.91 GPa and $C_{44}$ = 121.49 GPa, see Table 11. We see that the difference here does not exceed 10% relative to those determined from *ab initio* calculations in [11], i.e., $C_{11}$=208.2 GPa, $C_{12}$=134.5 GPa and $C_{44}$=118.4 GPa. This confirms, of course, the good quality of the interatomic potential used, but also the correctness of our molecular statics calculations. NiAl monocrystal is strongly anisotropic, i.e., the universal elastic anisotropy index $A^U$ = 3.92, so naturally the representations of the stiffness tensor must differ, see Tables 11–15. However, when we calculate for these three orientations such quantities as the isotropised bulk $B$, the shear $G$, Young's modulus $E$, Poisson's ratio $\nu$ and $A^U$ , we see that they are identical, i.e., $B$ = 158.90 GPa, $G$ = 64.37 GPa, $E$ = 170.13 GPa, $\nu$ = 0.32 and $A^U$ = 3.92. It is worth mentioning here that these values are in good agreement with those from experiments with polycrystalline NiAl [53], where $B$ = 163 GPa, $G$ = 71 GPa, $E$ = 186 GPa, $\nu$ = 0.31. We will next look at the elastic properties of amorphous NiAl depicted in Fig. 5d). The elasticity tensor, see Table 14, is nearly isotropic with $A^U$=0.32. Analyzing the isotropised moduli, i.e., $B$ = 135.30 GPa, $G$ = 25.24 GPa and $E$ = 71.28 GPa, we see that they are lower than those for monocrystalline NiAl, while $\nu$= 0.41 has increased.

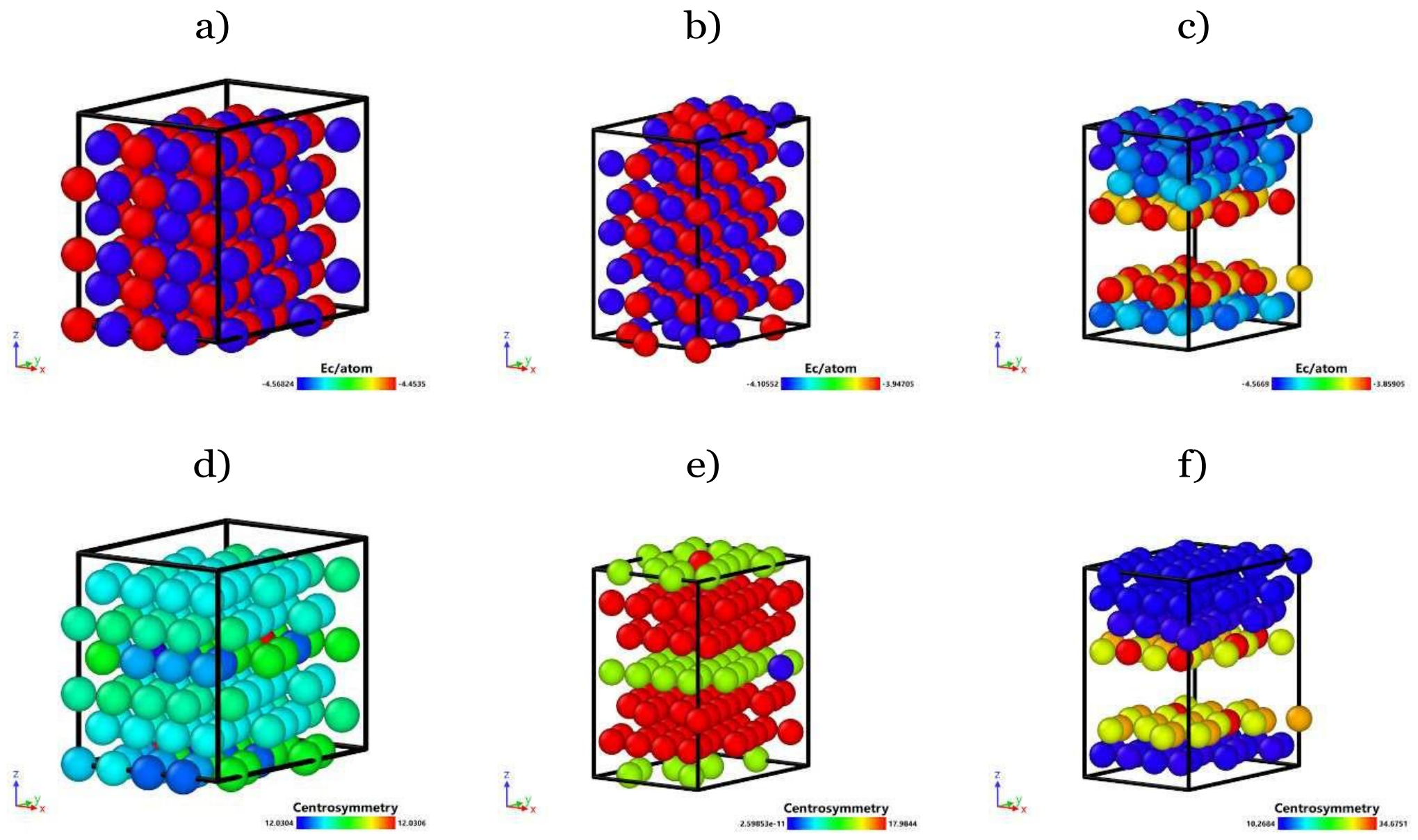

Figure 9: NiAl monocrystal X=[111] Y=[-1-12] Z=[1-10]: a) the cohesive energy per atom ($E_c$/atom) for $\varepsilon_{zz}$=0, b) $E_c$/atom for $\varepsilon_{zz}$=0.312, c) $E_c$/atom for $\varepsilon_{zz}$=0.328, d) the centrosymmetry parameter (CSP) for $\varepsilon_{zz}$=0, e) CSP for $\varepsilon_{zz}$=0.312 and f) CSP for $\varepsilon_{zz}$=0.328.

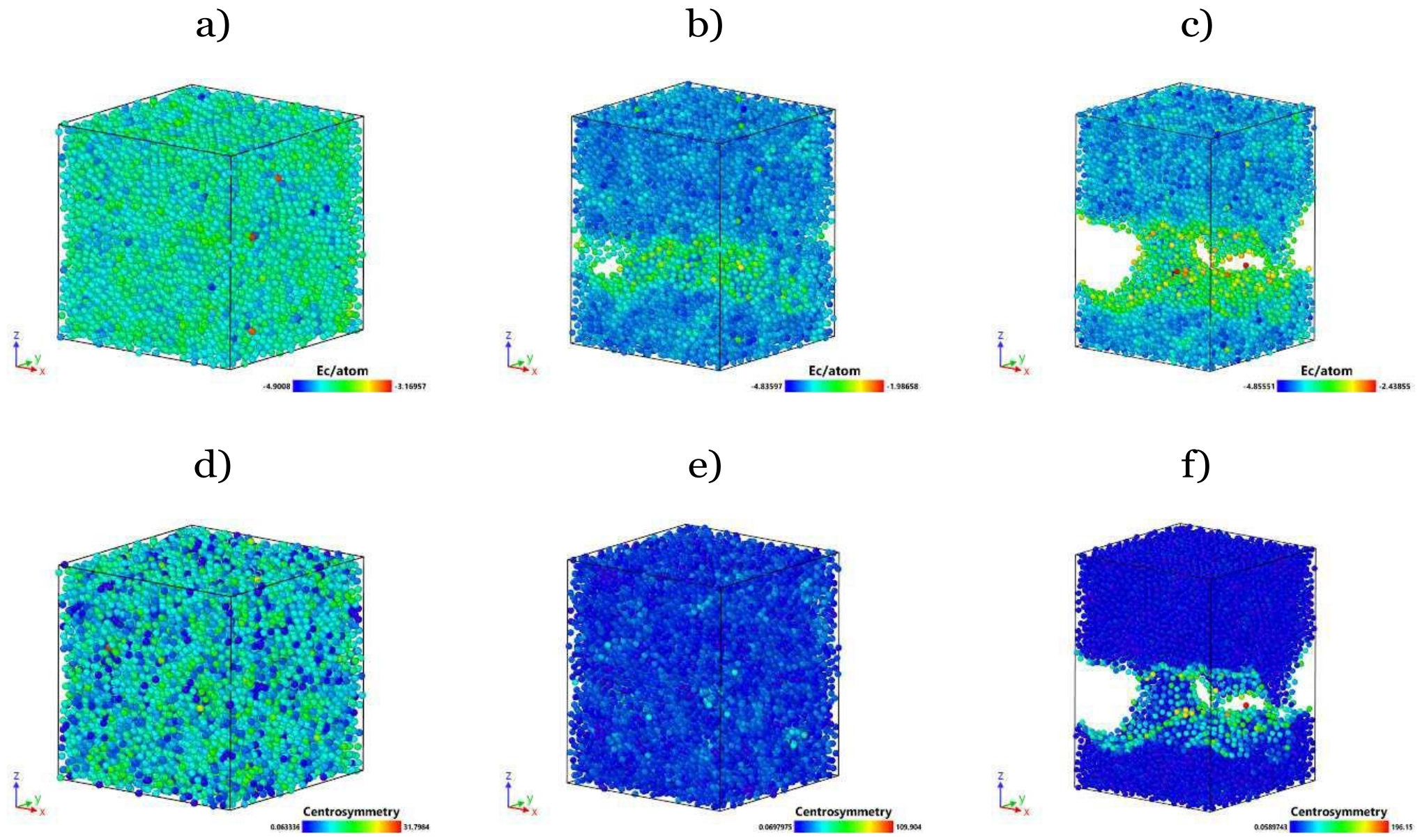

Figure 10: NiAl amorphous: a) the cohesive energy per atom ($E_c$/atom) for $\varepsilon_{zz}$=0, b) $E_c$/atom for $\varepsilon_{zz}$=0.24, c) $E_c$/atom for $\varepsilon_{zz}$=0.4, d) the centrosymmetry parameter (CSP) for $\varepsilon_{zz}$=0, e) CSP for $\varepsilon_{zz}$=0.184 and f) CSP for $\varepsilon_{zz}$=0.4.

- Stiffness tensor: NiAl oriented $X$ =[100] $Y$ =[010] $Z$ =[001]

$$[C_{IJ}] \rightarrow \begin{pmatrix} 190.87 & 142.91 & 142.91 & 0. & 0. & 0. \\ 142.91 & 190.87 & 142.91 & 0. & 0. & 0. \\ 142.91 & 142.91 & 190.87 & 0. & 0. & 0. \\ 0. & 0. & 0. & 121.49 & 0. & 0. \\ 0. & 0. & 0. & 0. & 121.49 & 0. \\ 0. & 0. & 0. & 0. & 0. & 121.49 \end{pmatrix} \text{[GPa]}. \tag{11}$$

$B$ = 158.90 GPa, $G$ = 64.37 GPa, $E$ = 170.13 GPa, $\nu$ = 0.32 and $A^U$ = 3.92.

- Stiffness tensor: NiAl oriented $X$ =[110] $Y$ =[-110] $Z$ =[001]

$$
[C_{IJ}] \rightarrow \begin{bmatrix} 288.37 & 45.40 & 142.91 & 0. & 0. & 0. \\ 45.40 & 288.37 & 142.91 & 0. & 0. & 0. \\ 142.91 & 142.91 & 190.87 & 0. & 0. & 0. \\ 0. & 0. & 0. & 121.49 & 0. & 0. \\ 0. & 0. & 0. & 0. & 121.49 & 0. \\ 0. & 0. & 0. & 0. & 0. & 23.98 \end{bmatrix} \text{[GPa]}. \qquad (12)
$$

$B$ = 158.90 GPa, $G$ = 64.37 GPa, $E$ = 170.13 GPa, $\nu$ = 0.32 and $A^U$ = 3.92.

- Stiffness tensor: NiAl oriented $X$ =[111] $Y$ =[-1-12] $Z$ =[1-10]

$$
[C_{IJ}] \rightarrow \begin{bmatrix} 320.88 & 77.91 & 77.91 & 0. & 0. & 0. \\ 77.91 & 288.38 & 110.41 & 0. & 0. & -45.96 \\ 77.91 & 110.41 & 288.37 & 0. & 0. & 45.96 \\ 0. & 0. & 0. & 88.98 & 45.96 & 0. \\ 0. & 0. & 0. & 45.96 & 56.48 & 0. \\ 0. & -45.96 & 45.96 & 0. & 0. & 56.48 \end{bmatrix} \text{[GPa]}. \qquad (13)
$$

$B$ = 158.90 GPa, $G$ = 64.37 GPa, $E$ = 170.13 GPa, $\nu$ = 0.32 and $A^U$ = 3.92.

- Stiffness tensor: NiAl amorphous direct simulation result:

$$[C_{IJ}] \rightarrow \begin{bmatrix} 170.03 & 115.17 & 126.48 & -5.17 & -0.08 & 7.10 \\ 115.17 & 168.41 & 114.76 & 1.26 & -5.85 & 2.99 \\ 126.48 & 114.76 & 171.48 & 1.69 & 0.18 & 1.50 \\ -5.17 & 1.26 & 1.69 & 23.57 & -2.85 & 2.53 \\ -0.08 & -5.85 & 0.18 & -2.85 & 32.26 & 0.45 \\ 7.10 & 2.99 & 1.50 & 2.53 & 0.45 & 23.00 \end{bmatrix} \text{[GPa]}. \quad (14)$$

$B$ = 135.30 GPa, $G$ = 25.24 GPa, $E$ = 71.28 GPa, $\nu$ = 0.41 and $A^U$ = 0.32,

reduction to isotropy:

$$[C_{IJ}] \rightarrow \begin{bmatrix} 169.97 & 118.80 & 118.80 & 0. & 0. & 0. \\ 118.80 & 169.97 & 118.80 & 0. & 0. & 0. \\ 118.80 & 118.80 & 169.97 & 0. & 0. & 0. \\ 0. & 0. & 0. & 26.28 & 0. & 0. \\ 0. & 0. & 0. & 0. & 26.28 & 0. \\ 0. & 0. & 0. & 0. & 0. & 26.28 \end{bmatrix} \text{[GPa]}. \quad (15)$$

$B$ = 135.86 GPa, $G$ = 26.00 GPa, $E$ = 73.32 GPa, $\nu$ = 0.41 and $A^U$ = 0.00085.

### *3.2. Ceramic reinforcement*

By analyzing the stress–strain relations for $Al_2O_3$ in Figs. 11a)–c) it can be seen that the behavior of the material is greatly affected by whether it is crystalline or amorphous. Only for amorphous $Al_2O_3$, depicted in Fig. 6c), and during tension, was damage of the material obtained, see Fig. 11a): the maximum stress was about 60 GPa at a strain of 0.22. Similar results from molecular calculations in the uniaxial tensile test, i.e., 50.7±4.4 GPa at a strain of 0.24, were obtained in [54]. When compressing an $Al_2O_3$ monocrystal, for $\varepsilon_{ZZ}^{C} \approx 0.3$ we have a stress jump, see Fig. 11b). For such a shortening of the lattice constant $c$, we are most likely dealing with a phase transformation from hexagonal $\alpha$-$Al_2O_3$ to $\delta$-$Al_2O_3$, see [55]. Atomistic analysis of the damage of amorphous $Al_2O_3$ shows a significant increase in the cohesive energy per atom and the centrosymmetry parameter for a strain around $\varepsilon_{ZZ}^{T} \approx 0.24$ and gradual bond breakage, see 12a)-f). Again, for an amorphous crystal, we have ductile behavior.

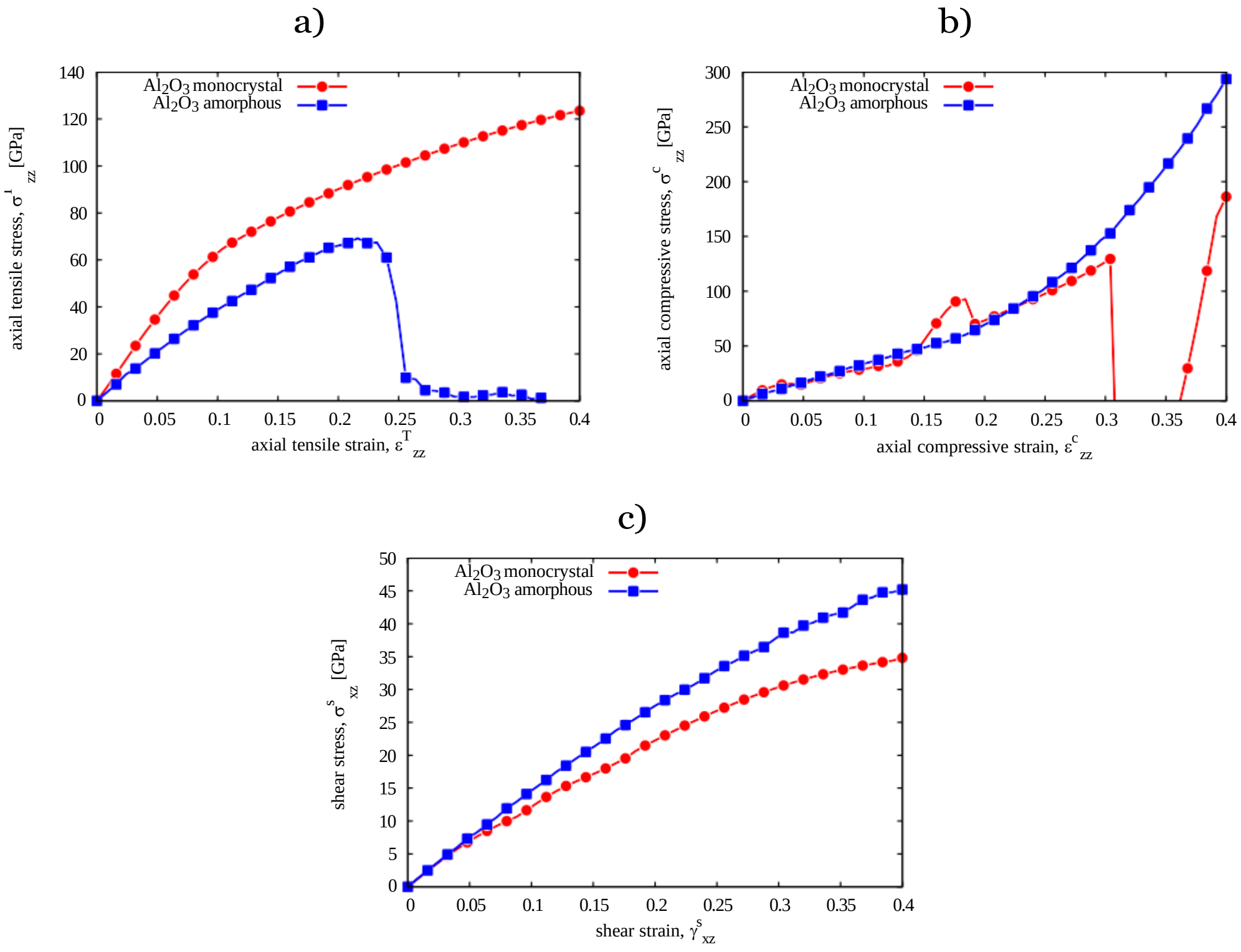

Figure 11: Stress (Cauchy) – strain (Biot) results of: a) uniaxial tensile test, b) uniaxial compression test and c) simple shear test of $Al_2O_3$ mono- and amorphous.

The stiffness tensors for monocrystalline and amorphous corundum are collected in Tables 16–18. Corundum monocrystal is less anisotropic than NiAl, i.e., the universal elastic anisotropy index $A^U = 2.03$. The calculated isotropised bulk $B = 242.15$ GPa, shear $G = 131.11$ GPa, Young's modulus $E = 333.20$ GPa and Poisson's ratio $\nu = 0.27$ are in good agreement with those from the experiment for polycrystalline $Al_2O_3$ [53] and other *ab initio*/molecular calculations [54]. The spread of these results is much larger than for NiAl and $B = 228$–$253$ GPa, $G = 119$–$162$ GPa, $E = 304$–$401$ and

$\nu$ = 0.22–0.27

We will next look at the elastic properties of amorphous $Al_2O_3$ depicted in Fig. 6c). The elasticity tensor, see Table 17, is nearly isotropic with $A^U$ = 0.27. Analyzing the isotropised moduli, i.e., $B$ = 201.97 GPa, $G$ = 121.34 GPa, $E$ = 303.28 GPa and $\nu$ = 0.25, we see that they are only slightly lower than those for monocrystalline corundum.

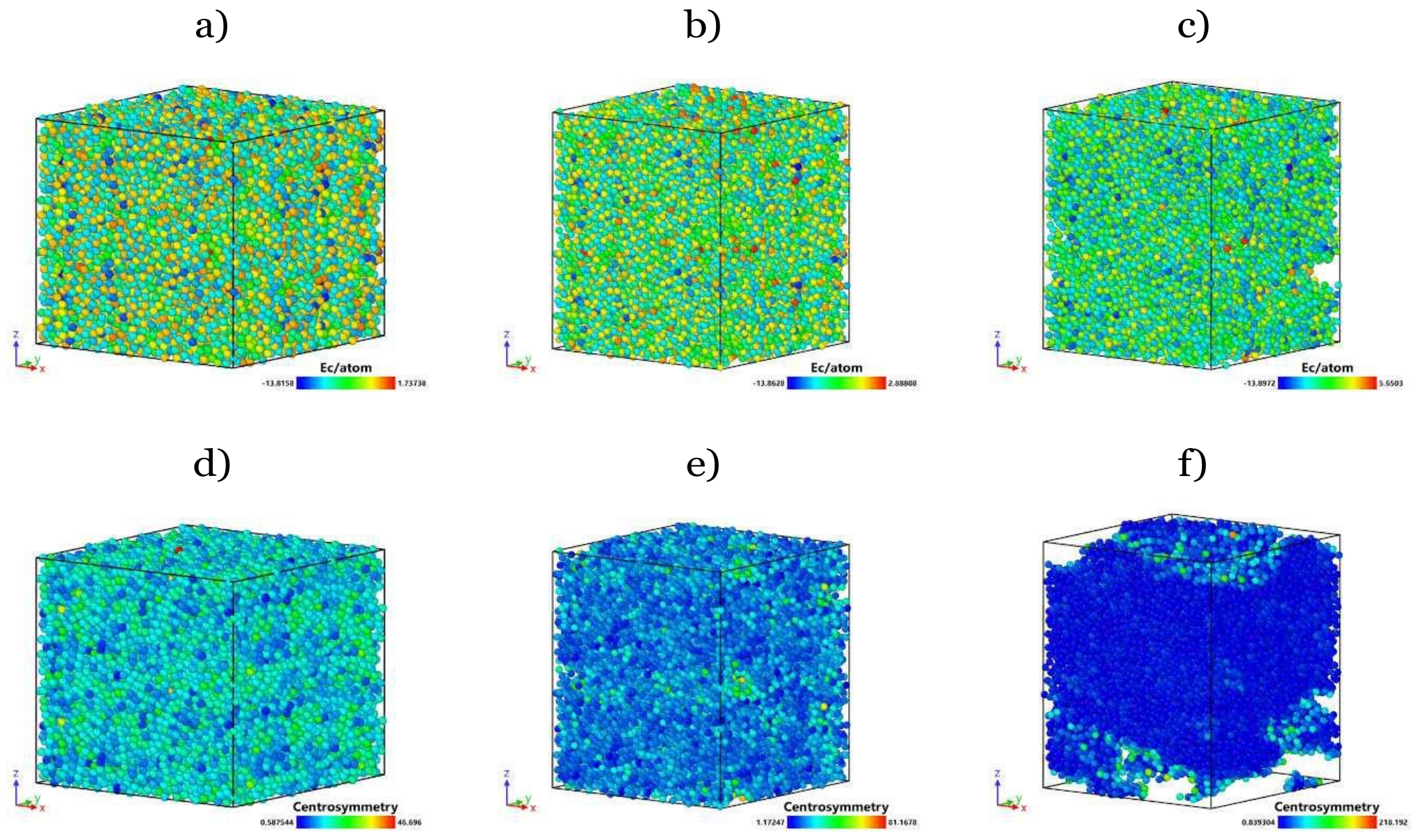

Figure 12: $Al_2O_3$ amorphous: a) the cohesive energy per atom ($E_c$/atom) for $\varepsilon_{zz}$=0, b) $E_c$/atom for $\varepsilon_{zz}$=0.232, c) $E_c$/atom for $\varepsilon_{zz}$=0.248, d) the centrosymmetry parameter (CSP) for $\varepsilon_{zz}$=0, e) CSP for $\varepsilon_{zz}$=0.232 and f) CSP for $\varepsilon_{zz}$=0.248.

- Stiffness tensor: $Al_2O_3$ oriented $X = [100]$ $Y = [-1\ \sqrt{3}\ 0]$ $Z = [001]$

$$
[C_{IJ}] \rightarrow \begin{bmatrix} 540.69 & 186.42 & 77.72 & 61.09 & 0. & 0. \\ 186.42 & 540.69 & 77.72 & -61.09 & 0. & 0. \\ 77.72 & 77.72 & 445.92 & 0. & 0. & 0. \\ 61.09 & -61.09 & 0. & 96.29 & 0. & 0. \\ 0. & 0. & 0. & 0. & 96.29 & 61.09 \\ 0. & 0. & 0. & 0. & 61.09 & 177.13 \end{bmatrix} [\text{GPa}]. \tag{16}
$$

$B = 242.15$ GPa, $G = 131.11$ GPa, $E = 333.20$ GPa, $\nu = 0.27$ and $A^U = 2.03$.

- Stiffness tensor: $Al_2O_3$ amorphous

direct simulation results:

$$
[C_{IJ}] \rightarrow \begin{bmatrix} 394.87 & 117.36 & 90.03 & -28.94 & -25.17 & -18.67 \\ 117.36 & 403.08 & 141.56 & -15.8 & -12.01 & -7.31 \\ 90.03 & 141.56 & 370.45 & -12.05 & -15.72 & -23.02 \\ -28.94 & -15.8 & -12.05 & 122.23 & 5.07 & -17.76 \\ -25.17 & -12.01 & -15.72 & 5.07 & 112.79 & -7.63 \\ -18.67 & -7.31 & -23.02 & -17.76 & -7.63 & 111.37 \end{bmatrix} [\text{GPa}]. \tag{17}
$$

$B = 201.97$ GPa, $G = 121.34$ GPa, $E = 303.28$ GPa, $\nu = 0.25$ and $A^U = 0.27$,

reduction to isotropy:

$$[C_{IJ}] \rightarrow \begin{bmatrix} 389.47 & 116.32 & 116.32 & 0. & 0. & 0. \\ 116.32 & 389.47 & 116.32 & 0. & 0. & 0. \\ 116.32 & 116.32 & 389.47 & 0. & 0. & 0. \\ 0. & 0. & 0. & 115.46 & 0. & 0. \\ 0. & 0. & 0. & 0. & 115.46 & 0. \\ 0. & 0. & 0. & 0. & 0. & 115.46 \end{bmatrix} \text{[GPa]}. \tag{18}$$

$B = 207.37$ GPa, $G = 123.49$ GPa, $E = 309.11$ GPa, $\nu = 0.25$ and $A^U = 0.0339$.

### 3.3. Metal–ceramic interface

Examining the stress–strain curves for the NiAl–$Al_2O_3$ interface in Figs. 13a)–b), it can be seen that the behavior of the material does not greatly depend on the mutual orientation of the crystals and whether it is crystalline or amorphous. For all four interfaces depicted in Fig. 7a)–d), during tension, damage of the material was obtained, see Fig. 13a), the maximum stress was in the range of 13–17 GPa at a strain of about 0.10. Very close results from *ab initio* calculations for a case similar to the one in the Fig. 7b) in the uniaxial tensile test, i.e., the maximum stress was 12.84 GPa at a strain of 0.1042, were obtained in [8].

a) b)

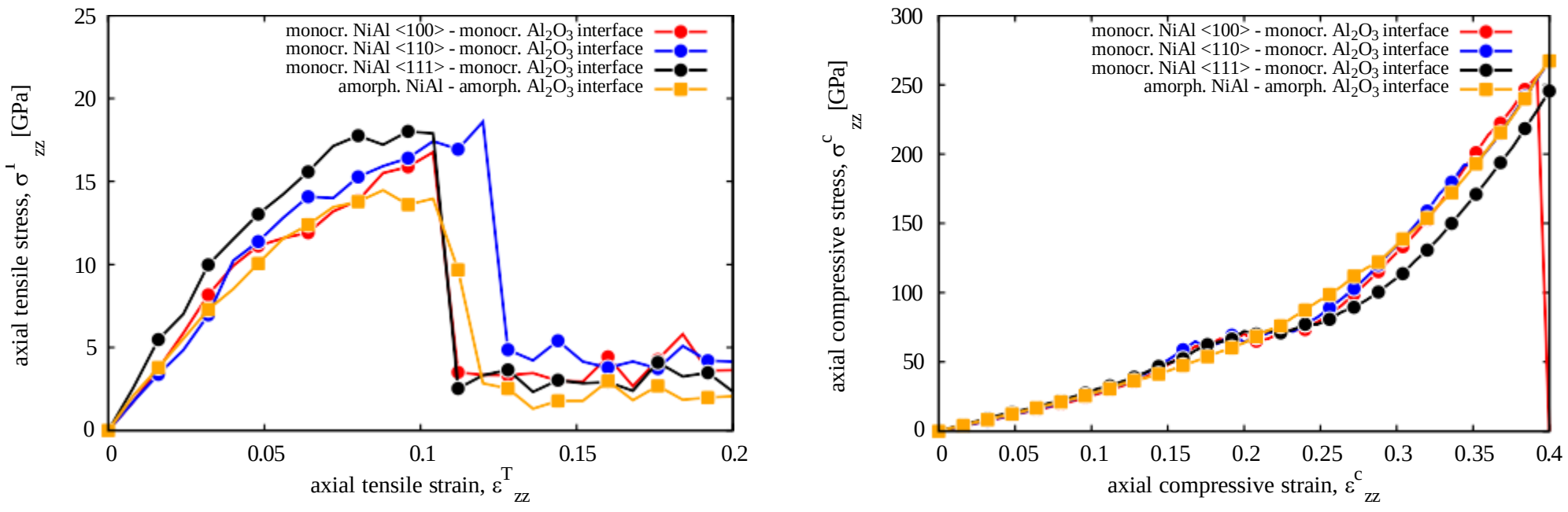

Figure 13: Stress (Cauchy) - strain (Biot) results of: a) uniaxial tensile test and b) simple shear test of NiAl - $Al_2O_3$ monocrystal and amorphous.

Atomistic analysis of the damage for all four interfaces studied shows a similar mechanism: we can observe a significant increase in the cohesive energy per atom and the centrosymmetry parameter for a strain around $\varepsilon^T_{ZZ} \approx 0.1$ and sudden bond breaking between NiAl–corundum, see Fig. 14a)–f), Fig. 15a)–f), Fig. 16a)–f) and Fig. 17a)–f). Unlike for NiAl and corundum, even for the interface of amorphous crystals we have brittle behavior.

the Stiffness tensors for all four NiAl–$Al_2O_3$ interfaces depicted in Fig. 7a)–d) are collected in Tables 19–23. We can see that the calculated initial stiffness moduli greatly depend on the mutual orientation of the NiAl crystal and the corundum. Thus, $B$ = 110.58–185.99 GPa, $G$ = 59.51–84.19 GPa, $E$ = 151.38–211.66 and $\nu$ = 0.26–0.32, with a fairly similar elastic anisotropy $A^U$ = 1.45–1.61. The Young's modulus, $E$, for the clean NiAl–corundum interface from *ab initio* calculations in [8] was equal to 172.93 GPa. The elasticity tensor of amorphous NiAl - amorphous $Al_2O_3$ interface, see Table 22, is nearly isotropic with $A^U = 0.46$. Analyzing isotropised moduli, i.e. $B$ = 155.11 GPa, $G$ = 89.40 GPa, $E$ = 224.97 GPa and $\nu$ = 0.26, we see that they are either intermediate or even higher than those for crystalline interfaces. For a micro-composite produced by sintering and consisting of 50% NiAl and 50% $Al_2O_3$, in [53] was obtained $B \approx 185$ GPa, $G \approx 100$ GPa, $E \approx 250$ and $\nu \approx 0.26$, which is very similar to the values calculated here.

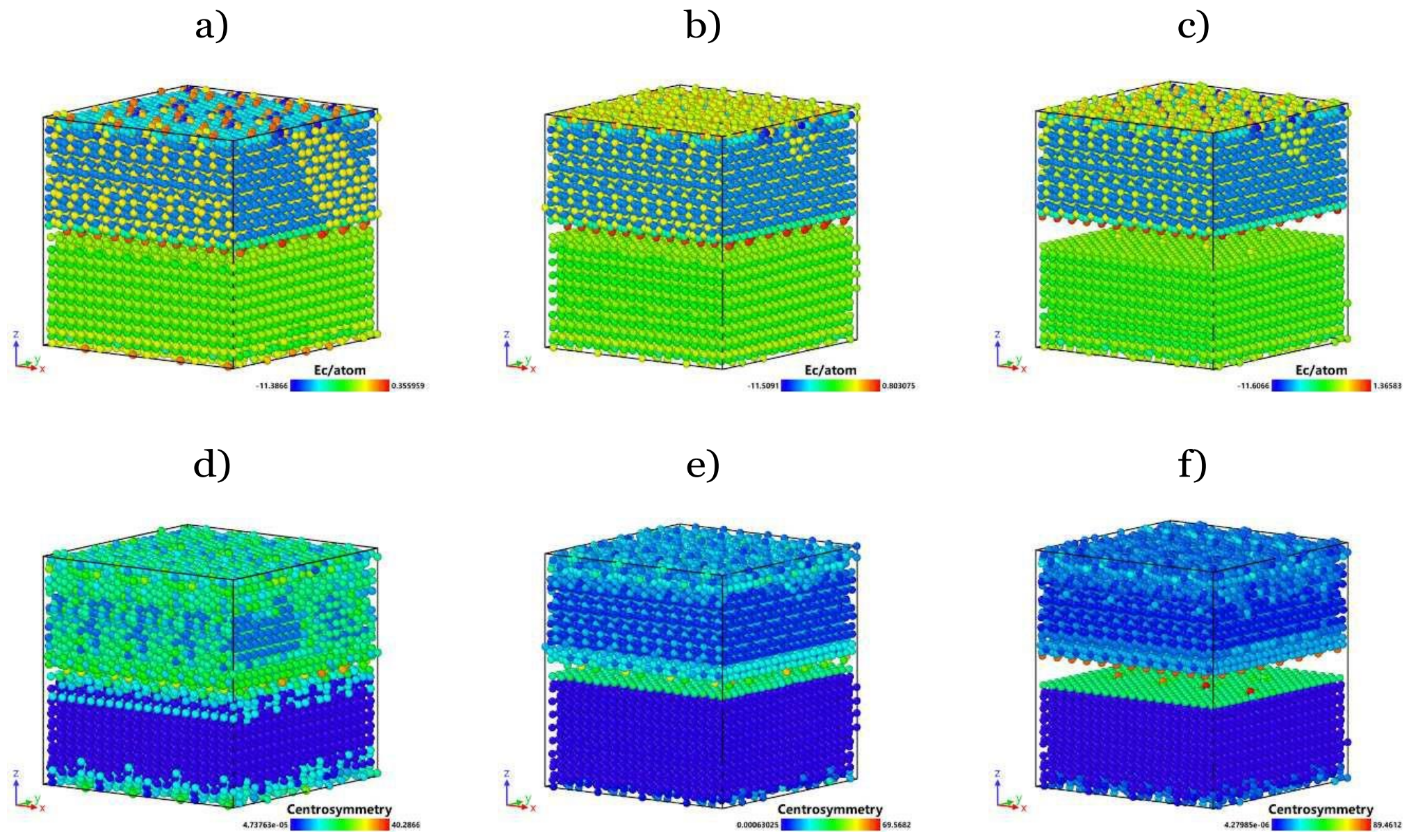

Figure 14: $Al_2O_3$-NiAl 12×7×4 orthorhombic $Al_2O_3$ basic cells and 20×20×18 NiAl basic cells $X$ =[100] $Y$ =[010] $Z$ =[001]: a) the cohesive energy per atom ($E_c$/atom) for $\varepsilon_{zz}$=0, b) $E_c$/atom for $\varepsilon_{zz}$=0.096, c) $E_c$/atom for $\varepsilon_{zz}$=0.112, d) the centrosymmetry parameter (CSP) for $\varepsilon_{zz}$=0, e) CSP for $\varepsilon_{zz}$=0.096 and f) CSP for $\varepsilon_{zz}$=0.112.

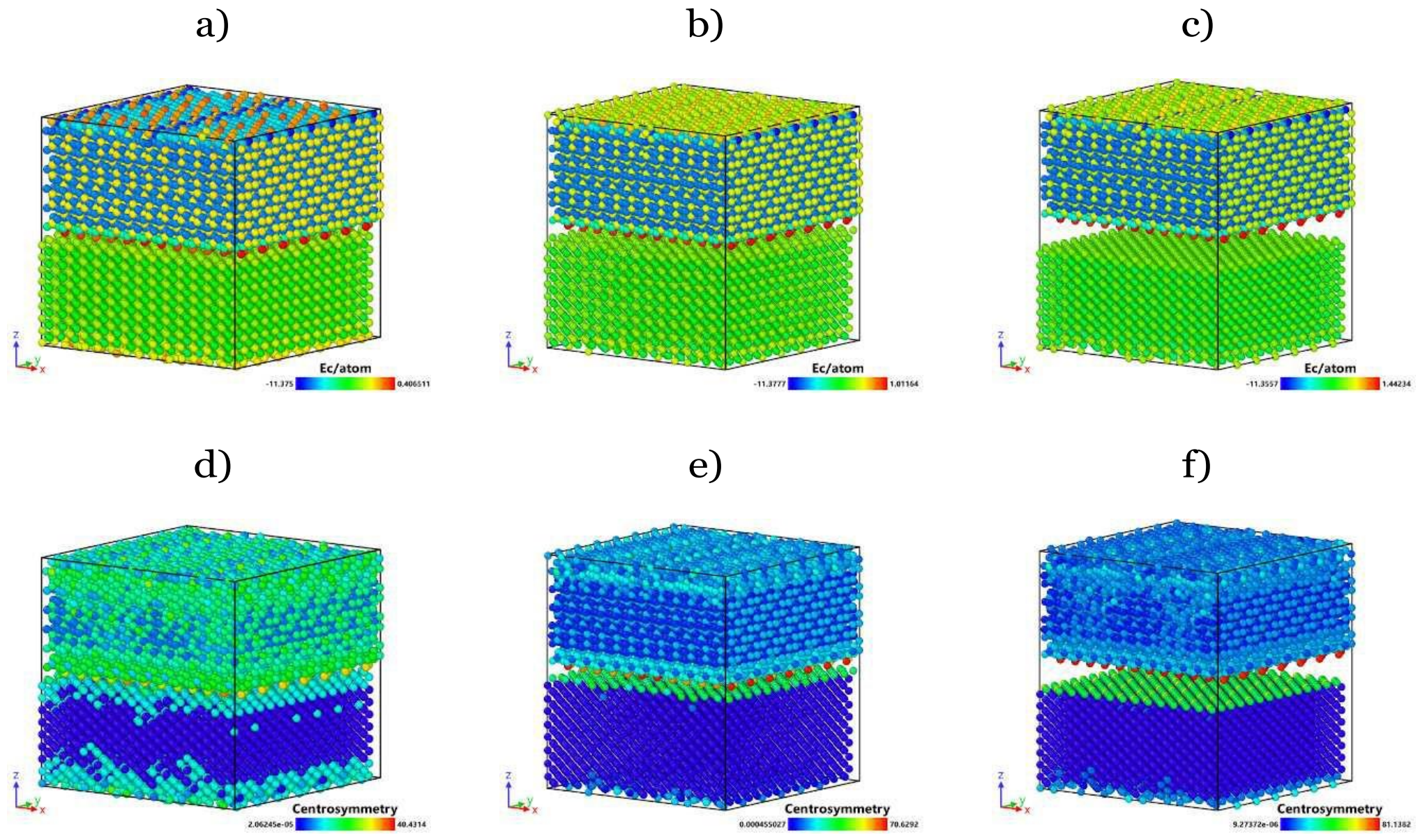

Figure 15: $Al_2O_3$-NiAl 12×7×4 orthorhombic $Al_2O_3$ basic cells and 14×14×18 NiAl basic cells $X$ =[110] $Y$ =[-110] $Z$ =[001]: a) the cohesive energy per atom ($E_c$/atom) for $\varepsilon_{zz}$=0, b) $E_c$/atom for $\varepsilon_{zz}$=0.112, c) $E_c$/atom for $\varepsilon_{zz}$=0.128, d) the centrosymmetry parameter (CSP) for $\varepsilon_{zz}$=0, e) CSP for $\varepsilon_{zz}$=0.112 and f) CSP for $\varepsilon_{zz}$=0.128.

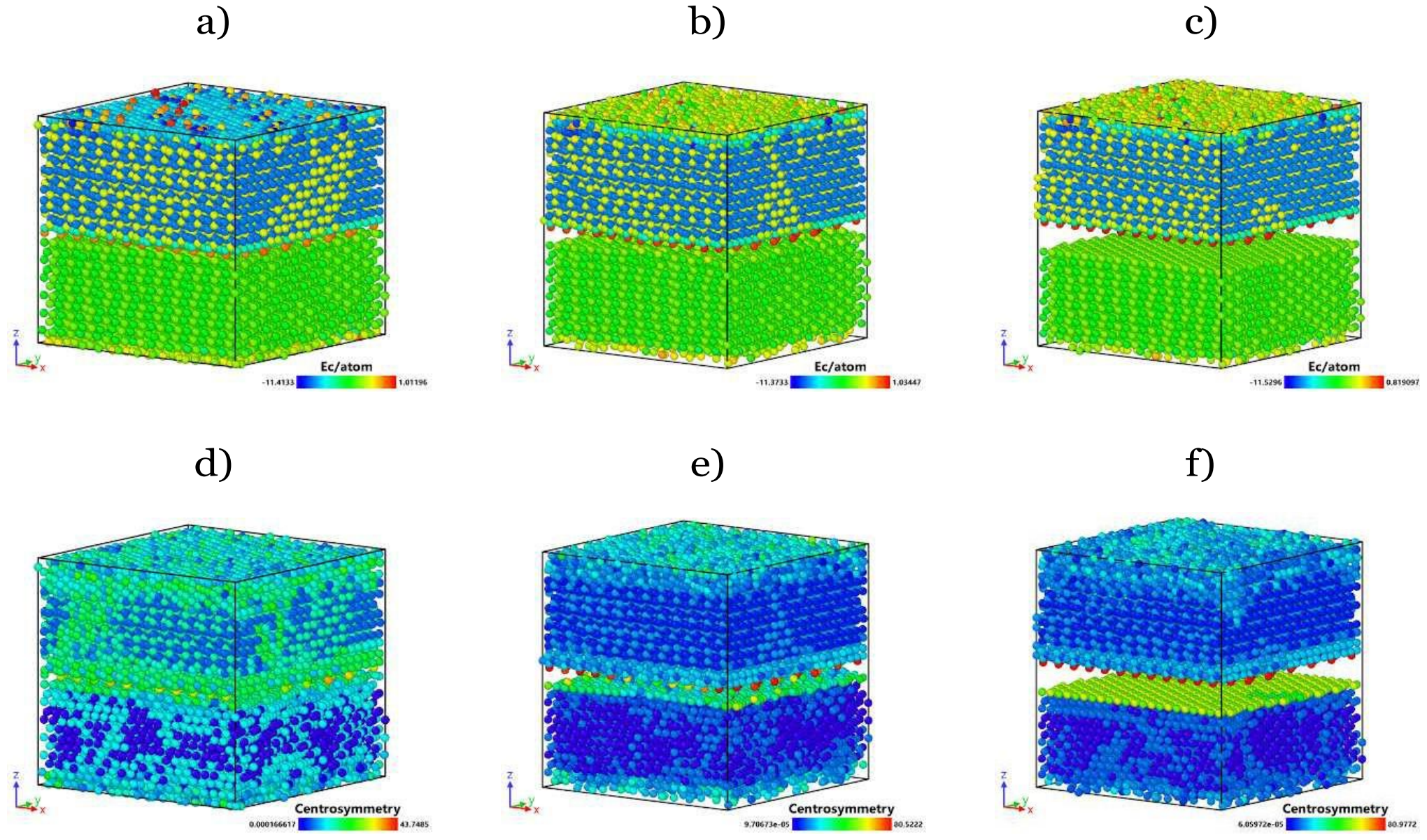

Figure 16: $Al_2O_3$-NiAl 12×7×4 orthorhombic $Al_2O_3$ basic cells and 11×8×13 NiAl basic cells $X$ =[111] $Y$ =[-1-12] $Z$ =[1-10]: a) the cohesive energy per atom ($E_c$/atom) for $\varepsilon_{zz}$=0, b) $E_c$/atom for $\varepsilon_{zz}$=0.104, c) $E_c$/atom for $\varepsilon_{zz}$=0.112, d) the centrosymmetry parameter (CSP) for $\varepsilon_{zz}$=0, e) CSP for $\varepsilon_{zz}$=0.104 and f) CSP for $\varepsilon_{zz}$=0.112.

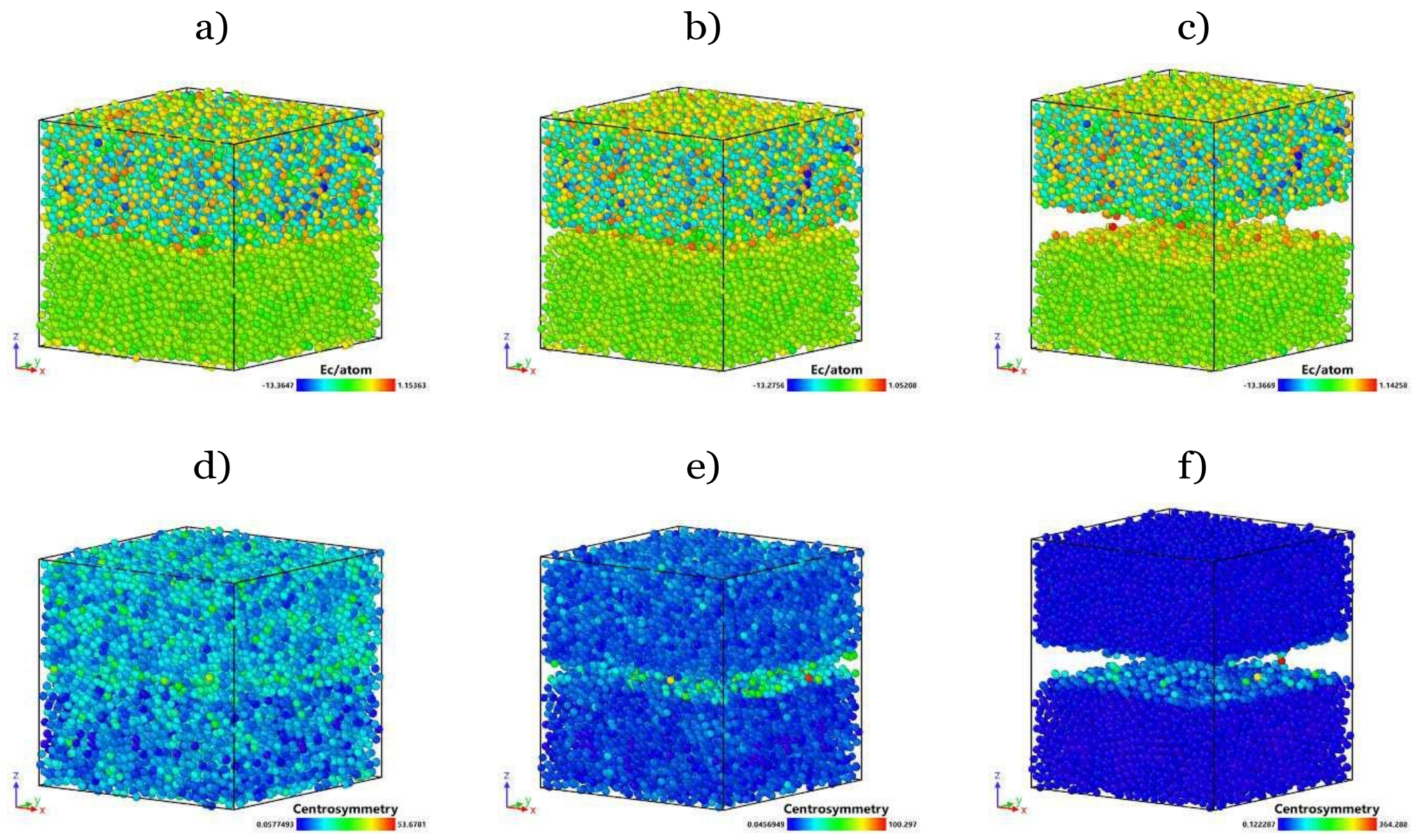

Figure 17: $Al_2O_3$ amorphous and NiAl amorphous: a) the cohesive energy per atom ($E_c$/atom) for $\varepsilon_{zz}$=0, b) $E_c$/atom for $\varepsilon_{zz}$=0.088, c) $E_c$/atom for $\varepsilon_{zz}$=0.168, d) the centrosymmetry parameter (CSP) for $\varepsilon_{zz}$=0, e) CSP for $\varepsilon_{zz}$=0.088 and f) CSP for $\varepsilon_{zz}$=0.168.

- Stiffness tensor: 12×7×4 orthorhombic $Al_2O_3$ basic cells and 20×20×18 NiAl basic cells $X$ =[100] $Y$ =[010] $Z$ =[001]

$$[C_{IJ}] \rightarrow \begin{bmatrix} 296.57 & 144.76 & 125.5 & -35.27 & -2.5 & 3.45 \\ 144.76 & 273.54 & 74.42 & 17.96 & -4.93 & 1.37 \\ 125.5 & 74.42 & 169.18 & -39.37 & -18.81 & 9.45 \\ -35.27 & 17.96 & -39.37 & 110.56 & 0.02 & 0.17 \\ -2.5 & -4.93 & -18.81 & 0.02 & 113.03 & -31.15 \\ 3.45 & 1.37 & 9.45 & 0.17 & -31.15 & 112.41 \end{bmatrix} [\mathrm{GPa}]. \quad (19)$$

$B$ = 145.19 GPa, $G$ = 84.19 GPa, $E$ = 211.66 GPa, $\nu$ = 0.26 and $A^U$

= 1.45.

- Stiffness tensor: 12×7×4 orthorhombic $Al_2O_3$ basic cells and 14×14×18 NiAl basic cells $X$ =[110] $Y$ =[-110] $Z$ =[001]

$$[C_{IJ}] \rightarrow \begin{bmatrix} 328.94 & 115.36 & 108.47 & -72.04 & -7.95 & -15.51 \\ 115.36 & 335.65 & 121.99 & 35.07 & -0.6 & 4.53 \\ 108.47 & 121.99 & 167.61 & -82.61 & -15.04 & -33.39 \\ -72.04 & 35.07 & -82.61 & 125.71 & -4.94 & 4.32 \\ -7.95 & -0.6 & -15.04 & -4.94 & 112.47 & -30.68 \\ -15.51 & 4.53 & -33.39 & 4.32 & -30.68 & 69.87 \end{bmatrix} \text{[GPa]}. \quad (20)$$

$B$ = 110.58 GPa, $G$ = 59.51 GPa, $E$ = 151.38 GPa, $\nu$ = 0.27 and $A^U$ = 1.61.

- Stiffness tensor: 12×7×4 orthorhombic $Al_2O_3$ basic cells and 11×8×13 NiAl basic cells $X$ =[111] $Y$ =[-1-12] $Z$ =[1-10]

$$[C_{IJ}] \rightarrow \begin{bmatrix} 336.98 & 126.06 & 127.63 & -31.82 & -17.25 & 15.94 \\ 126.06 & 342.81 & 114.04 & 21.32 & 5.29 & -23.55 \\ 127.63 & 114.04 & 273.92 & -12.22 & 12.38 & 10.75 \\ -31.82 & 21.32 & -12.22 & 99.41 & 30.38 & -6.19 \\ -17.25 & 5.29 & 12.38 & 30.38 & 60.32 & -18.76 \\ 15.94 & -23.55 & 10.75 & -6.19 & -18.76 & 63.29 \end{bmatrix} \text{[GPa]}. \quad (21)$$

$B$ = 185.99 GPa, $G$ = 74.19 GPa, $E$ = 196.45 GPa, $\nu$ = 0.32 and $A^U$ = 1.48.

- Stiffness tensor: $Al_2O_3$ amorphous and NiAl amorphous

direct simulation results:

$$[C_{IJ}] \to \begin{bmatrix} 306.42 & 101.08 & 53.46 & -5.47 & -7.41 & -1.79 \\ 101.08 & 314.69 & 115.43 & 6.84 & -9.35 & 2.66 \\ 53.46 & 115.43 & 264.65 & 9.78 & -29.48 & 1.81 \\ -5.47 & 6.84 & 9.78 & 59.12 & 3.86 & -5.97 \\ -7.41 & -9.35 & -29.48 & 3.86 & 98.99 & -0.61 \\ -1.79 & 2.66 & 1.81 & -5.97 & -0.61 & 101.55 \end{bmatrix} [\text{GPa}]. \tag{22}$$

$B$ = 155.11 GPa, $G$ = 89.40 GPa, $E$ = 224.97 GPa, $\nu$ = 0.26 and $A^U$ = 0.46,

reduction to isotropy:

$$[C_{IJ}] \to \begin{bmatrix} 295.25 & 89.99 & 89.99 & 0. & 0. & 0. \\ 89.99 & 295.25 & 89.99 & 0. & 0. & 0. \\ 89.99 & 89.99 & 295.25 & 0. & 0. & 0. \\ 0. & 0. & 0. & 86.55 & 0. & 0. \\ 0. & 0. & 0. & 0. & 86.55 & 0. \\ 0. & 0. & 0. & 0. & 0. & 86.55 \end{bmatrix} [\text{GPa}]. \tag{23}$$

$B$ = 158.41 GPa, $G$ = 92.66 GPa, $E$ = 232.63 GPa, $\nu$ = 0.26 and $A^U$ = 0.035.

### *3.4. Discussion*

As was discussed in Section 2.1, the macroscopic strength of the composite is a complex combination of the strength of the matrix, the strength of the reinforcement, the strength of the interface, and the residual stresses induced by thermal expansion mismatch [56]. The deformation and damage mechanism of the individual components of NiAl–$Al_2O_3$, presented in the previous section, can be presented in the light of existing research related to the fracture mechanics of the studied composite. The numerical results can be supportive and suggest the fracture/damage mode of the composite.

First of all, the NiAl–$Al_2O_3$ interface seems to be a key factor in the context of composite failure. A strong and well-bonded interface can hinder crack growth, while a weak interface can promote crack initiation. As was reported in [57], the structure of the NiAl–$Al_2O_3$ interface appears to be devoid of any additional phases that might have arisen during the sintering process. This was confirmed by nanoanalysis using a transmission electron microscope (TEM) equipped with an energy-dispersive spectroscopy (EDS) detector conducted along the designated line traversing the interface, revealing variations in the Ni, O, and Al content. These variations, as depicted in Fig. 18, did not suggest the presence of any transitional phases. The TEM examinations affirmed the robust and adhesive nature of the bond at the NiAl–$Al_2O_3$ interface. Furthermore, alterations in contrast at the interface indicated the absence the formation of any diffusive-type interface layer. [57].

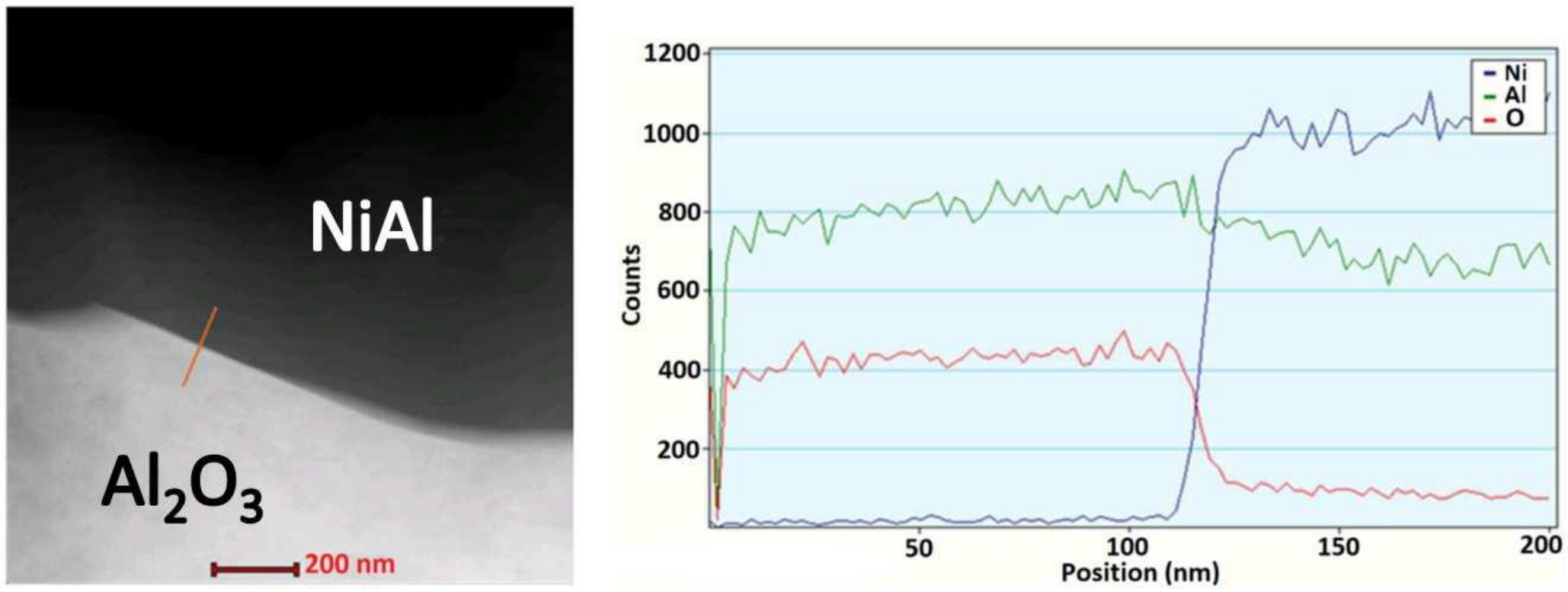

Figure 18: Transmission electron microscopy image and the evolution of vol. content of Al, O, and Ni along the NiAl-$Al_2O_3$ interface [57].

The real structure of the NiAl–$Al_2O_3$ interface has been represented within the molecular dynamics framework. A relatively sharp transition between the intermetallic and ceramic phases (Fig. 18), indicating the adhesive type of bonding, has been generated (Fig. 7) and simulated in the context of deformation and failure. Comparing the tensile/compressive/shear strengths of the individual components—NiAl monocrystal along different orientations, $Al_2O_3$ monocrystal, amorphous NiAl (as a representative of NiAl grain boundary), amorphous $Al_2O_3$ (as a representative of $Al_2O_3$ grain boundary) and the NiAl–$Al_2O_3$ interface (the amorphous one and along different orientations)—it should be pointed out that the lowest values were obtained for the intermetallic–ceramic interface (Fig. 13). Even though the tensile/compressive strength is quite close to those of the NiAl amorphous sample, the fracture strain for this interface indicates its having a much lower value, making it the first composite component to fail.

In contrast with the various metal–ceramic interfaces with high cohesion energy resulting in relatively considerable strength, the NiAl–$Al_2O_3$ interface is quite weak. The presented atomistic results are confirmed by several experimental analyses identifying the main failure mechanism of the NiAl–$Al_2O_3$ composite as interface failure [56, 58].

On the one hand, the presence of the $Al_2O_3$ reinforcement forces the crack to follow a tortuous path through the ceramic material, significantly extending its route and consequently enhancing the strength of the composite [57]. On the other hand, this is only in the case of the optimal amount of the ceramic phase, which must evenly occupy all the inter-grain boundaries in the material, and avoid causing any agglomeration of the $Al_2O_3$. Exceeding this value leads to weakening of the structure and consequently to poorer mechanical properties.

The weak bonding between the NiAl and the $Al_2O_3$ does not allow taking full advantage of such a toughening mechanism. Figure 19 confirms which is the primary damage mode (interface failure) by revealing the fracture surface of sintered NiAl–$Al_2O_3$ composites with 20% vol. content of ceramic reinforcement. As can be seen, most of the studied fracture surface consist of the voids remaining after the pull out effect of the ceramic inclusions. The weakness of the NiAl–$Al_2O_3$ interface failure leads to voids, separation, or regions of discontinuity along the reinforcement–matrix interface. The observed failure mechanism of NiAl–$Al_2O_3$ is in line with the numerical results obtained from molecular dynamics simulations presented in the previous sections.

Generally, it is rare that fractures are exclusively transgranular or intergranular or due to interfacial failure: a mixture of these modes often occurs [2]. Confirming this statement, Fig. 19 reveals the failure via the NiAl matrix,

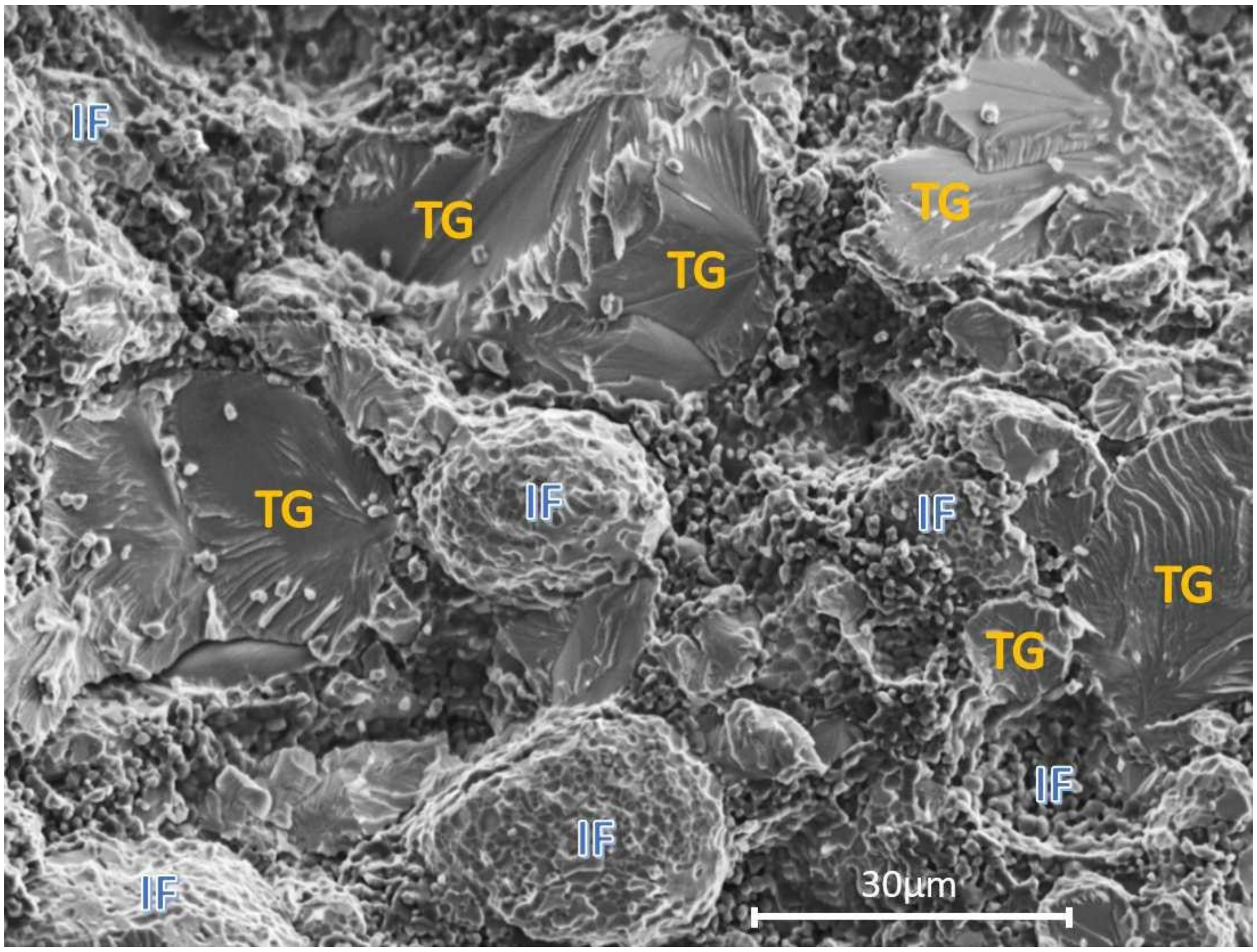

Figure 19: SEM image of fracture surface of NiAl-20%$Al_2O_3$ composites with selection of metal–ceramic interface failure (IF) and transgranular fracture (TG) via NiAl grains.

whereas cleavage facets and a lack of plastic deformation might be observed. Such conclusion are in line with literature data. At room temperature, the plastic deformation of NiAl is in the range of 0 to a maximum 4% [59]. The brittleness of the NiAl phase can be indicated quantitatively by the parameter $K_{IC}$, which is from 4–7 MPa$\sqrt{m}$ for polycrystalline [60] and sintered NiAl [61], and 4–10 MPa$\sqrt{m}$ for single crystals depending on the crystallographic direction [62]. The above results do not differ significantly from the values of $K_{IC}$ obtained for polycrystalline ceramic materials, e.g., for $Al_2O_3$, $K_{IC}$ =

5–6 MPa$\sqrt{m}$ [63].

The low fracture toughness and low ductility of NiAl is associated with a limited number of slip systems. Much research has been devoted to understanding the main sliding mechanism in both monocrystalline [64] and polycrystalline NiAl [65]. The brittle deformation of the NiAl matrix has been also confirmed by our molecular calculations (Section 3.1). Both differently orientated monocrystals and amorphous NiAl, representing the averaged mechanical response of the grain boundaries, demonstrate the relatively linear stress–strain dependence with local fluctuations up to maximum stress (Fig. 8). For various types of mechanical tests (tensile, compressive, shear), the amorphous NiAl sample had the lowest strength compared to monocrystals. This may suggest the intergranular fracture mode via grain boundaries of pure NiAl polycrystalline.

Experimental studies of NiAl grain boundaries carried out after compression and tensile tests confirm that the low ductility is the result of inconsistency in shape changes of neighboring grains caused by a limited number of slip systems [64, 65]. A detailed analysis by Auger spectroscopy confirmed that it was at the intergrain limits in NiAl alloys there are no impurities that could influence the mechanical properties at room temperature [66]. Due to the interconnection of the NiAl grains, it is possible for them to bridge the surfaces of the crack [56] (Fig. 20).

The atomistic results about NiAl monocrystals with various orientations with respect to the loading direction have shown that there is a high anisotropy effect regardless of the type of mechanical test (Fig. 8). The highest stiffness and fracture strength of NiAl monocrystal can be observed in the ⟨111⟩ orientations, while the (110) and (100) orientations have a relatively unique character. The presence of 'hard' and 'soft' orientations can be proved by experimental studies on the structure of the electron bands of stoichiometric NiAl. This confirms the presence of stronger Ni d – Al p hybridization along the [1 1 1] direction between adjacent pairs of Ni–Al atoms [67]. It has also been found [68] that electron depletion of both the Ni and Al lattices occur along the [1 0 0] direction. For this reason, the [1 1 1] direction displays increasing electron density. Strong Ni d–Al p hybridization with increasing electron density indicates that there are strong covalent bonds in the [1 1 1] direction between the nearest pairs of Ni–Al atoms. Experimental observations also suggest the presence of weak ionic interactions between the 'second' nearest atoms in the [1 0 0] direction. The presence of said directional bonds is superimposed on the presence of metallic bonds. Strong atomic bonds along the [1 1 1] direction and weak bonds along the [1 0 0] direction cause an anisotropy of the elastic properties of NiAl, which has been proved by atomistic calculations and seen in the different form of the NiAl stiffness tensor for various orientations (Section 3.1).

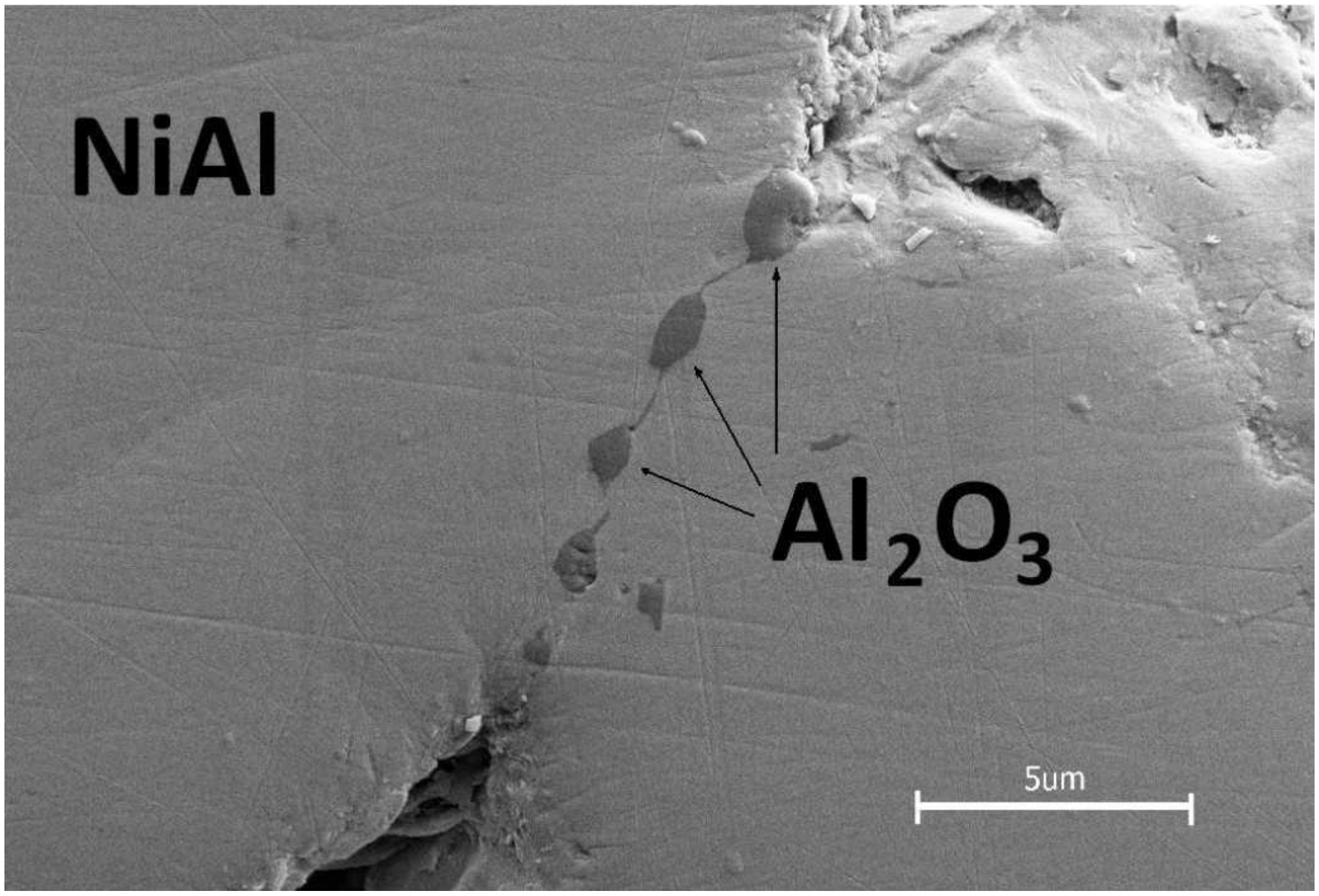

Figure 20: Intergranular fracture via NiAl grain boundary.

Based on the literature data, NiAl monocrystal is deformed by displacement of dislocation planes according to the Burgers vector b = ⟨100⟩ in the whole range of temperatures. The exceptions are single crystals oriented in the [1 0 0] direction. A shift along the plane (100) has also been theoretically confirmed [64, 69]. Due to the presence of a major slip plane (100), there are only three possible independent slip systems [64], which translates into low ductility and low fracture toughness at room temperature.

## 4. Conclusions

The presented work can be summarized in the following remarks:

1. An atomistic study of the deformation and failure behavior of crystalline and amorphous components of NiAl–$Al_2O_3$ composite has been performed. The molecular statics framework has been employed to calculate the upper scale parameters, the elastic constants and strength, of each composite element: the metal and ceramic monocrystals, their grain boundaries, and the metal/ceramic interface.
2. NiAl monocrystal has been simulated under different lattice orientations to reveal the effect of its anisotropy. Amorphous samples have

been generated as the averaged representations of the grain boundaries. The mechanical properties of the metal–ceramic interface have been investigated as the combination of two monocrystals and alternatively two amorphous forms. The obtained samples were tested via three main strength tests: uniaxial tensile, uniaxial compressive, and simple shear.

3. Based on the stress–strain curves obtained from the atomistic simulations, it can be stated that the NiAl–$Al_2O_3$ interface has been revealed as the weakest element of the composite. Regardless of its form (whether crystalline or amorphous), the metal–ceramic bonding shows the lower tensile strength compared to the other components due to its adhesive structure. This conclusion has been confirmed by fractographic analysis of the surface of the NiAl–$Al_2O_3$ composite after failure. It was shown that cracks initiate at a weak NiAl–$Al_2O_3$ interface, propagate through the matrix, and transition to intergranular mode when encountering a grain boundary.
4. Atomistic simulations of NiAl monocrystals found a large anisotropy effect regardless of the type of mechanical test. As is confirmed in literature data, the soft (⟨100⟩ and ⟨110⟩) and hard (⟨111⟩) orientations are different. Moreover, all of the components of the NiAl–$Al_2O_3$ have a brittle character during deformation, with a lack of plasticity. This effect has been confirmed by the application of local measures of lattice disorder—the cohesive energy per atom and the centrosymmetry parameter—making it possible to explain the reason for the brittle behavior of the material at the atomistic level.

5. The atomistic calculations confirmed the well-known experimental fact that corundum is much stiffer elastically and has a higher strength than NiAl. This effect can be confirmed by fractographic analysis, which excludes ceramic particle cracking as the damage mechanism.
6. The molecular statics framework proved that the lower-scale mechanical parameters can be successfully evaluated from atomistic simulations and furthermore transferred to upper-scale models of the deformation and damage of the metal matrix composite.

## CRediT authorship contribution statement

Marcin Maździarz: Conceptualization, Methodology, Resources, Software, Visualization, Writing – original draft.
Szymon Nosewicz: Conceptualization, Methodology, Visualization, Writing – original draft, Writing – review & editing, Funding acquisition.

## Declaration of Competing Interest

The authors declare that they have no known competing financial interests or personal relationships that could have appeared to influence the work reported in this paper.

## ACKNOWLEDGMENTS

The authors would like to acknowledge the financial support of the National Science Centre, Poland, under Grant Agreement No. OPUS 2020/37/B/ST8/03907 for project ”Multiscale investigation of deformation and damage behavior of novel hybrid metal matrix composites. Experimental studies and numerical modeling". Additional support for the work was provided by the computing cluster GRAFEN at Biocentrum Ochota, the Interdisciplinary Centre for Mathematical and Computational Modelling of Warsaw University (ICM UW) and Poznań Supercomputing and Networking Center (PSNC).

## Appendix A. Supplementary material

The following is available online at Supplementary data

Supplementary data
# Atomistic investigation of deformation and fracture of individual structural components of metal matrix composites

Marcin Maździarz, Szymon Nosewicz*

*Institute of Fundamental Technological Research Polish Academy of Sciences, Pawińskiego 5B, 02-106 Warsaw, Poland*

## Abstract

Collected all mechanical data of the NiAl-$Al_2O_3$ interfaces and its components: stiffness tensors and deformation-stress characteristics obtained from molecular statics calculations.

## Deformations

To obtain stress-strain profiles three numerical molecular homogeneous deformation tests were performed using molecular statics (MS) approach, these selected tests are namely uniaxial strain (US) in Z direction, simple shear (SS) in XZ direction and in YZ direction. Each test was divided into 50 steps, where the results were recorded after minimizing energy and forces. The deformation gradient F for uniaxial strain in Z direction without per-

*Corresponding author

*Email address:* snosew@ippt.pan.pl (Szymon Nosewicz)

pendicular deformations is defined by

$$F_{Z}^{US} \rightarrow \begin{bmatrix} 1 & 0 & 0 \\ 0 & 1 & 0 \\ 0 & 0 & \lambda \end{bmatrix}, \tag{1}$$

where $\lambda = L/L_0$ is the principal stretch/compression ratio. The simulation box was stretched by 40%, returned along the same path to the initial configuration, then compressed 40%, and again returned along the same path to the initial configuration.

The deformation gradient F for simple shear in XZ direction can be written as

$$F_{XZ}^{SS} \rightarrow \begin{bmatrix} 1 & 0 & \gamma \\ 0 & 1 & 0 \\ 0 & 0 & 1 \end{bmatrix}, \tag{2}$$

whereas the deformation gradient F for simple shear in YZ direction as

$$F_{YZ}^{SS} \rightarrow \begin{bmatrix} 1 & 0 & 0 \\ 0 & 1 & \gamma \\ 0 & 0 & 1 \end{bmatrix}, \tag{3}$$

where $\gamma = \tan(\Phi)$, $\Phi$ is the angular change. The simulation box was sheared by $\gamma$=40% and returned along the same path to the initial configuration.

## Results

### *0.1. NiAl*

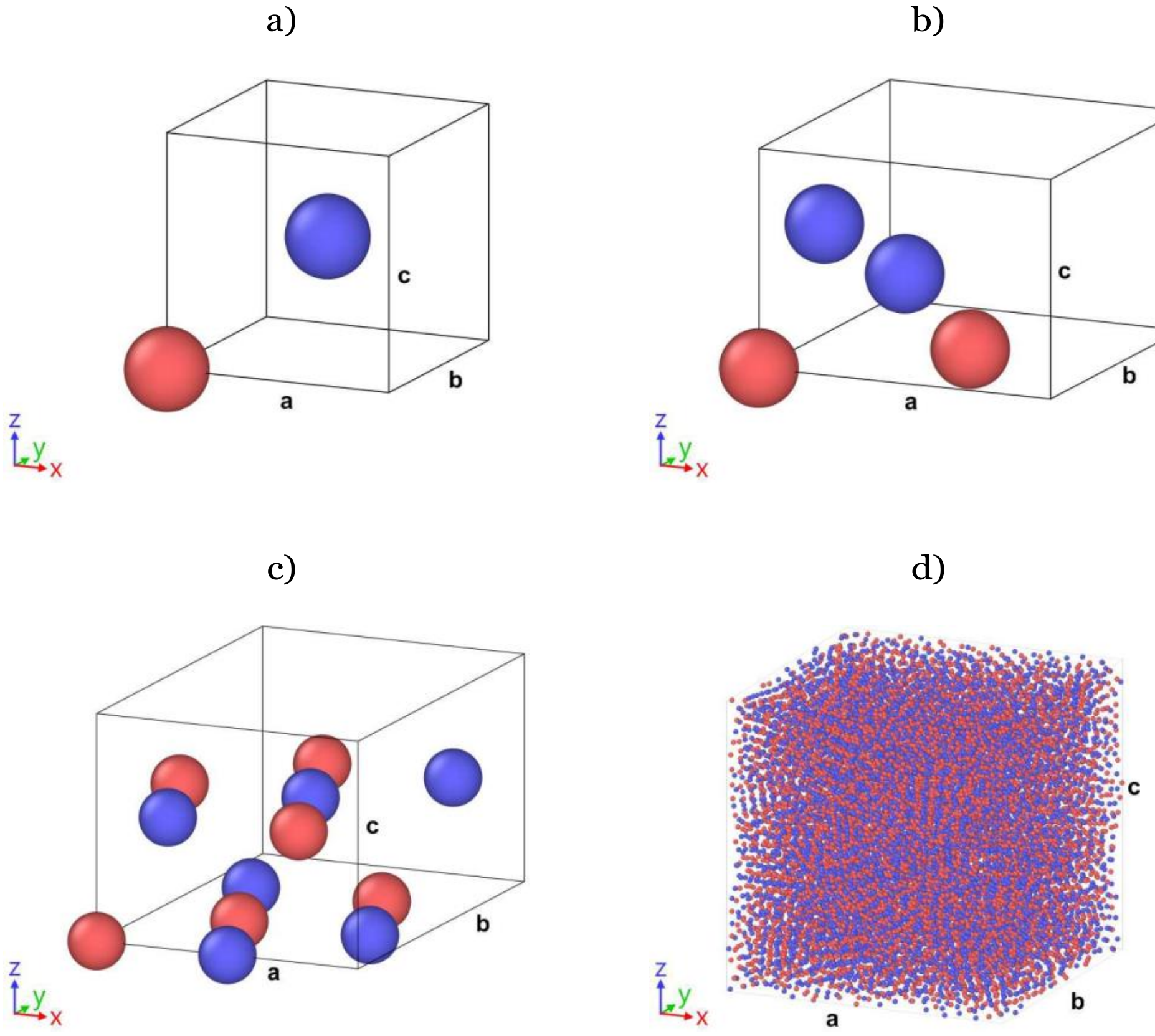

Figure 1: NiAl: a) basic cell X=[100] Y=[010] Z=[001], b) basic cell X=[110] Y=[-110] Z=[001], c) basic cell X=[111] Y=[-1-12] Z=[1-10], d) amorphous (The red and blue atoms represent Ni and Al, respectively).

*0.1.1. Stiffness tensors*

- Stiffness tensor: NiAl oriented X=[100] Y=[010] Z=[001]

$$[C_{IJ}] \rightarrow \begin{bmatrix} 190.87 & 142.91 & 142.91 & 0. & 0. & 0. \\ 142.91 & 190.87 & 142.91 & 0. & 0. & 0. \\ 142.91 & 142.91 & 190.87 & 0. & 0. & 0. \\ 0. & 0. & 0. & 121.49 & 0. & 0. \\ 0. & 0. & 0. & 0. & 121.49 & 0. \\ 0. & 0. & 0. & 0. & 0. & 121.49 \end{bmatrix} [\text{GPa}]. \quad (4)$$

$B$ = 158.90 GPa, $G$ = 64.37 GPa, $E$ = 170.13 GPa, $\nu$ = 0.32 and $A^U$ = 3.92.

- Stiffness tensor: NiAl oriented X=[110] Y=[-110] Z=[001]

$$[C_{IJ}] \rightarrow \begin{bmatrix} 288.37 & 45.40 & 142.91 & 0. & 0. & 0. \\ 45.40 & 288.37 & 142.91 & 0. & 0. & 0. \\ 142.91 & 142.91 & 190.87 & 0. & 0. & 0. \\ 0. & 0. & 0. & 121.49 & 0. & 0. \\ 0. & 0. & 0. & 0. & 121.49 & 0. \\ 0. & 0. & 0. & 0. & 0. & 23.98 \end{bmatrix} [\text{GPa}]. \quad (5)$$

$B$ = 158.90 GPa, $G$ = 64.37 GPa, $E$ = 170.13 GPa, $\nu$ = 0.32 and $A^U$ = 3.92.

- Stiffness tensor: NiAl oriented X=[111] Y=[-1-12] Z=[1-10]

$$[C_{IJ}] \rightarrow \begin{bmatrix} 320.88 & 77.91 & 77.91 & 0. & 0. & 0. \\ 77.91 & 288.38 & 110.41 & 0. & 0. & -45.96 \\ 77.91 & 110.41 & 288.37 & 0. & 0. & 45.96 \\ 0. & 0. & 0. & 88.98 & 45.96 & 0. \\ 0. & 0. & 0. & 45.96 & 56.48 & 0. \\ 0. & -45.96 & 45.96 & 0. & 0. & 56.48 \end{bmatrix} [\text{GPa}]. \tag{6}$$

$B$ = 158.90 GPa, $G$ = 64.37 GPa, $E$ = 170.13 GPa, $v$ = 0.32 and $A^U$ = 3.92.

- Stiffness tensor: NiAl amorphous

direct simulation result:

$$[C_{IJ}] \rightarrow \begin{bmatrix} 170.03 & 115.17 & 126.48 & -5.17 & -0.08 & 7.10 \\ 115.17 & 168.41 & 114.76 & 1.26 & -5.85 & 2.99 \\ 126.48 & 114.76 & 171.48 & 1.69 & 0.18 & 1.50 \\ -5.17 & 1.26 & 1.69 & 23.57 & -2.85 & 2.53 \\ -0.08 & -5.85 & 0.18 & -2.85 & 32.26 & 0.45 \\ 7.10 & 2.99 & 1.50 & 2.53 & 0.45 & 23.00 \end{bmatrix} [\text{GPa}]. \tag{7}$$

$B$ = 135.30 GPa, $G$ = 25.24 GPa, $E$ = 71.28 GPa, $v$ = 0.41 and $A^U$ = 0.32,

reduction to isotropy:

$$[C_{IJ}] \rightarrow \begin{bmatrix} 169.97 & 118.80 & 118.80 & 0. & 0. & 0. \\ 118.80 & 169.97 & 118.80 & 0. & 0. & 0. \\ 118.80 & 118.80 & 169.97 & 0. & 0. & 0. \\ 0. & 0. & 0. & 26.28 & 0. & 0. \\ 0. & 0. & 0. & 0. & 26.28 & 0. \\ 0. & 0. & 0. & 0. & 0. & 26.28 \end{bmatrix} [\mathrm{GPa}]. \tag{8}$$

$B$ = 135.86 GPa, $G$ = 26.00 GPa, $E$ = 73.32 GPa, $v$ = 0.41 and $A^U$ = 0.00085.

### *0.1.2. Deformation-stress*

- NiAl oriented X=[100] Y=[010] Z=[001]

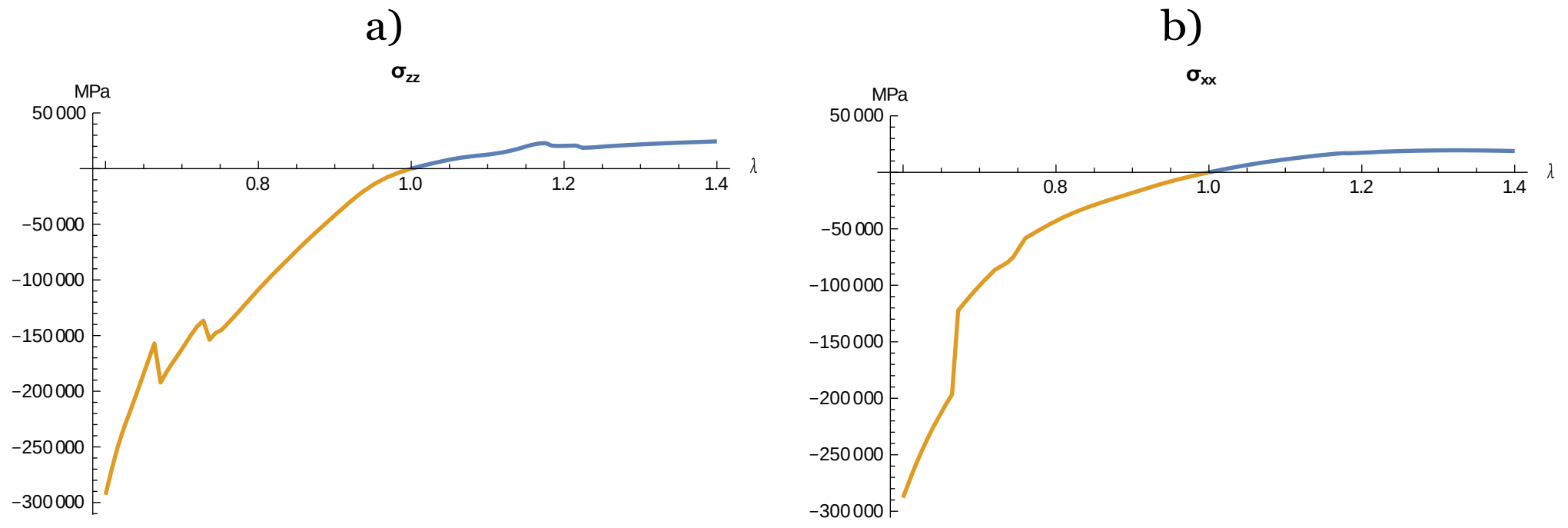

Figure 2: Uniaxial strain in z direction: a) $\sigma_{zz}$, b) $\sigma_{xx}$ ($\sigma_{xx}$=$\sigma_{yy}$, $\sigma_{xy}$=$\sigma_{xz}$=$\sigma_{yz}$=0)

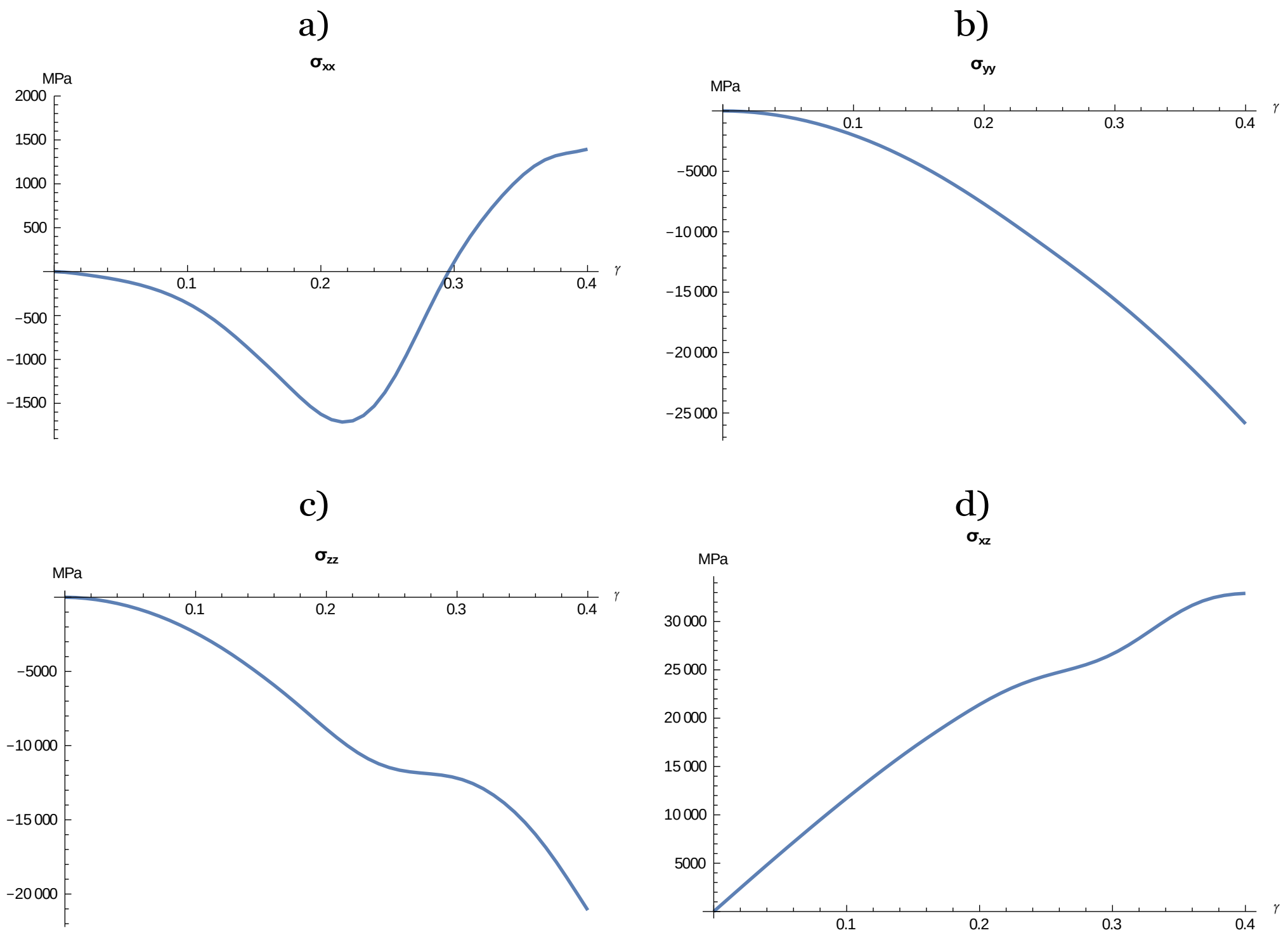

Figure 3: Simple shear in xz direction: a) $\sigma_{xx}$, b) $\sigma_{yy}$, c) $\sigma_{zz}$, d) $\sigma_{xz}$ ($\sigma_{xy}$=$\sigma_{yz}$=0)

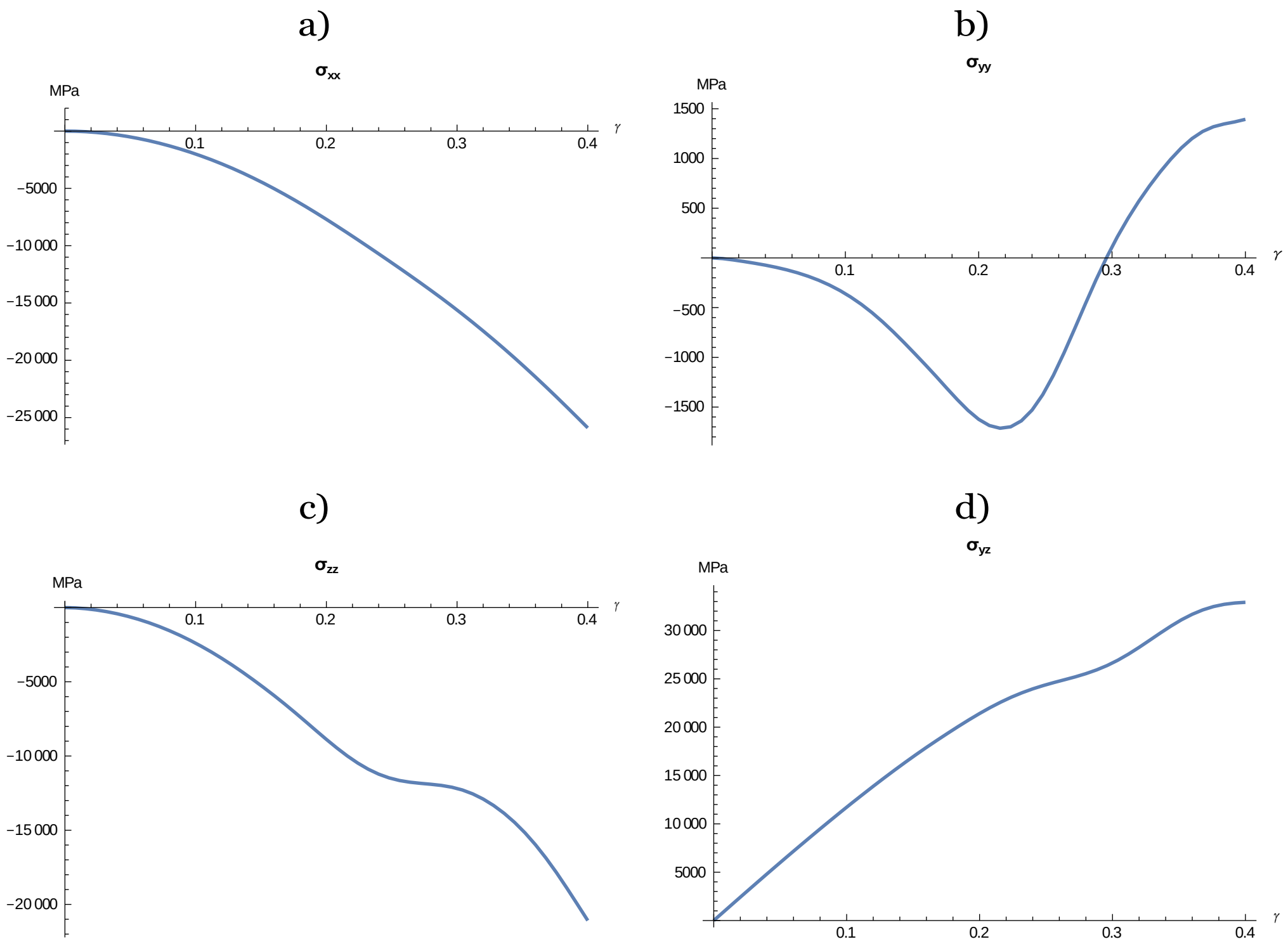

Figure 4: Simple shear in yz direction: a) $\sigma_{xx}$, b) $\sigma_{yy}$, c) $\sigma_{zz}$, d) $\sigma_{yz}$ ($\sigma_{xy}$=$\sigma_{xz}$=0)

## - NiAl oriented X=[110] Y=[-110] Z=[001]

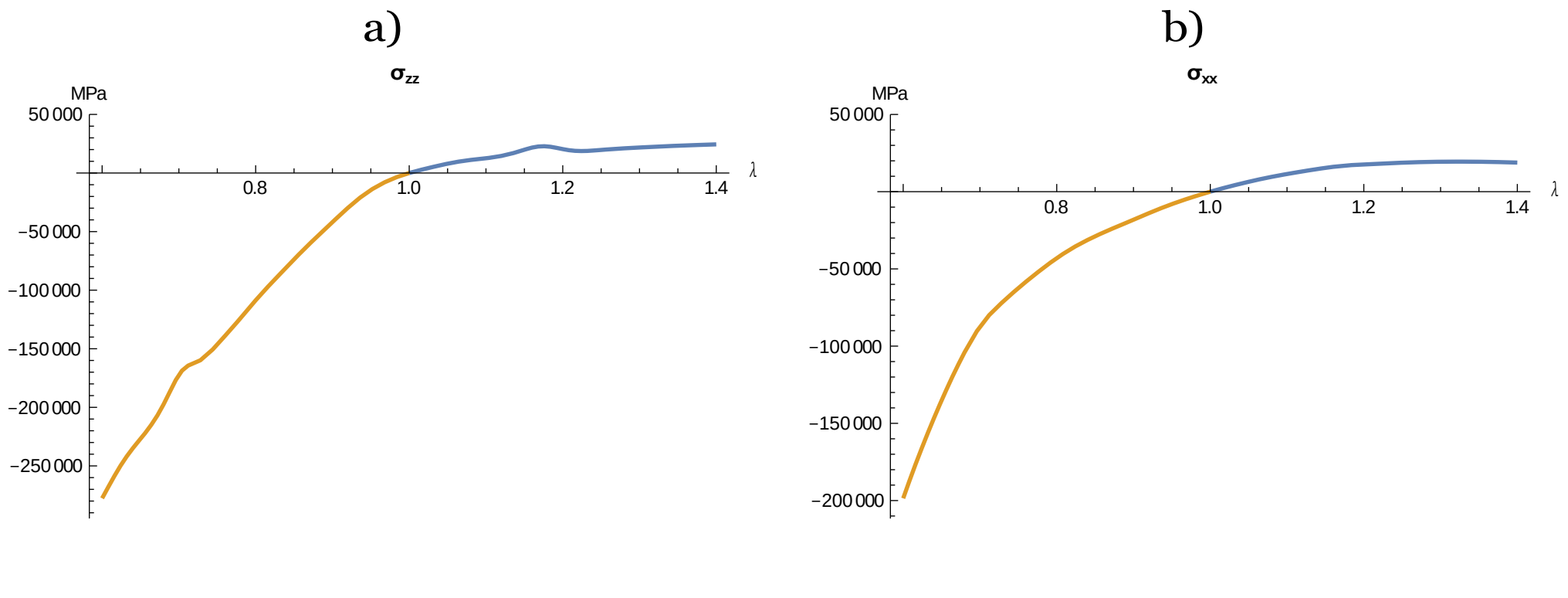

Figure 5: Uniaxial strain in z direction: a) $\sigma_{zz}$, b) $\sigma_{xx}$ ($\sigma_{xx}$=$\sigma_{yy}$, $\sigma_{xy}$=$\sigma_{xz}$=$\sigma_{yz}$=0)

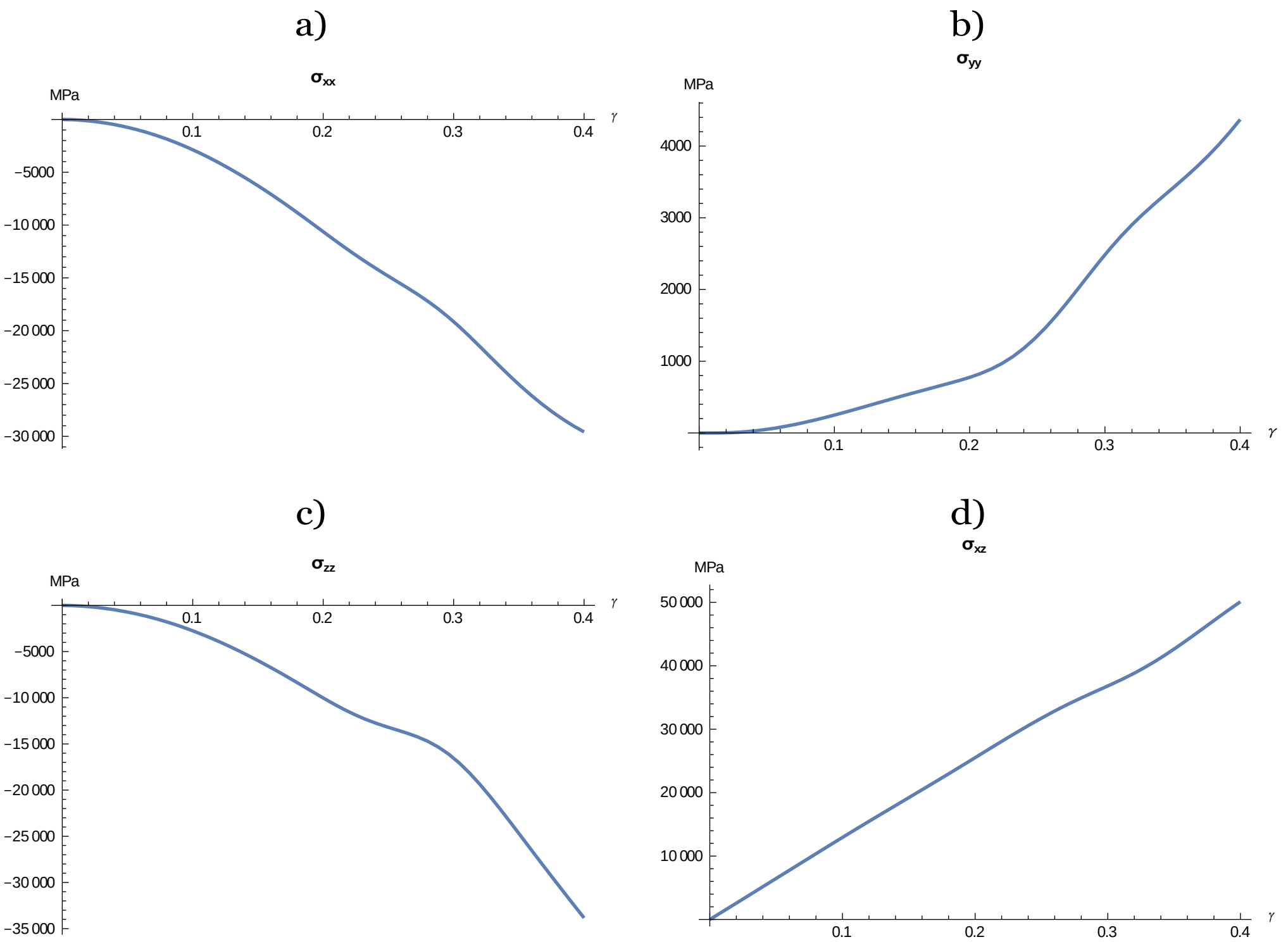

Figure 6: Simple shear in xz direction: a) $\sigma_{xx}$, b) $\sigma_{yy}$, c) $\sigma_{zz}$, d) $\sigma_{xz}$ ($\sigma_{xy}$=$\sigma_{yz}$=0)

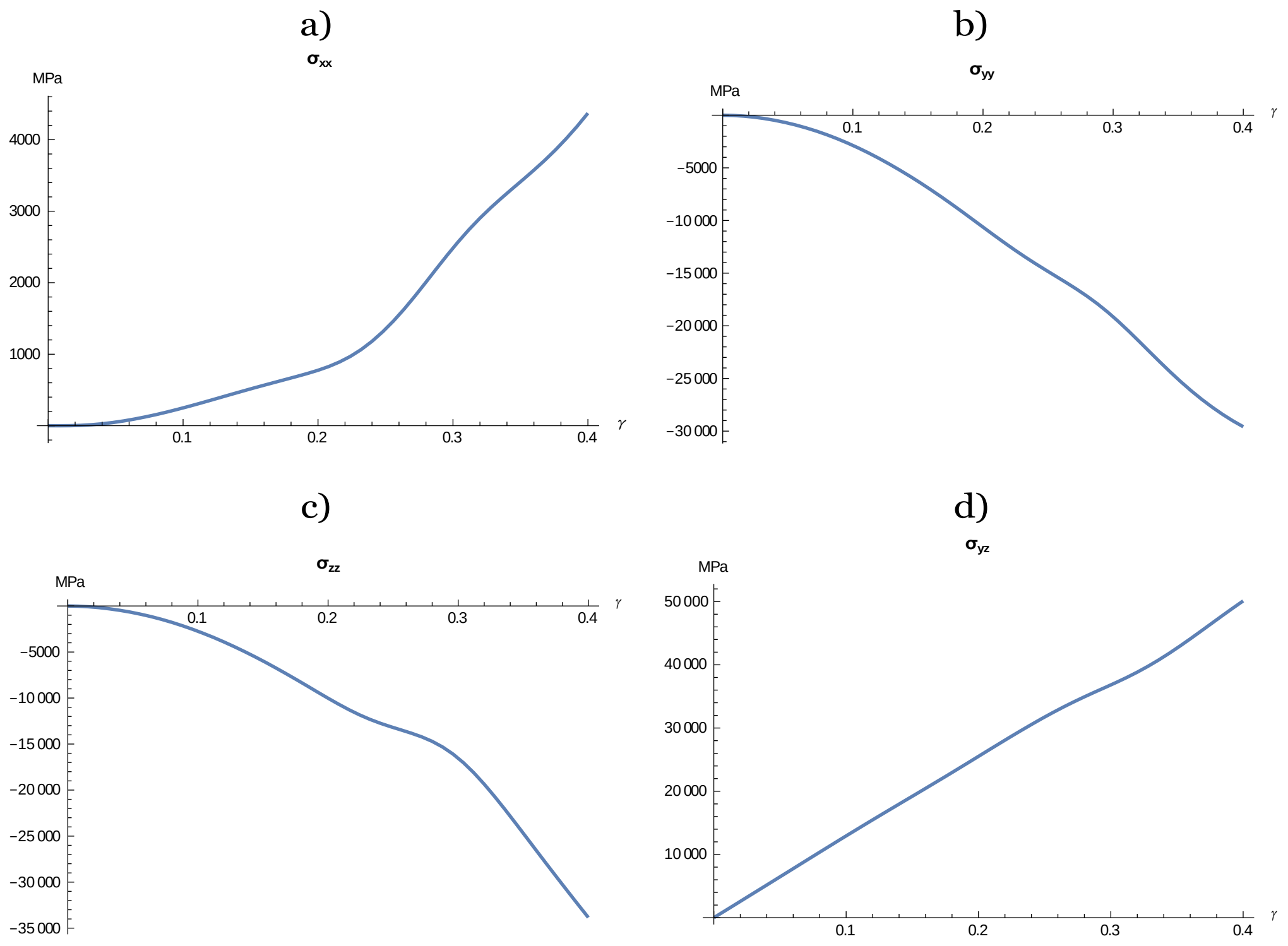

Figure 7: Simple shear in yz direction: a) $\sigma_{xx}$, b) $\sigma_{yy}$, c) $\sigma_{zz}$, d) $\sigma_{yz}$ ($\sigma_{xy}$=$\sigma_{xz}$=0)

- NiAl oriented X=[111] Y=[-1-12] Z=[1-10]

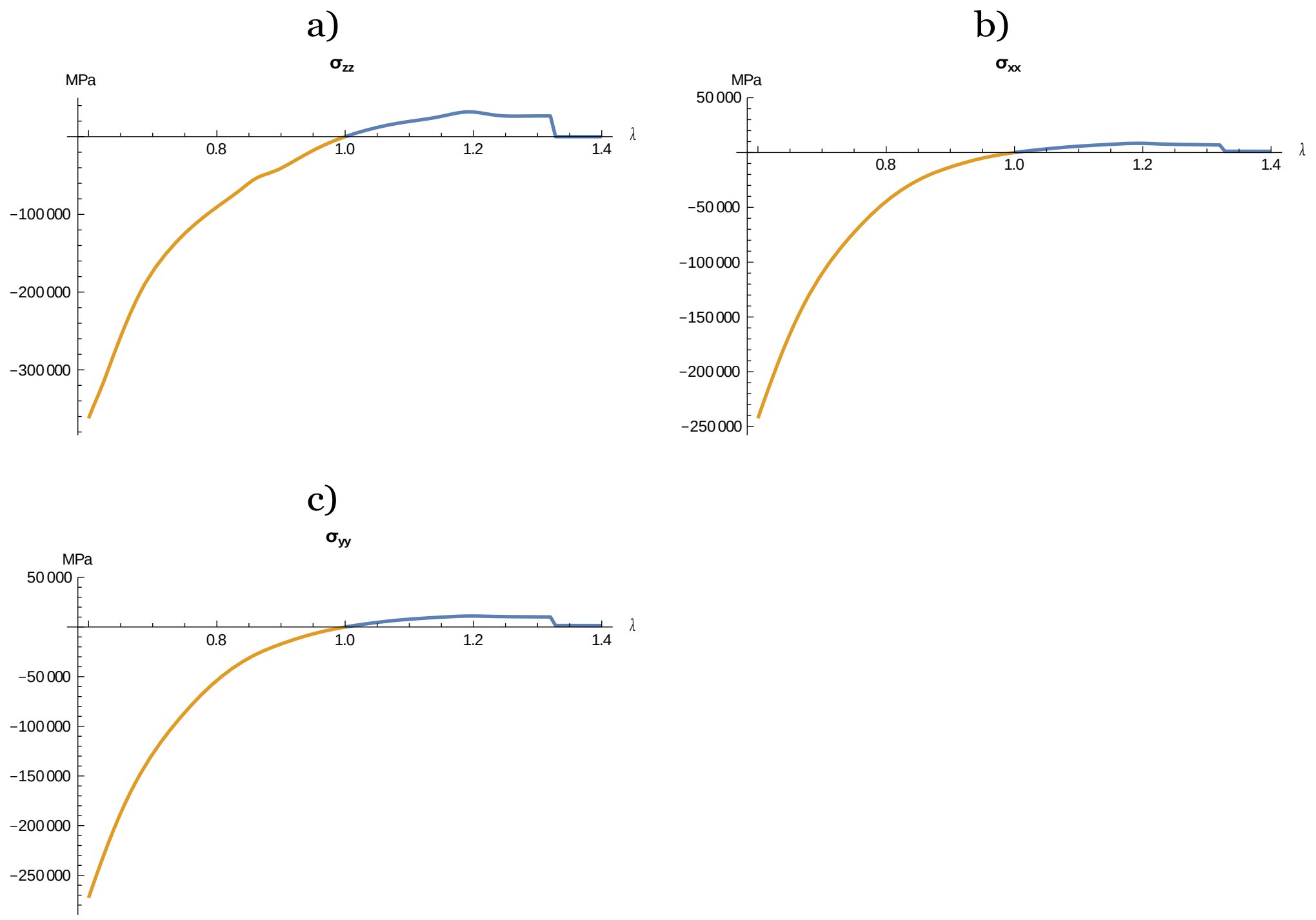

Figure 8: Uniaxial strain in z direction: a) $\sigma_{zz}$, b) $\sigma_{xx}$, c) $\sigma_{yy}$ ($\sigma_{xy}$=$\sigma_{xz}$=$\sigma_{yz}$=0)

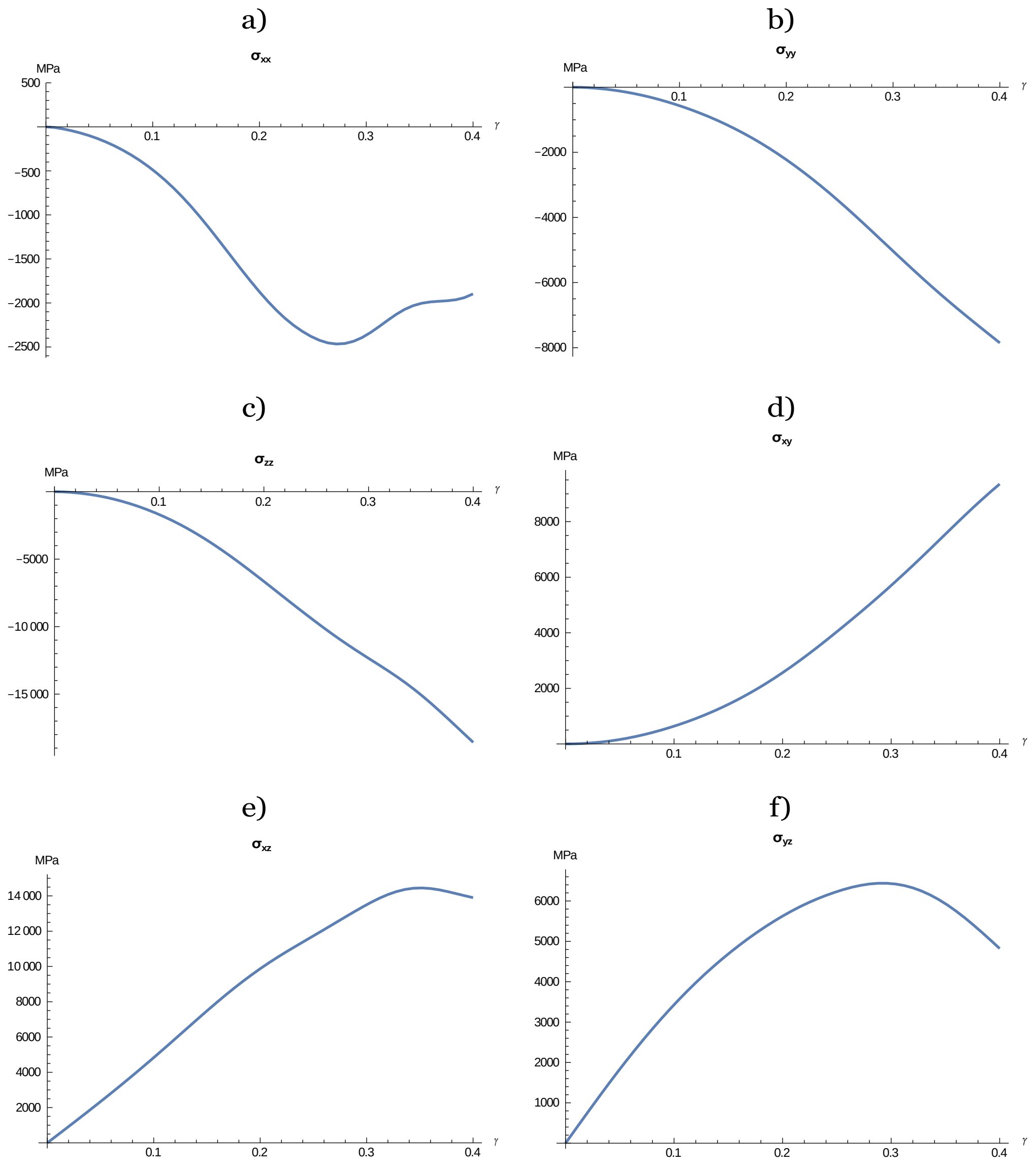

Figure 9: Simple shear in xz direction: a) $\sigma_{xx}$, b) $\sigma_{yy}$, c) $\sigma_{zz}$, d) $\sigma_{xy}$, e) $\sigma_{xz}$, f) $\sigma_{yz}$=0

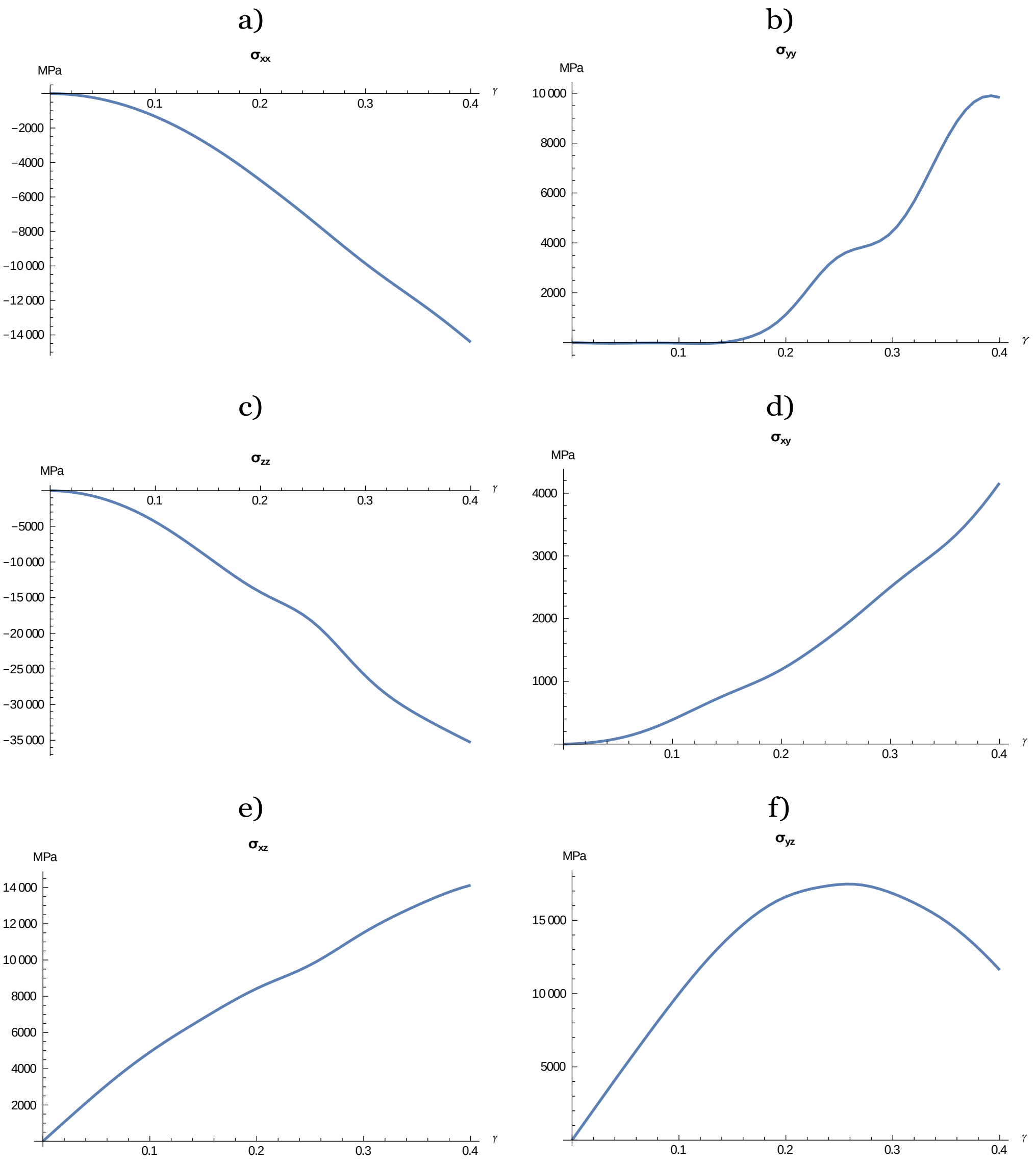

Figure 10: Simple shear in yz direction: a) $\sigma_{xx}$, b) $\sigma_{yy}$, c) $\sigma_{zz}$, d) $\sigma_{xy}$, e) $\sigma_{xz}$, f) $\sigma_{yz}$=0

- NiAl amorphous

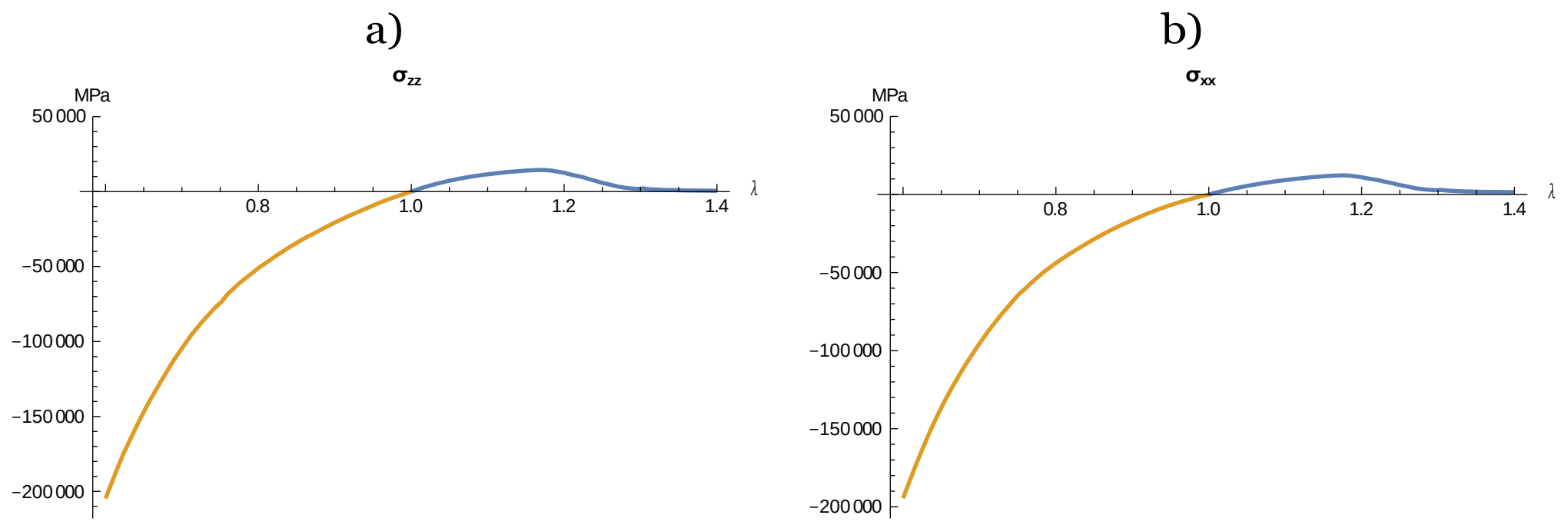

Figure 11: Uniaxial strain in z direction: a) $\sigma_{zz}$, b) $\sigma_{xx}$ ($\sigma_{xx}$=$\sigma_{yy}$, $\sigma_{xy}$=$\sigma_{xz}$=$\sigma_{yz}$≈0)

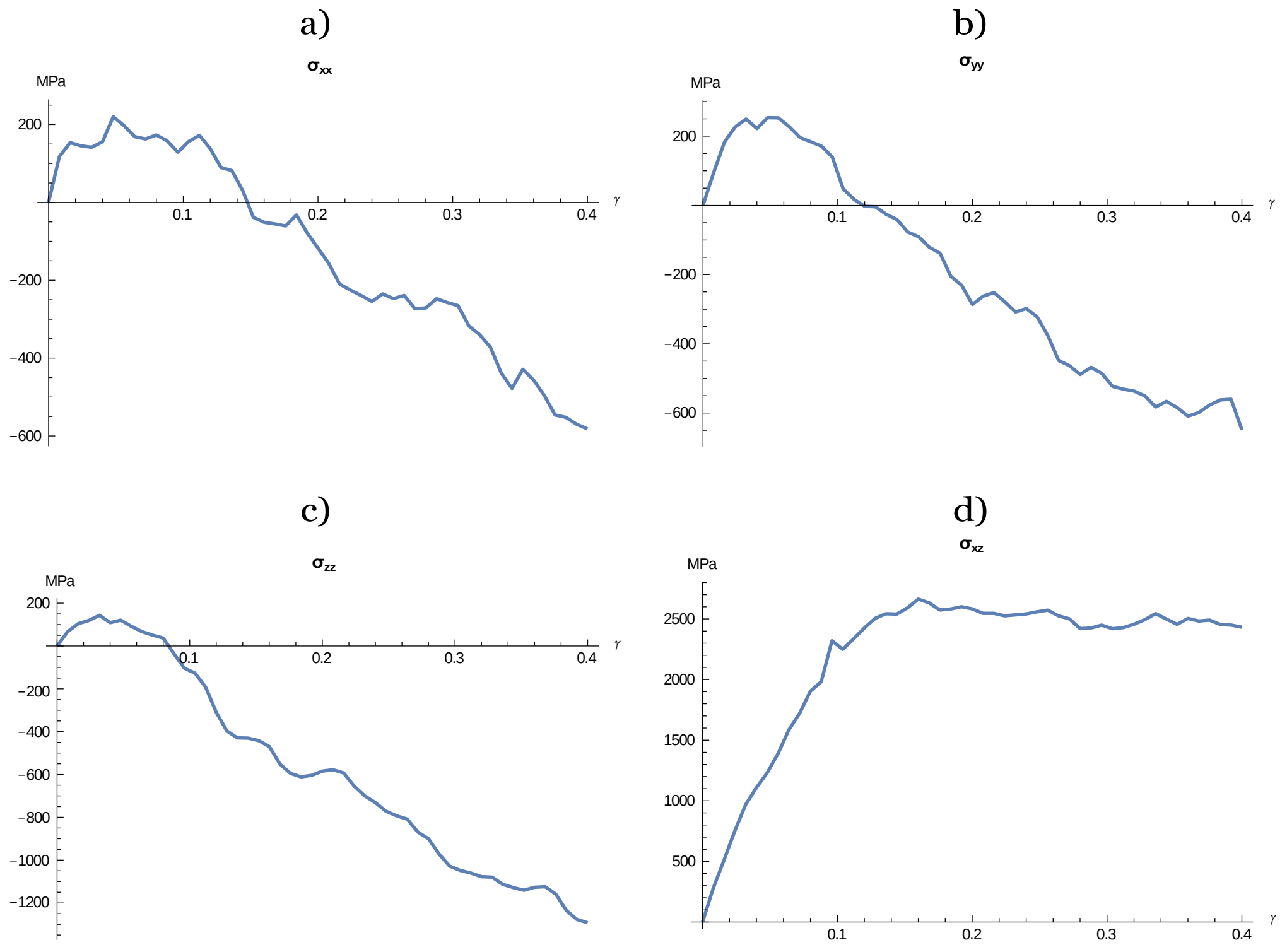

Figure 12: Simple shear in xz direction: a) $\sigma_{xx}$, b) $\sigma_{yy}$, c) $\sigma_{zz}$, d) $\sigma_{xz}$ ($\sigma_{xy}$=$\sigma_{yz}$≈0)

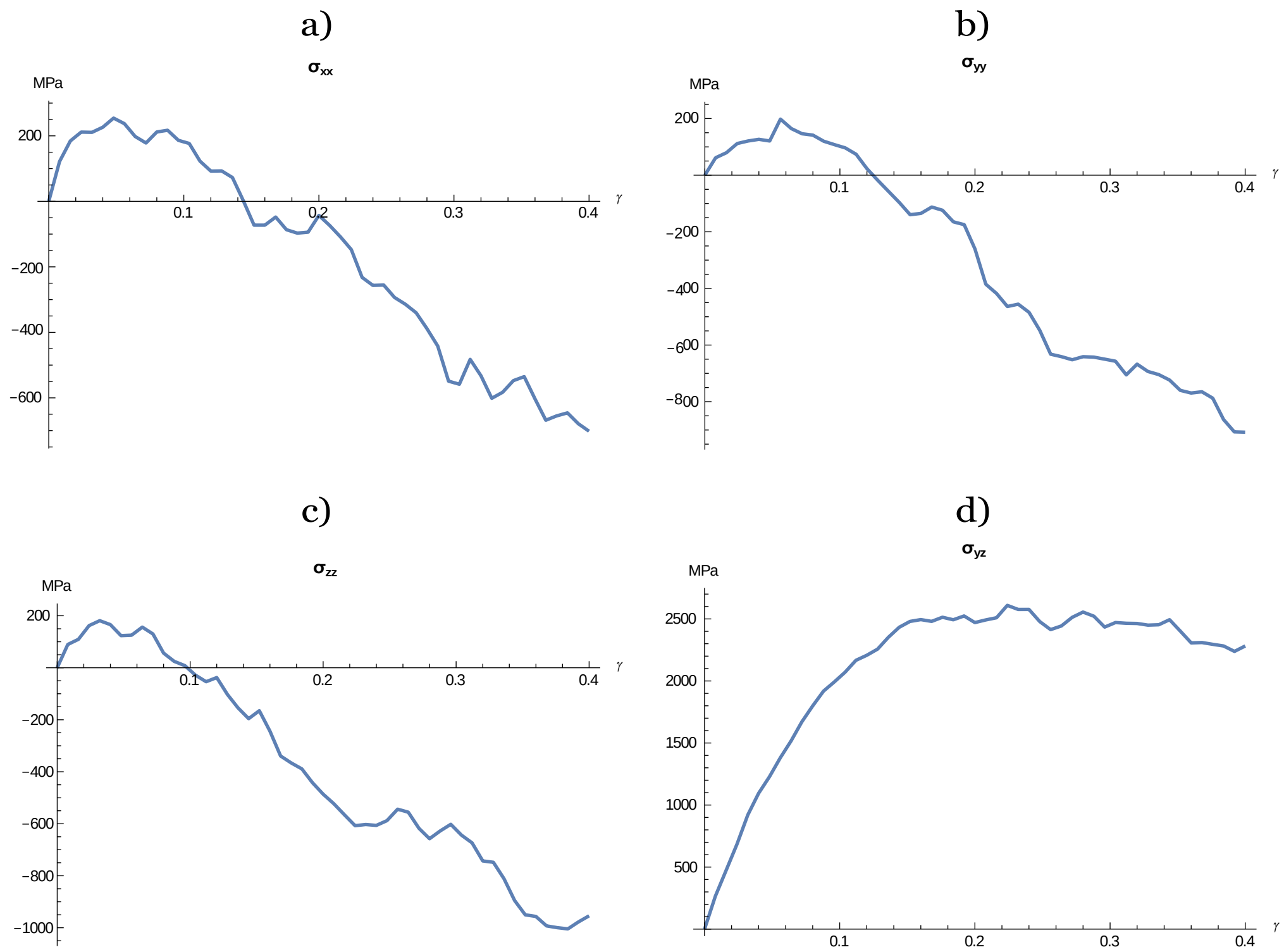

Figure 13: Simple shear in yz direction: a) $\sigma_{xx}$, b) $\sigma_{yy}$, c) $\sigma_{zz}$, d) $\sigma_{yz}$ ($\sigma_{xy}$=$\sigma_{xz}$≈0)

## *0.2. $Al_2O_3$*

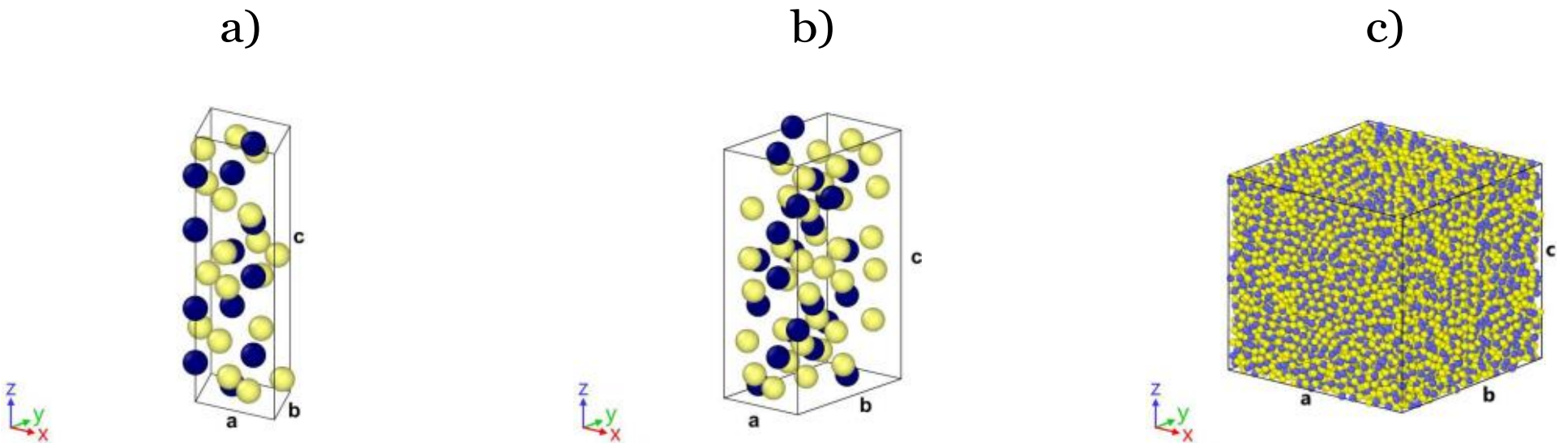

Figure 14: $Al_2O_3$: a) hexagonal, b) orthorhombic basic cell and c) amorphous (The yellow and blue atoms represent O and Al, respectively).

### 0.2.1. *Stiffness tensors*

- Stiffness tensor: $Al_2O_3$ oriented X=[100] Y=[-1 $\sqrt{3}$ $\bar{0}$] Z=[001]

$$[C_{IJ}] \to \begin{bmatrix} 540.69 & 186.42 & 77.72 & 61.09 & 0. & 0. \\ 186.42 & 540.69 & 77.72 & -61.09 & 0. & 0. \\ 77.72 & 77.72 & 445.92 & 0. & 0. & 0. \\ 61.09 & -61.09 & 0. & 96.29 & 0. & 0. \\ 0. & 0. & 0. & 0. & 96.29 & 61.09 \\ 0. & 0. & 0. & 0. & 61.09 & 177.13 \end{bmatrix} [\text{GPa}]. \tag{9}$$

$B$ = 242.15 GPa, $G$ = 131.11 GPa, $E$ = 333.20 GPa, $v$ = 0.27 and $A^U$ = 2.03.

- Stiffness tensor: $Al_2O_3$ amorphous

direct simulation result:

$$[C_{IJ}] \to \begin{bmatrix} 394.87 & 117.36 & 90.03 & -28.94 & -25.17 & -18.67 \\ 117.36 & 403.08 & 141.56 & -15.8 & -12.01 & -7.31 \\ 90.03 & 141.56 & 370.45 & -12.05 & -15.72 & -23.02 \\ -28.94 & -15.8 & -12.05 & 122.23 & 5.07 & -17.76 \\ -25.17 & -12.01 & -15.72 & 5.07 & 112.79 & -7.63 \\ -18.67 & -7.31 & -23.02 & -17.76 & -7.63 & 111.37 \end{bmatrix} [\text{GPa}]. \tag{10}$$

$B$ = 201.97 GPa, $G$ = 121.34 GPa, $E$ = 303.28 GPa, $v$ = 0.25 and $A^U$ = 0.27,

reduction to isotropy:

$$[C_{IJ}] \rightarrow \begin{bmatrix} 389.47 & 116.32 & 116.32 & 0. & 0. & 0. \\ 116.32 & 389.47 & 116.32 & 0. & 0. & 0. \\ 116.32 & 116.32 & 389.47 & 0. & 0. & 0. \\ 0. & 0. & 0. & 115.46 & 0. & 0. \\ 0. & 0. & 0. & 0. & 115.46 & 0. \\ 0. & 0. & 0. & 0. & 0. & 115.46 \end{bmatrix} [\text{GPa}]. \tag{11}$$

$B = 207.37$ GPa, $G = 123.49$ GPa, $E = 309.11$ GPa, $v = 0.25$ and $A^U = 0.0339$.

### *0.2.2. Deformation-stress*

– $Al_2O_3$ oriented $X=[100]$ $Y=[-1\sqrt{3}0]$ $Z=[001]$

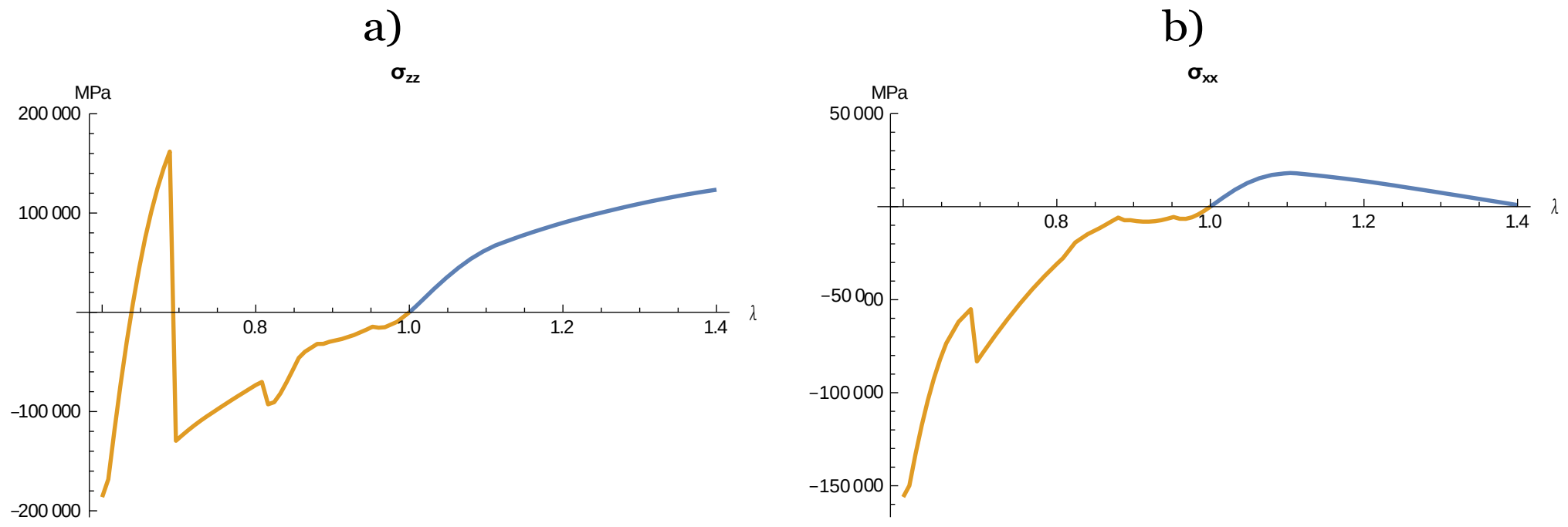

Figure 15: Uniaxial strain in z direction: a) $\sigma_{zz}$, b) $\sigma_{xx}$ ($\sigma_{xx}$=$\sigma_{yy}$, $\sigma_{xy}$=$\sigma_{xz}$=$\sigma_{yz}$=0)

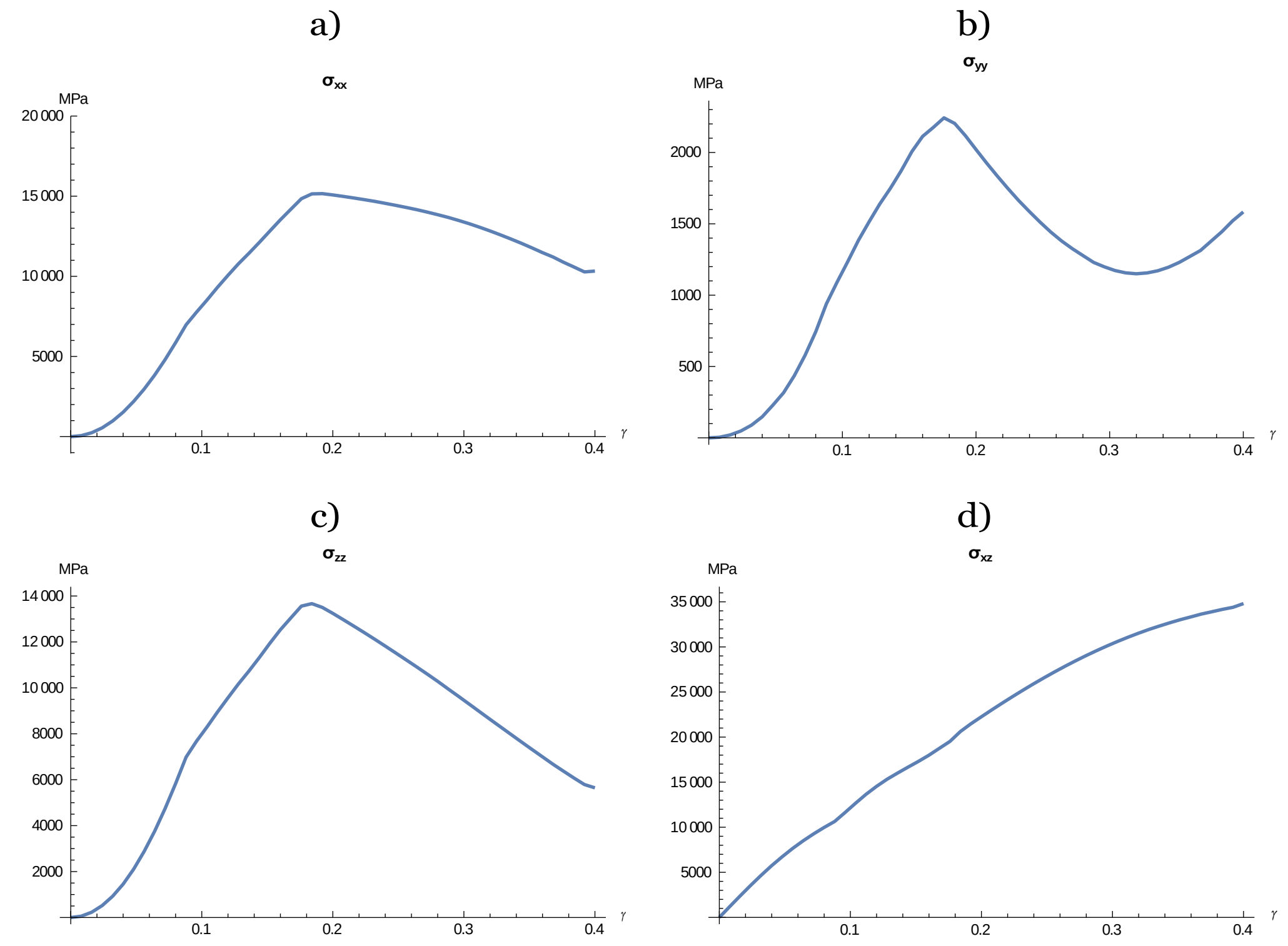

Figure 16: Simple shear in xz direction: a) $\sigma_{xx}$, b) $\sigma_{yy}$, c) $\sigma_{zz}$, d) $\sigma_{xz}$ ($\sigma_{xy}$=$\sigma_{yz}$=0)

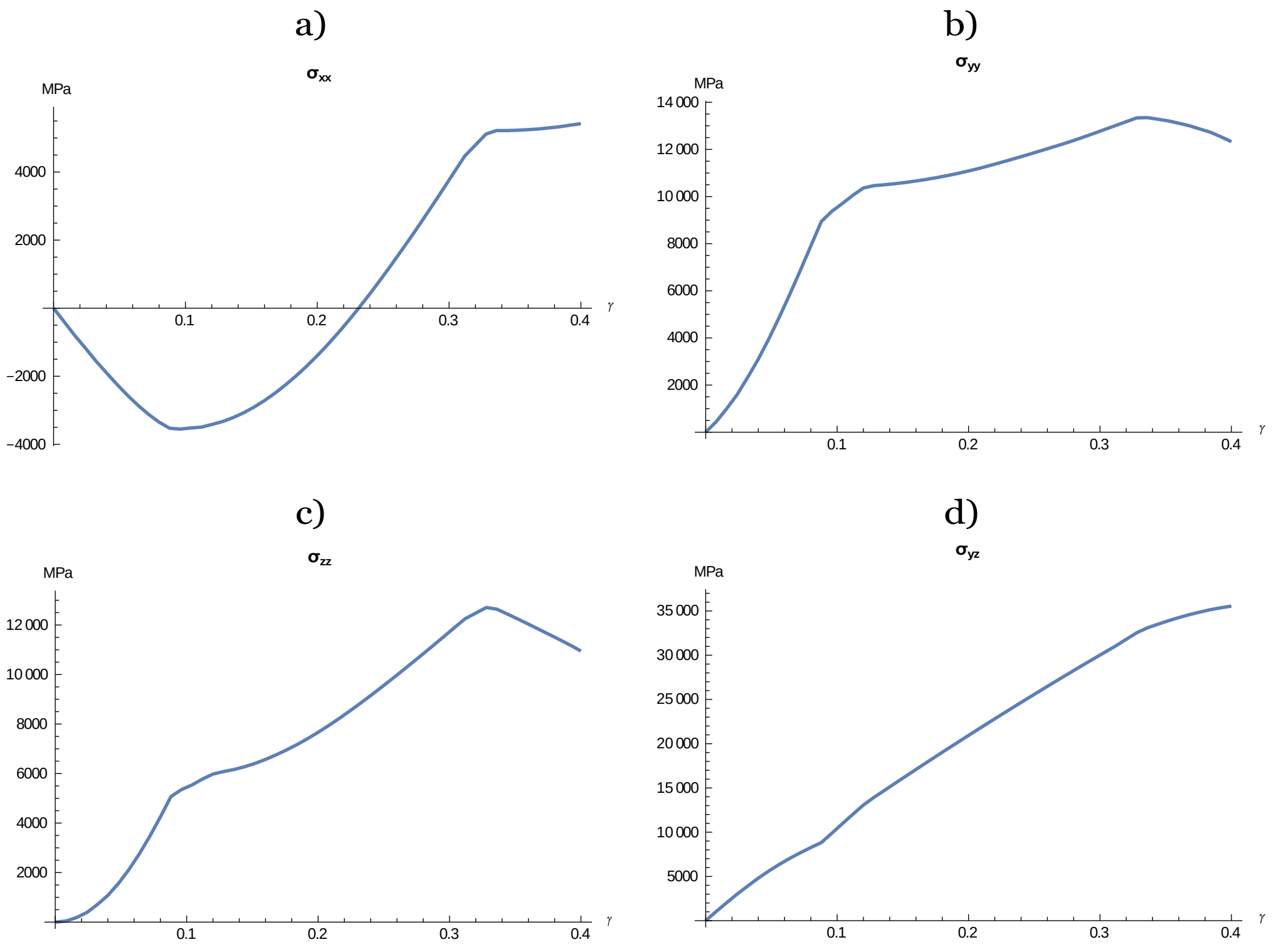

Figure 17: Simple shear in yz direction: a) $\sigma_{xx}$, b) $\sigma_{yy}$, c) $\sigma_{zz}$, d) $\sigma_{yz}$ ($\sigma_{xy}$=$\sigma_{xz}$=0)

## – $Al_2O_3$ amorphous

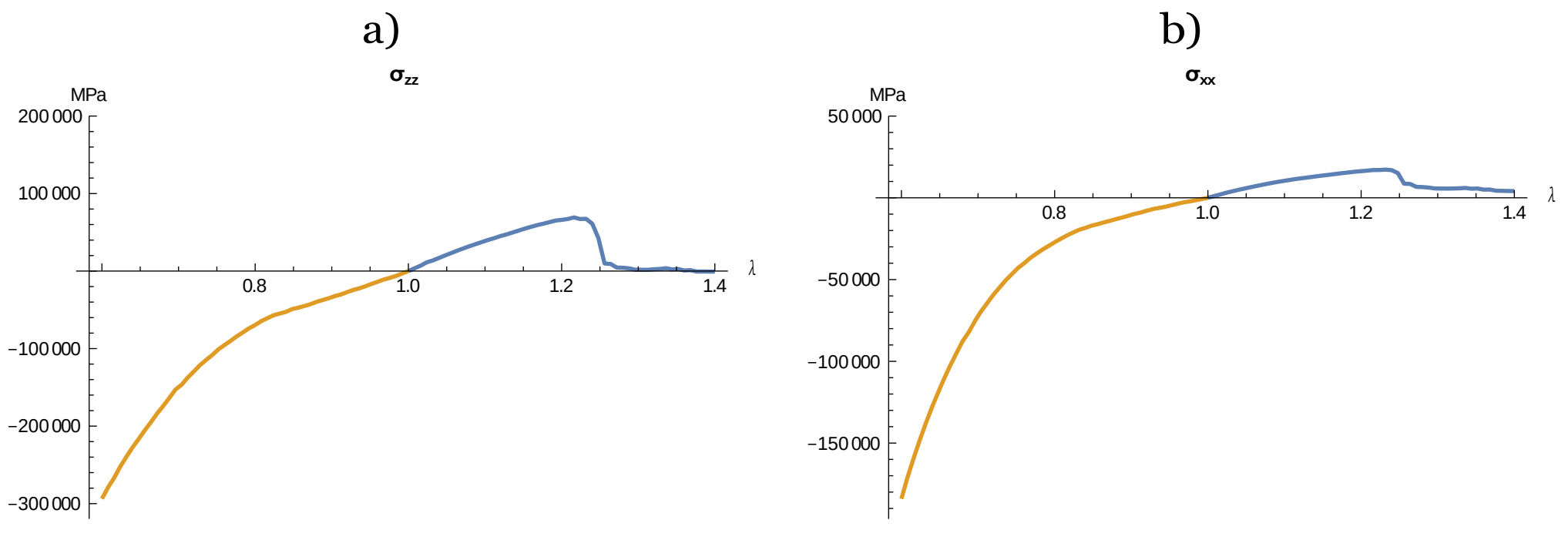

Figure 18: Uniaxial strain in z direction: a) $\sigma_{zz}$, b) $\sigma_{xx}$ ($\sigma_{xx}$=$\sigma_{yy}$, $\sigma_{xy}$=$\sigma_{xz}$=$\sigma_{yz}\approx$0)

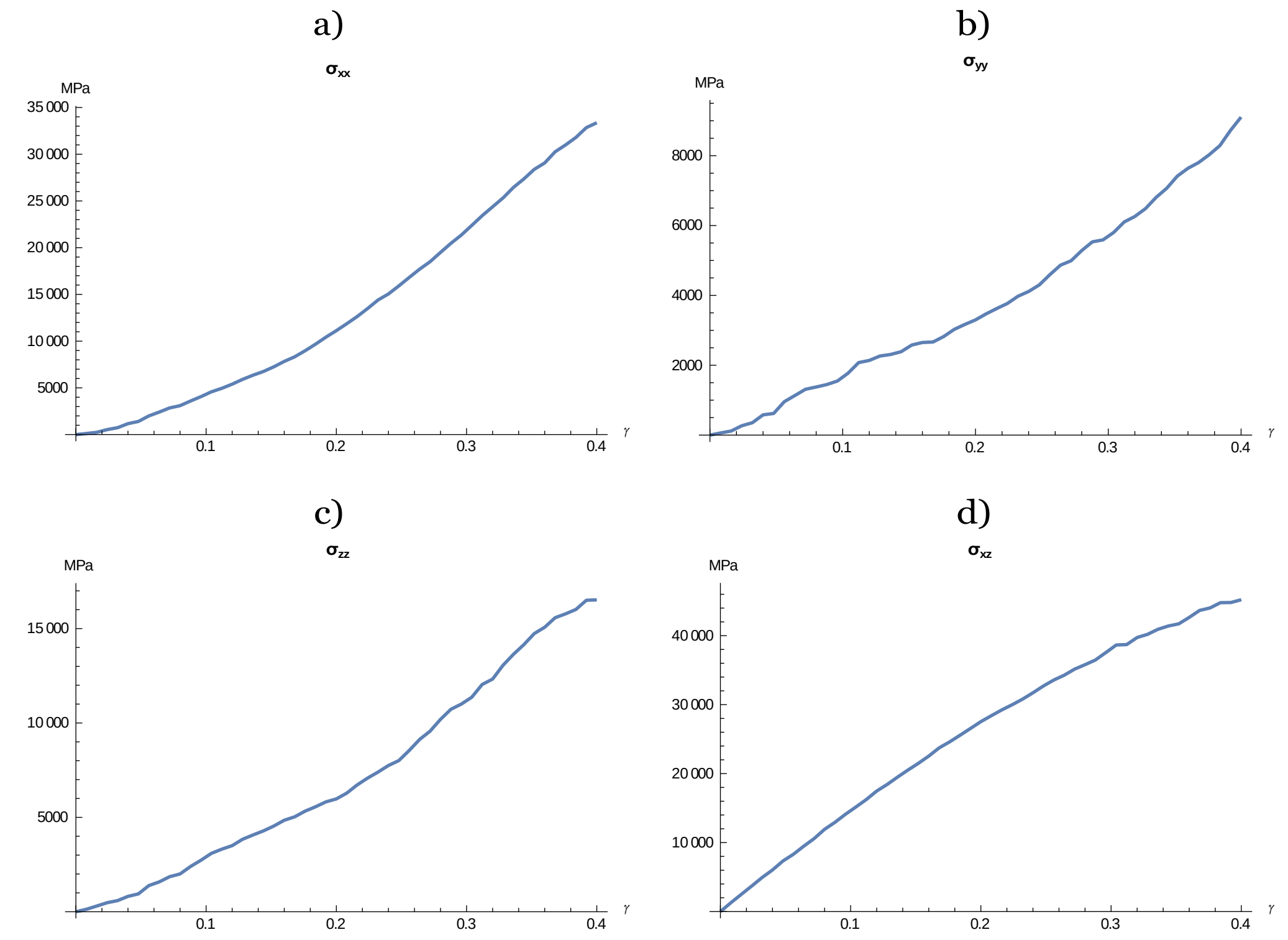

Figure 19: Simple shear in xz direction: a) $\sigma_{xx}$, b) $\sigma_{yy}$, c) $\sigma_{zz}$, d) $\sigma_{xz}$ ($\sigma_{xy}$=$\sigma_{yz}$≈0)

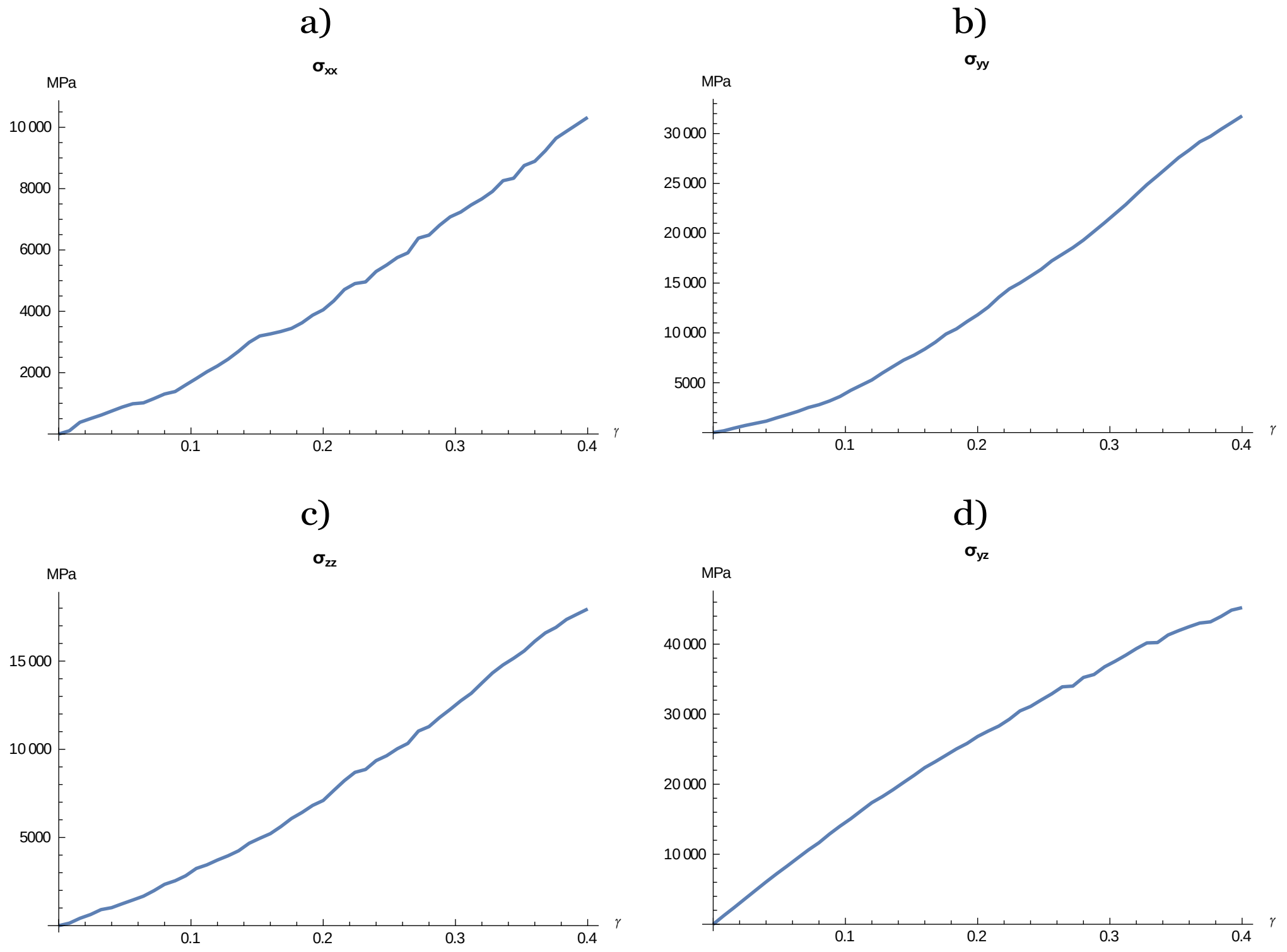

Figure 20: Simple shear in yz direction: a) $\sigma_{xx}$, b) $\sigma_{yy}$, c) $\sigma_{zz}$, d) $\sigma_{yz}$ ($\sigma_{xy}$=$\sigma_{xz}$≈0)

## 0.3. $Al_2O_3$-NiAl

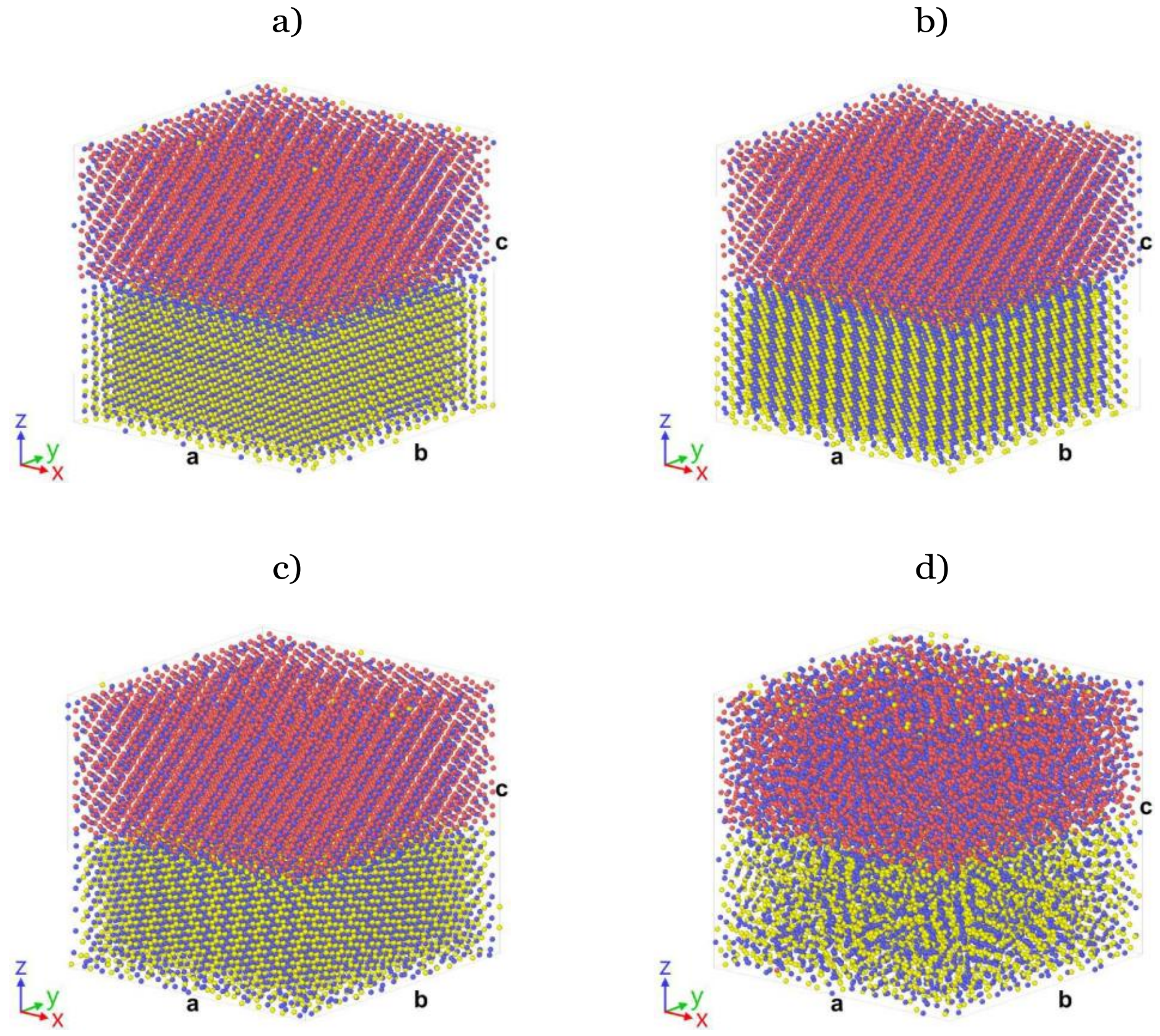

Figure 21: $Al_2O_3$-NiAl: a) 12×7×4 orthorhombic $Al_2O_3$ basic cells and 20×20×18 NiAl basic cells X=[100] Y=[010] Z=[001], b) 12×7×4 orthorhombic $Al_2O_3$ basic cells and 14×14×18 NiAl basic cells X=[110] Y=[-110] Z=[001], c) 12×7×4 orthorhombic $Al_2O_3$ basic cells and 11×8×13 NiAl basic cells X=[111] Y=[-1-12] Z=[1-10], d) $Al_2O_3$ amorphous and NiAl amorphous (The red, yellow and blue atoms represent Ni, O and Al, respectively).

*0.3.1. Stiffness tensors*

- Stiffness tensor: 12×7×4 orthorhombic $Al_2O_3$ basic cells and 20×20×18 NiAl basic cells X=[100] Y=[010] Z=[001]

$$[C_{IJ}] \rightarrow \begin{bmatrix} 296.57 & 144.76 & 125.5 & -35.27 & -2.5 & 3.45 \\ 144.76 & 273.54 & 74.42 & 17.96 & -4.93 & 1.37 \\ 125.5 & 74.42 & 169.18 & -39.37 & -18.81 & 9.45 \\ -35.27 & 17.96 & -39.37 & 110.56 & 0.02 & 0.17 \\ -2.5 & -4.93 & -18.81 & 0.02 & 113.03 & -31.15 \\ 3.45 & 1.37 & 9.45 & 0.17 & -31.15 & 112.41 \end{bmatrix} [\text{GPa}]. \quad (12)$$

$B$ = 145.19 GPa, $G$ = 84.19 GPa, $E$ = 211.66 GPa, $v$ = 0.26 and $A^U$ = 1.45.

- Stiffness tensor: 12×7×4 orthorhombic $Al_2O_3$ basic cells and 14×14×18 NiAl basic cells X=[110] Y=[-110] Z=[001]

$$[C_{IJ}] \rightarrow \begin{bmatrix} 328.94 & 115.36 & 108.47 & -72.04 & -7.95 & -15.51 \\ 115.36 & 335.65 & 121.99 & 35.07 & -0.6 & 4.53 \\ 108.47 & 121.99 & 167.61 & -82.61 & -15.04 & -33.39 \\ -72.04 & 35.07 & -82.61 & 125.71 & -4.94 & 4.32 \\ -7.95 & -0.6 & -15.04 & -4.94 & 112.47 & -30.68 \\ -15.51 & 4.53 & -33.39 & 4.32 & -30.68 & 69.87 \end{bmatrix} [\text{GPa}]. \quad (13)$$

$B$ = 110.58 GPa, $G$ = 59.51 GPa, $E$ = 151.38 GPa, $v$ = 0.27 and $A^U$ = 1.61.

- Stiffness tensor: 12×7×4 orthorhombic $Al_2O_3$ basic cells and

11×8×13 NiAl basic cells X=[111] Y=[-1-12] Z=[1-10]

$$[C_{IJ}] \rightarrow \begin{bmatrix} 336.98 & 126.06 & 127.63 & -31.82 & -17.25 & 15.94 \\ 126.06 & 342.81 & 114.04 & 21.32 & 5.29 & -23.55 \\ 127.63 & 114.04 & 273.92 & -12.22 & 12.38 & 10.75 \\ -31.82 & 21.32 & -12.22 & 99.41 & 30.38 & -6.19 \\ -17.25 & 5.29 & 12.38 & 30.38 & 60.32 & -18.76 \\ 15.94 & -23.55 & 10.75 & -6.19 & -18.76 & 63.29 \end{bmatrix} [\text{GPa}]. \quad (14)$$

$B$ = 185.99 GPa, $G$ = 74.19 GPa, $E$ = 196.45 GPa, $v$ = 0.32 and $A^U$ = 1.48.

- Stiffness tensor: $Al_2O_3$ amorphous and NiAl amorphous direct simulation result:

$$[C_{IJ}] \rightarrow \begin{bmatrix} 306.42 & 101.08 & 53.46 & -5.47 & -7.41 & -1.79 \\ 101.08 & 314.69 & 115.43 & 6.84 & -9.35 & 2.66 \\ 53.46 & 115.43 & 264.65 & 9.78 & -29.48 & 1.81 \\ -5.47 & 6.84 & 9.78 & 59.12 & 3.86 & -5.97 \\ -7.41 & -9.35 & -29.48 & 3.86 & 98.99 & -0.61 \\ -1.79 & 2.66 & 1.81 & -5.97 & -0.61 & 101.55 \end{bmatrix} [\text{GPa}]. \quad (15)$$

$B$ = 155.11 GPa, $G$ = 89.40 GPa, $E$ = 224.97 GPa, $v$ = 0.26 and $A^U$ = 0.46,

reduction to isotropy:

$$[C_{IJ}] \to \begin{bmatrix} 295.253 & 89.99 & 89.99 & 0. & 0. & 0. \\ 89.99 & 295.253 & 89.99 & 0. & 0. & 0. \\ 89.99 & 89.99 & 295.253 & 0. & 0. & 0. \\ 0. & 0. & 0. & 86.5533 & 0. & 0. \\ 0. & 0. & 0. & 0. & 86.5533 & 0. \\ 0. & 0. & 0. & 0. & 0. & 86.5533 \end{bmatrix} [\mathrm{GPa}]. \quad (16)$$

$B$ = 158.41 GPa, $G$ = 92.66 GPa, $E$ = 232.63 GPa, $v$ = 0.26 and $A^U$ = 0.035.

- $Al_2O_3$-NiAl 12×7×4 orthorhombic $Al_2O_3$ basic cells and 20×20×18 NiAl basic cells X=[100] Y=[010] Z=[001]

a) b)

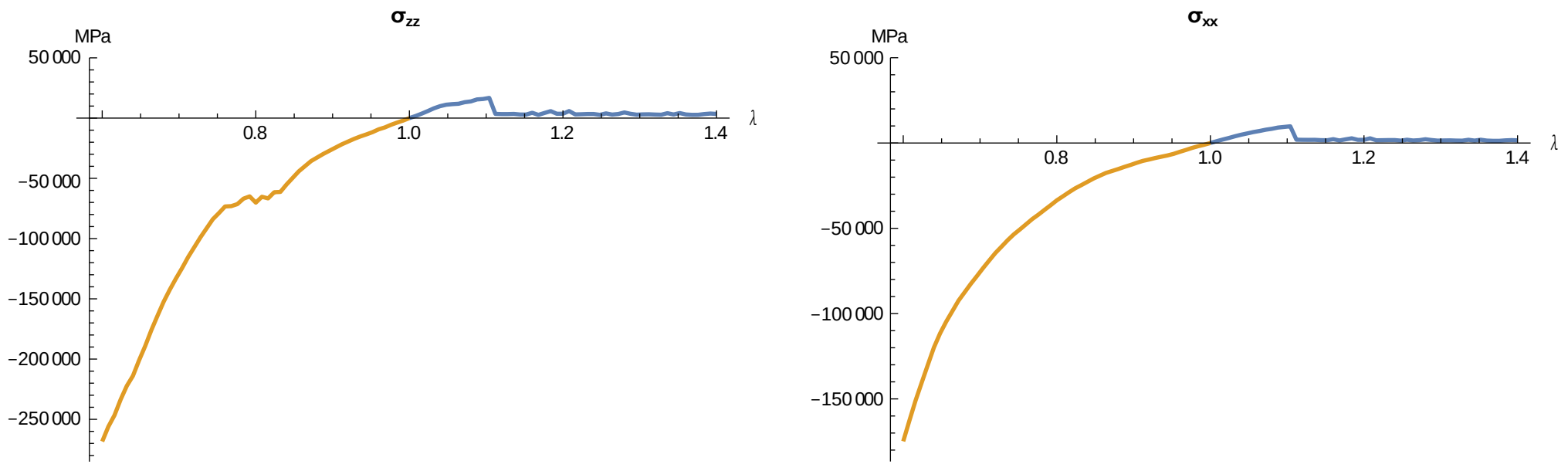

Figure 22: Uniaxial strain in z direction: a) $\sigma_{zz}$, b) $\sigma_{xx}$ ($\sigma_{xx}$=$\sigma_{yy}$, $\sigma_{xy}$=$\sigma_{xz}$=$\sigma_{yz}$≈0)

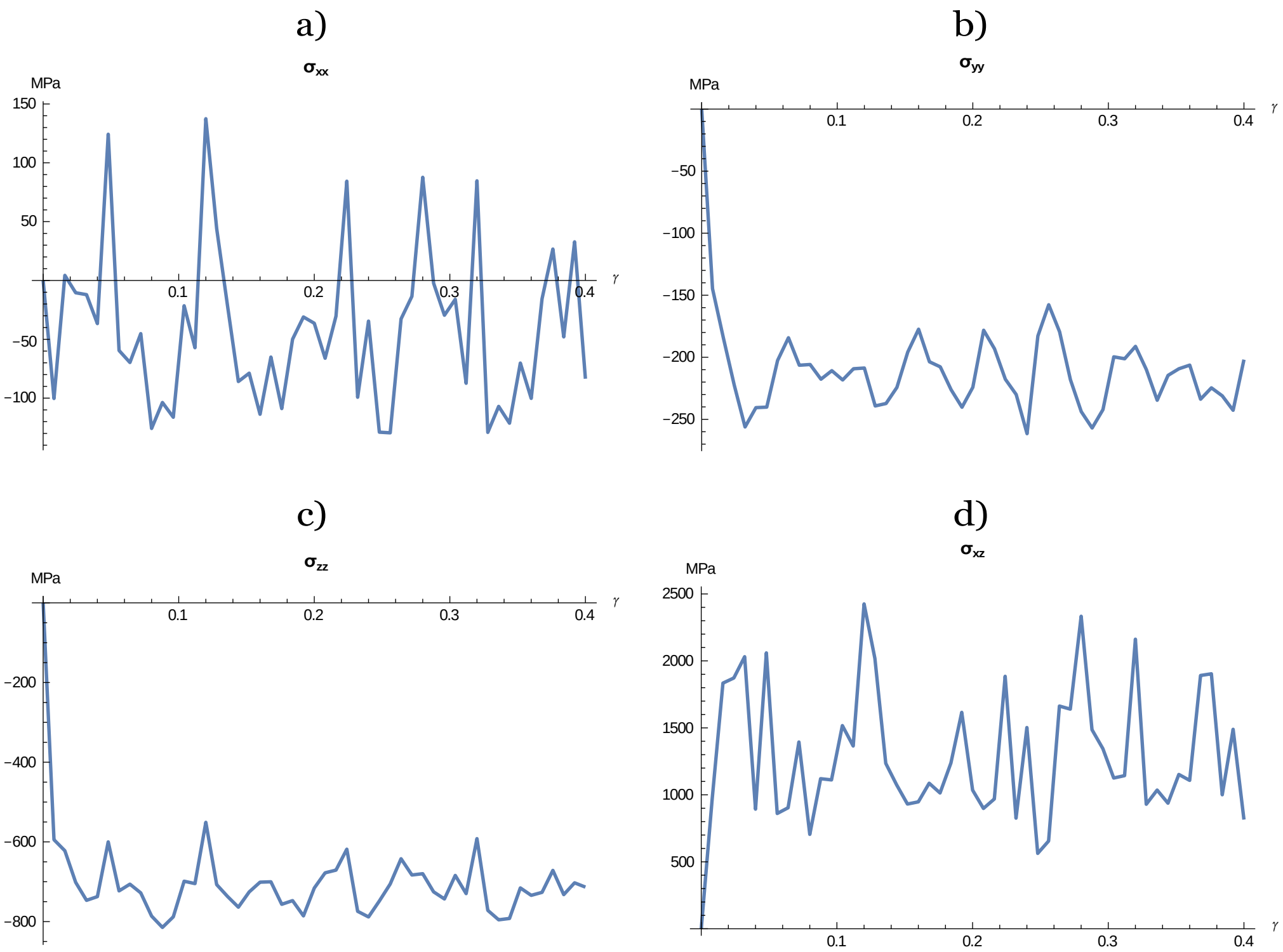

Figure 23: Simple shear in xz direction: a) $\sigma_{xx}$, b) $\sigma_{yy}$, c) $\sigma_{zz}$, d) $\sigma_{xz}$ ($\sigma_{xy}$=$\sigma_{yz}$≈0)

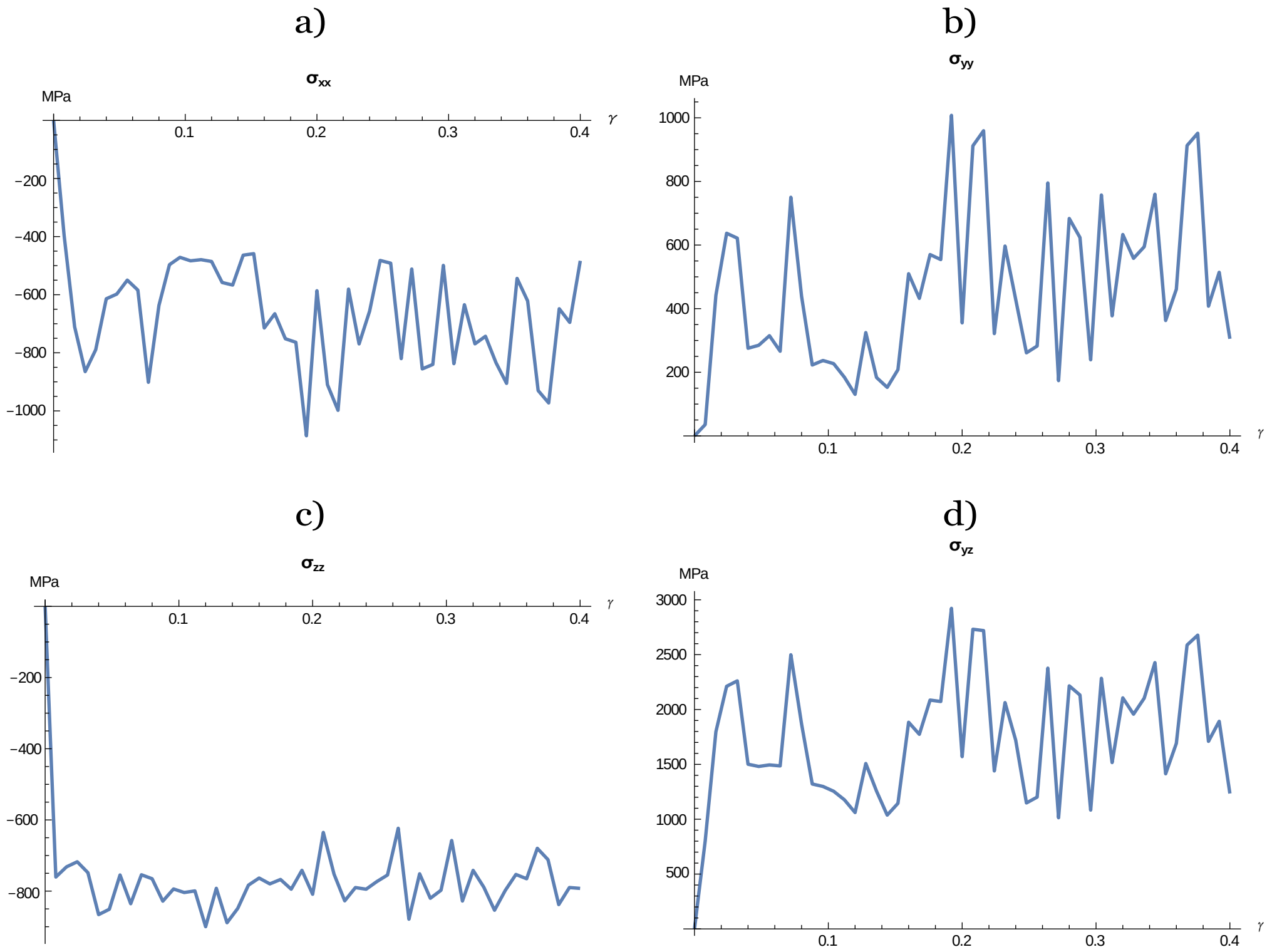

Figure 24: Simple shear in yz direction: a) $\sigma_{xx}$, b) $\sigma_{yy}$, c) $\sigma_{zz}$, d) $\sigma_{yz}$ ($\sigma_{xy}$=$\sigma_{xz}$≈0)

- $Al_2O_3$-NiAl 12×7×4 orthorhombic $Al_2O_3$ basic cells and 14×14×18 NiAl basic cells X=[110] Y=[-110] Z=[001]

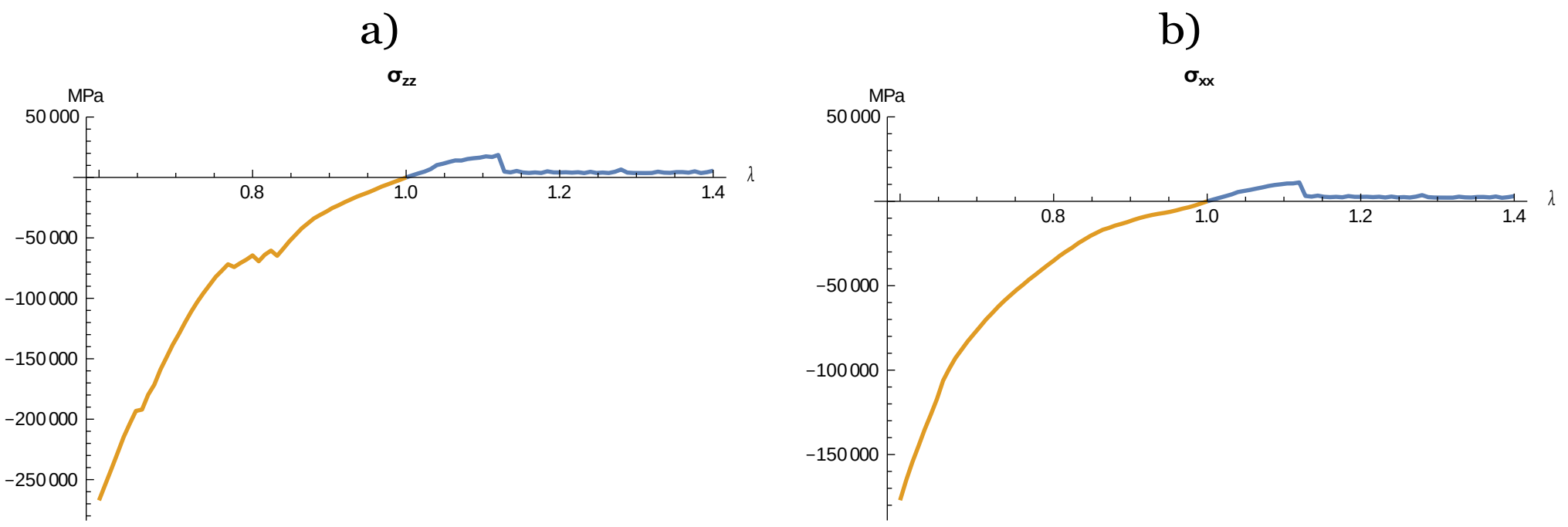

Figure 25: Uniaxial strain in z direction: a) $\sigma_{zz}$, b) $\sigma_{xx}$ ($\sigma_{xx}$=$\sigma_{yy}$, $\sigma_{xy}$=$\sigma_{xz}$=$\sigma_{yz}$≈0)

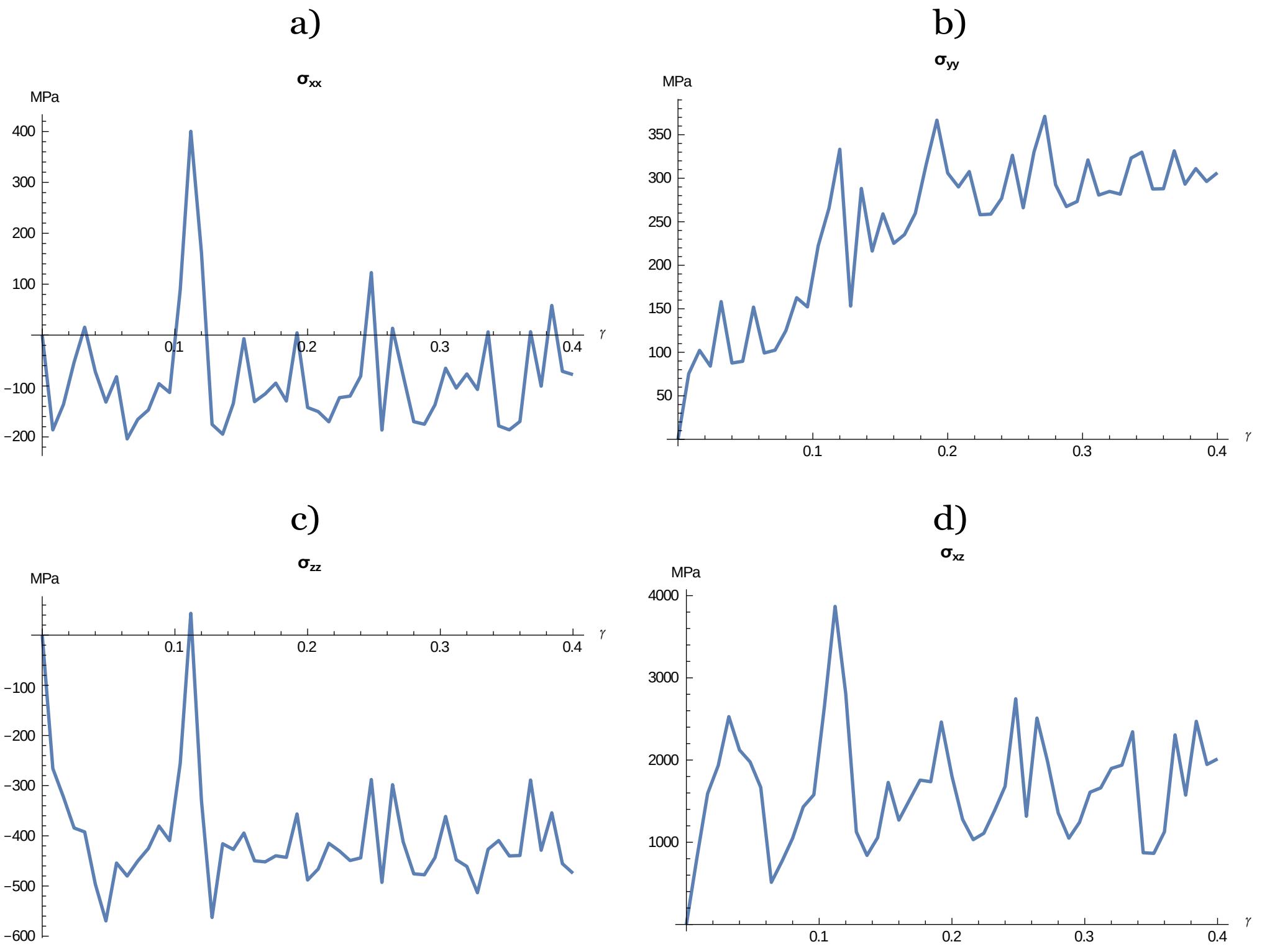

Figure 26: Simple shear in xz direction: a) $\sigma_{xx}$, b) $\sigma_{yy}$, c) $\sigma_{zz}$, d) $\sigma_{xz}$ ($\sigma_{xy}$=$\sigma_{yz}$≈0)

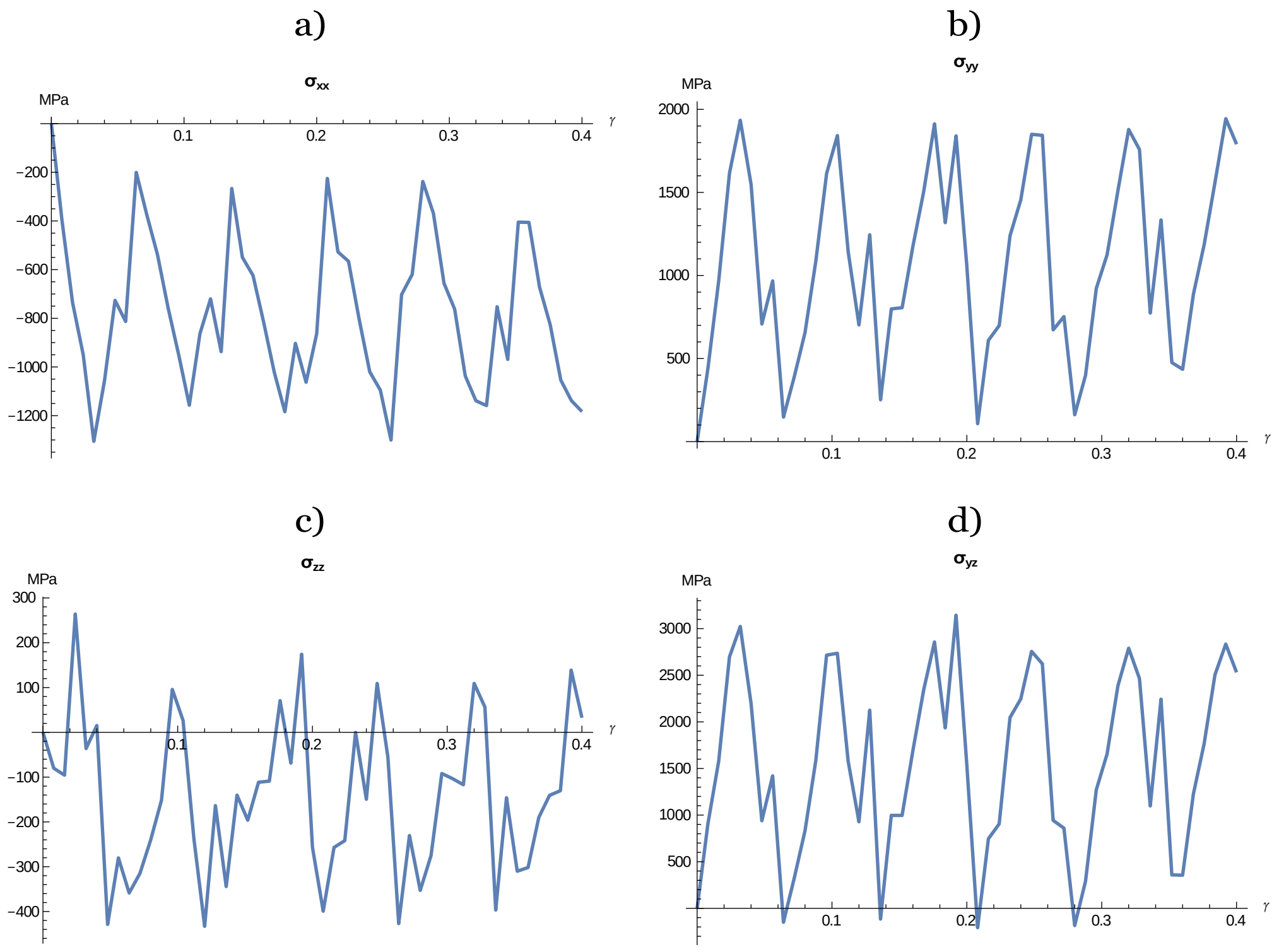

Figure 27: Simple shear in yz direction: a) $\sigma_{xx}$, b) $\sigma_{yy}$, c) $\sigma_{zz}$, d) $\sigma_{yz}$ ($\sigma_{xy}$=$\sigma_{xz}$≈0)

- $Al_2O_3$-NiAl 12×7×4 orthorhombic $Al_2O_3$ basic cells and 11×8×13 NiAl basic cells X=[111] Y=[-1-12] Z=[1-10]

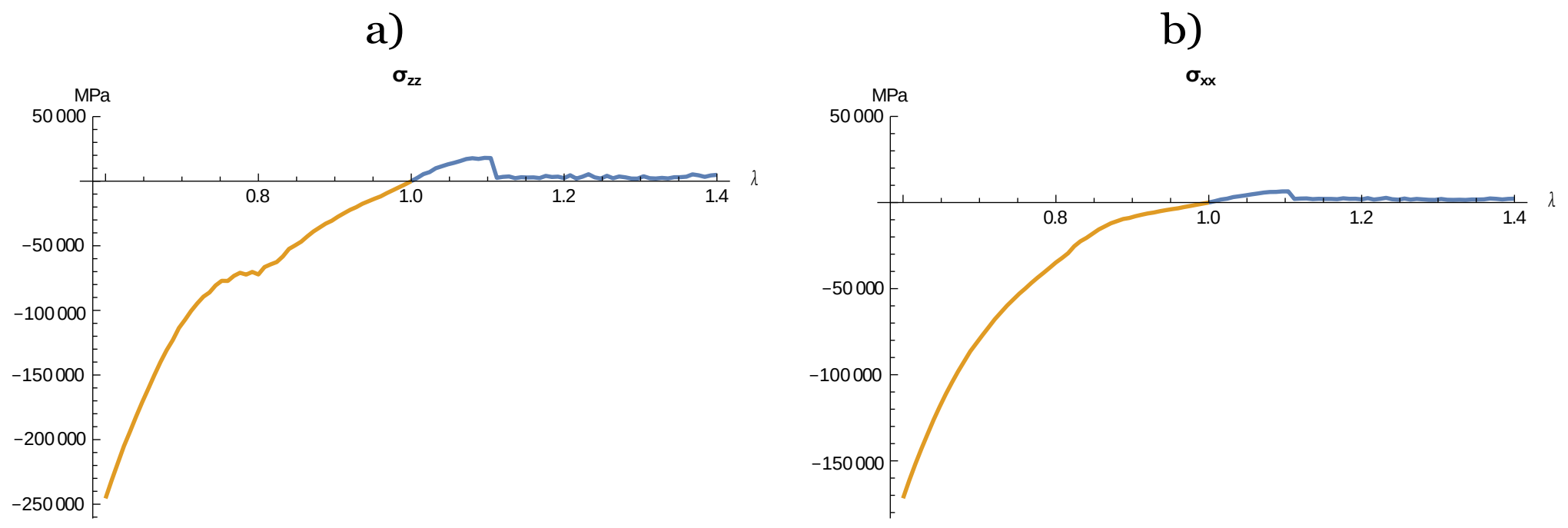

Figure 28: Uniaxial strain in z direction: a) $\sigma_{zz}$, b) $\sigma_{xx}$ ($\sigma_{xx}$=$\sigma_{yy}$, $\sigma_{xy}$=$\sigma_{xz}$=$\sigma_{yz}$≈0)

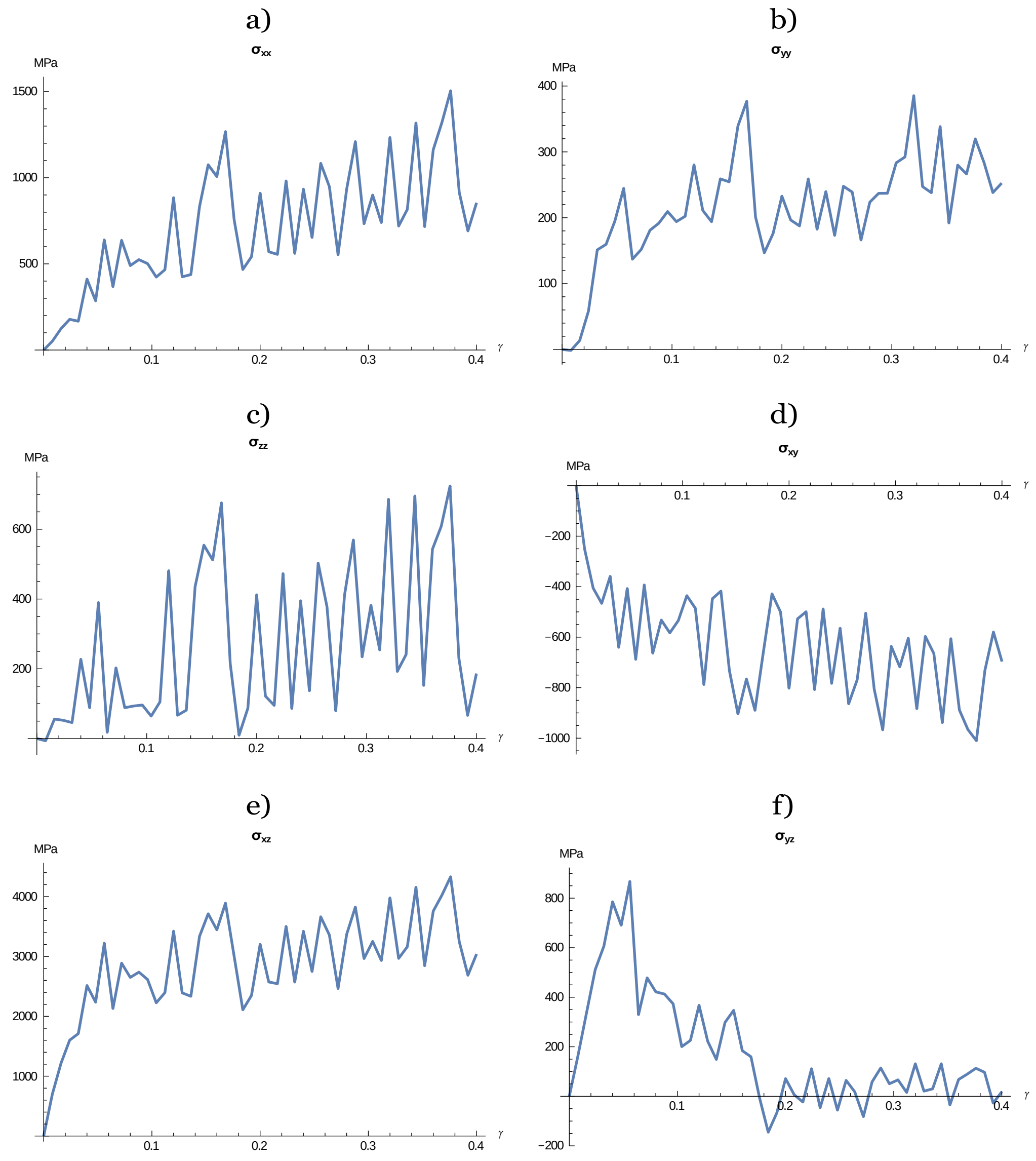

Figure 29: Simple shear in xz direction: a) $\sigma_{xx}$, b) $\sigma_{yy}$, c) $\sigma_{zz}$, d) $\sigma_{xy}$, e) $\sigma_{xz}$, f) $\sigma_{yz}$

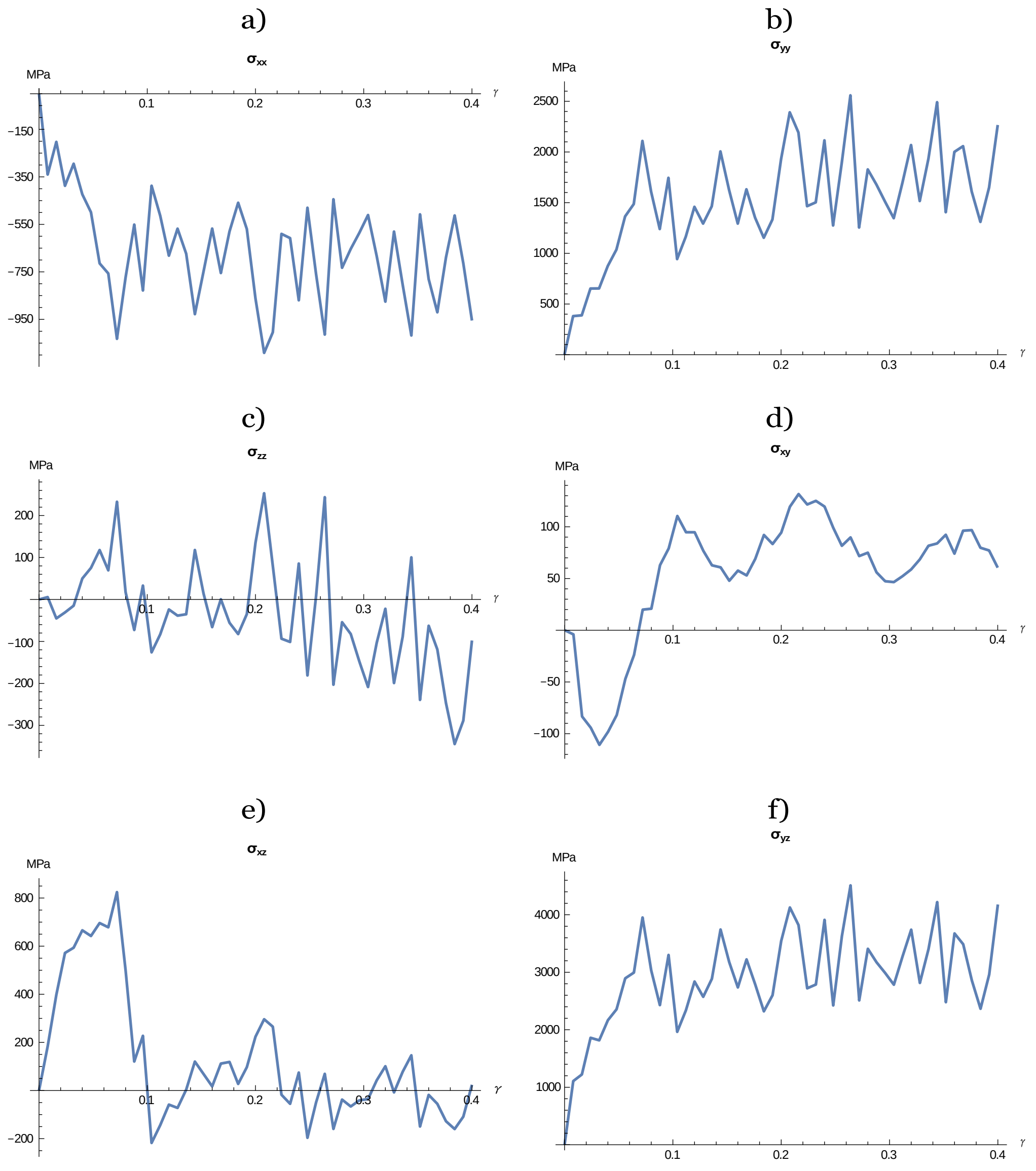

Figure 30: Simple shear in yz direction: a) $\sigma_{xx}$, b) $\sigma_{yy}$, c) $\sigma_{zz}$, d) $\sigma_{xy}$, e) $\sigma_{xz}$, f) $\sigma_{yz}$

- $Al_2O_3$-NiAl $Al_2O_3$ amorphous and NiAl amorphous

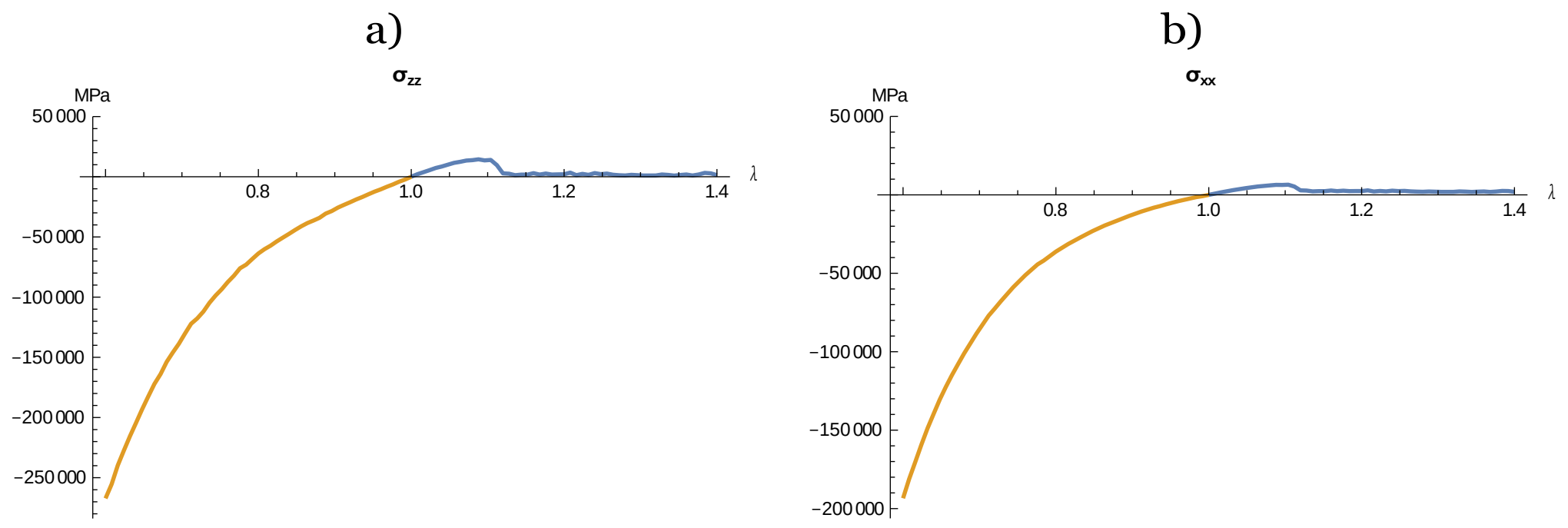

Figure 31: Uniaxial strain in z direction: a) $\sigma_{zz}$, b) $\sigma_{xx}$ ($\sigma_{xx}$=$\sigma_{yy}$, $\sigma_{xy}$=$\sigma_{xz}$=$\sigma_{yz}$≈0)

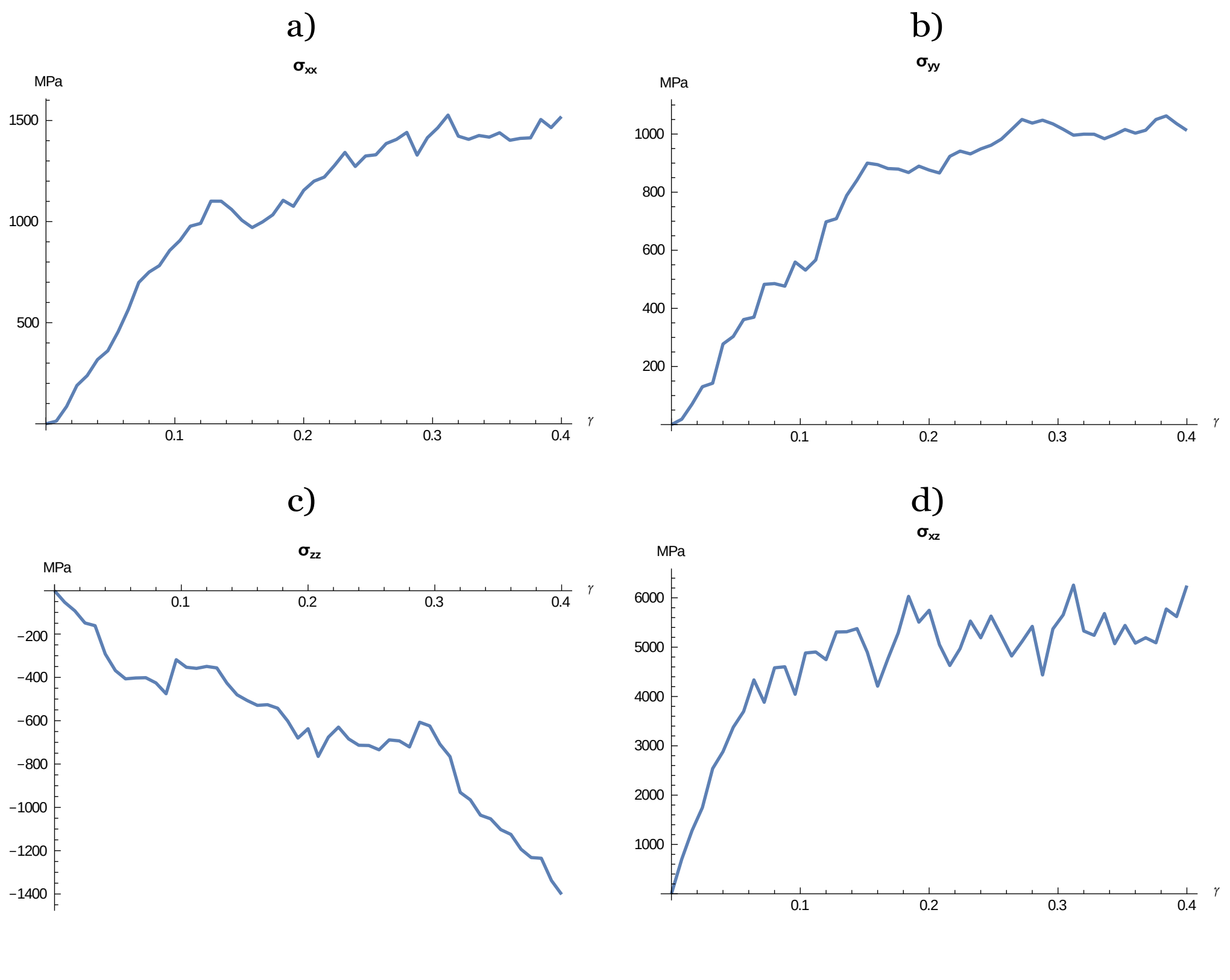

Figure 32: Simple shear in xz direction: a) $\sigma_{xx}$, b) $\sigma_{yy}$, c) $\sigma_{zz}$, d) $\sigma_{xz}$ ($\sigma_{xy}$=$\sigma_{yz}$≈0)

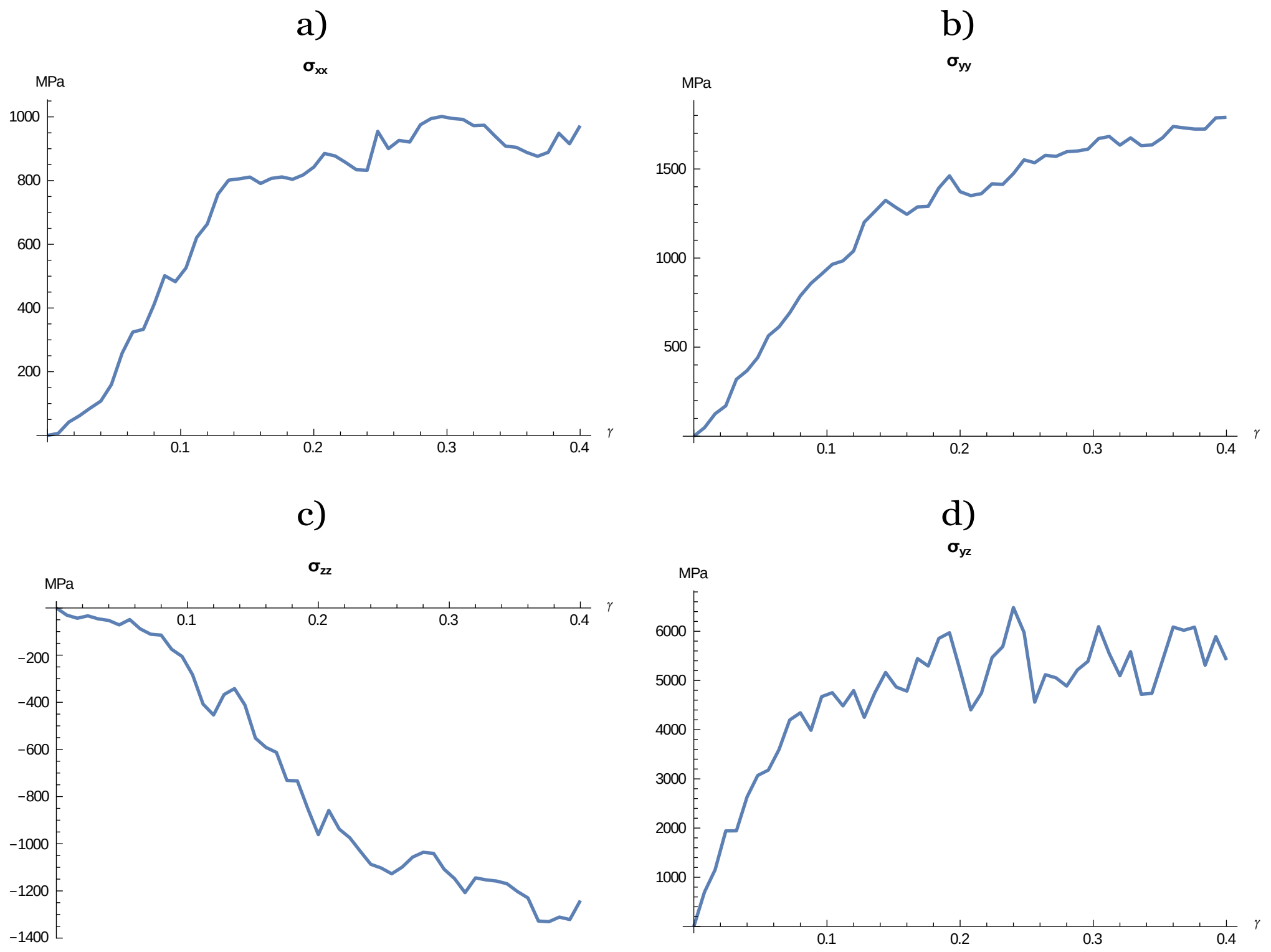

Figure 33: Simple shear in yz direction: a) $\sigma_{xx}$, b) $\sigma_{yy}$, c) $\sigma_{zz}$, d) $\sigma_{yz}$ ($\sigma_{xy}$=$\sigma_{xz}$≈0)